# Energy Efficiency in Wireless Sensor Networks

A thesis submitted in fulfilment of the requirements for the degree of
Doctor of Philosophy in the Faculty of Engineering and Information Technology at
The University of Technology Sydney

Najmeh Kamyab Pour

Supervised by
Professor Doan B. Hoang

December 2015

CERTIFICATE OF ORIGINAL AUTHORSHIP

I certify that the work in this thesis has not previously been submitted for a degree nor has it been submitted as part of requirements for a degree except as fully acknowledged within the text.

I also certify that the thesis has been written by me. Any help that I have received in my research work and the preparation of the thesis itself has been acknowledged. In addition, I certify that all information sources and literature used are indicated in the thesis.

Signature of Student:

Date:

# Abstract

Wireless sensor networks (WSNs), as distributed networks of sensors with the ability to sense, process and communicate, have been increasingly used in various fields including engineering, health and environment, to intelligently monitor remote locations at low cost. Sensors (a.k.a nodes) in such networks are responsible for four major tasks: data aggregation, sending and receiving data, and in-network data processing. This implies that they must effectively utilise their resources, including memory usage, CPU power and, more importantly, energy, to increase their lifetime and productivity. Besides harvesting energy, increasing the lifetime of sensors in the network by decreasing their energy consumption has become one of the main challenges of using WSNs in practical applications. In response to this challenge, over the last few years there have been increasing efforts to minimise energy consumption via new algorithms and techniques in different layers of the WSN, including the hardware layer (i.e., sensing, processing, transmission), network layer (i.e., protocols, routing) and application layer; most of these efforts have focused on specific and separate components of energy dissipation in WSNs. Due to the high integration of these components within a WSN, and therefore their interplay, each component cannot be treated independently without regard for other components; in another words, optimising the energy consumption of one component, e.g. MAC protocols, may increase the energy requirements of other components, such as routing. Therefore, minimising energy in one component may not guarantee optimisation of the overall energy usage of the network.

Unlike most of the current research that focuses on a single aspect of WSNs, we present an Energy Driven Architecture (EDA) as a new architecture for minimising the total energy consumption of WSNs. The architecture identifies generic and essential energy-consuming constituents of the network. EDA as a constituent-based architecture is used to deploy WSNs according to energy dissipation through their constituents. This view of overall energy consumption in WSNs can be applied to optimising and balancing energy consumption and increasing the network lifetime.

Based on the proposed architecture, we introduce a single overall model and propose a feasible formulation to express the overall energy consumption of a generic wireless sensor network application in terms of its energy constituents. The formulation offers a concrete expression for evaluating the performance of a wireless sensor network application, optimising its constituent's operations, and designing more energy-efficient applications. The ultimate aim is to produce an energy map architecture of a generic WSN application that comprises essential and definable energy constituents and the relationships between these constituents so that one can explore strategies for minimising the overall energy consumption of the application. Our architecture focuses on energy constituents rather than network layers or physical components. Importantly, it allows the identification and mapping of energy-consuming entities in a WSN application to energy constituents of the architecture.

Furthermore, we perform a comprehensive study of all possible tasks of a sensor in its embedded network and propose an energy management model. We categorise these tasks into five energy consuming constituents. The sensor's energy consumption (EC) is modelled based on its energy consuming constituents and their input parameters and tasks. The sensor's EC can thus be reduced by managing and executing efficiently the tasks of its constituents. The proposed approach can be effective for power management, and it also can be used to guide the design of energy efficient wireless sensor networks through network parameterisation and optimisation.

Later, parameters affecting energy consumption in WSNs are extracted. The dependency between these parameters and the average energy consumption of a specific application is then investigated. A few statistical tools are applied for parameter reduction, then random forest regression is employed to model energy consumption per delivered packet with and without parameter reduction to determine the reduction in accuracy due to reduction.

Finally, an energy-efficient dynamic topology management algorithm is proposed based on the EDA model and the prevalent parameters. The performance of the new topology management algorithm, which employs Dijkstra to find energy-efficient lowest cost paths among nodes, is compared to similar topology management algorithms. Extensive simulation tests on randomly simulated WSNs show the potential of the proposed topology management algorithm for identifying the lowest cost paths. The challenges of future research are revealed and their importance is explained.

# Acknowledgements

I first thank my supervisor Prof. Doan Hoang, for his full support and valuable advice during my PhD study. Without his help, it would have been impossible to complete the PhD journey. I also thank my spouse and my parents for their selfless care and warm encouragement. I would also like to thank my little kids who were born during my PhD and brought great inspiration and happiness to my life.

Finally, I thank my colleagues and friends in both school of Information Technology and iNext group for the happy time they have shared with me.

# List of publications

**Journal**

1. **N Kamyabpour**, DB Hoang, "Energy Efficient Parametric Topology Management and Routing in Wireless Sensor Networks", Ad Hoc Networks, elsevier (submitted).

**Conference**

2. DB Hoang, **N Kamyabpour**, "Energy-Constrained Paths for Optimization of Energy Consumption in Wireless Sensor Networks", *IEEE Fourth International Conference on Networking and Distributed Computing (ICNDC)*, 2013
3. **N Kamyabpour**, DB Hoang, "Statistical Analysis to Extract Prevalent parameters on Overall Energy Consumption of Wireless Sensor Network (WSN)", *IEEE 13th International Conference on Parallel and Distributed Computing, Applications and Technologies (PDCAT)*, 2012
4. DB Hoang, **N Kamyabpour**, "An energy driven architecture for wireless sensor networks", *IEEE 13th International Conference on Parallel and Distributed Computing, Applications and Technologies (PDCAT)*, 2012
5. **N Kamyabpour**, DB Hoang, "A Task Based Sensor-Centric Model for Overall Energy Consumption", *IEEE 12th International Conference on Parallel and Distributed Computing, Applications and Technologies (PDCAT)*, 2011
6. **N Kamyabpour**, DB Hoang, "A hierarchy energy driven architecture for wireless sensor networks", *2010 IEEE 24th International Conference on Advanced Information Networking and Applications Workshops (WAINA)*, 2010
7. **N Kamyabpour**, DB Hoang, "Modeling overall energy consumption in wireless sensor networks", *IEEE 11th International Conference on Parallel and Distributed Computing, Applications and Technologies (PDCAT)*, 2010

# List of Figures

# List of Tables

# Table of Contents

# Chapter 1. Introduction

The development of wireless sensor networks (WSNs) has recently opened up a new and interesting area for the creation of new types of applications. WSNs consist of a large number of small sensing nodes that monitor their environment, process data if necessary (using microprocessors) and send/receive processed data to/from other sensing nodes (Figure 1-1). These sensing nodes, distributed in the environment, are connected to a sink node – in centralised networks – or to other sensing nodes via a network. In centralised networks, the sink collects sensor data to be used by the end user. In many cases, the sink is also capable of activating sensing nodes via broadcasting, by sending network policy and control information (Le et al., 2008).

As with other networks, there are three common design challenges that highly influence the connectivity and productivity of the entire network: (1) using network protocols to minimise control and data packets, (2) selecting the best topology by positioning nodes in the right places, and (3) deploying a routing algorithm that effectively passes data through the network from the origin node to destination node/nodes.

Distribution of nodes in the environment can be non-structural or structural. The former is used when there is no control of nodes after distribution, and their only role is to monitor the environment, process the data and build the network by finding and connecting to their neighbours. In the latter, however, the position of each node (both sensing and sink) is clear in advance. As the nodes are under control, the communication between nodes is programmable and management and maintenance of the nodes is easier; also, because a lower number of nodes is used in the environment, the cost is much lower.

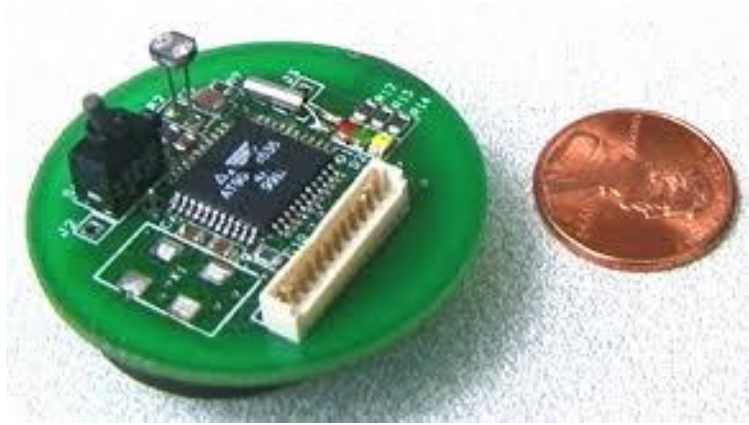

Figure 1-1. Wireless sensor

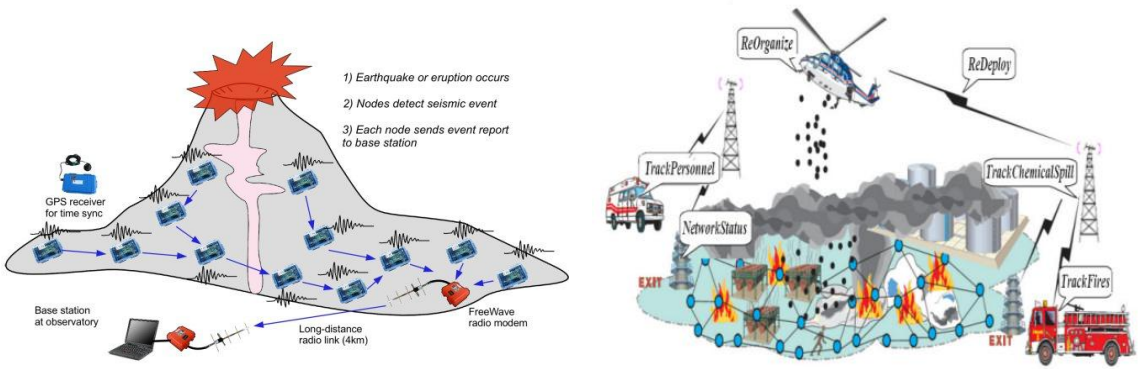

Figure 1-2. Two practical applications of a wireless sensor network.

Figure 1-2 shows two practical applications of WSNs: for volcano monitoring (Werner-Allen et al., 2006) and fire detection . As sensing nodes are generally used in non-accessible environments, they need to rely on their battery (and energy harvesting, e.g., solar cells); charging or changing of sensing nodes is not an option. Therefore, one of the biggest challenges in WSNs is saving energy; it is one of the main factors that determines the lifetime of the entire network.

The rest of the chapter is structured as follows. We first introduce WSN requirements in Section 1. Section 2 presents the motivation for the research and research issues, followed by a discussion of the research objectives and methodology for this project. In section 3, we summarise our work, and outline future research directions in Section 4.

## 1.1 Defining Wireless Sensor Network (WSN) requirements

There are a few requirements that apply to most sensor network applications (Rabaey et al., 2000, H.Edgar and Callaway, 2004, Akyildiz et al., 2002b, Pottie and Kaiser, 2000):

- Lifetime: it is desirable to prolong the lifetime of the network because sensors are not accessible after deployment.

- Network size: in most applications a larger network is of interest as it covers more area and therefore monitors more events.
- Minimise faults: a faulty network uses resources to generate incomplete data. At the sensor level, it means the monitoring of the environment is broken and many events may be missed. In transmission to the sink, it means packet loss is high; in both cases, the knowledge of the environment is incomplete and therefore the gathered data is not reliable. In other words, a reliable collective event-to-sink is vital in WSNs (Sankarasubramaniam et al., 2003).

These requirements dictate the following criteria in communication protocols:
- Lower energy consumption: as a direct consequence of the requirement for longer sensor lifetimes, the communication between these sensors (and sink) must slowly consume the available energy, as the majority of a sensor's energy is consumed in communication.
- Compatible with multi-hop communication: typically, sensors avoid direct communication with the sink (as energy usage is proportional to the square of distance); instead, it is preferred that sensors use other sensors as hops to communicate.
- Scalability: the communication protocol must be reliable in terms of establishing and keeping connectivity among sensors. This protocol must perform as normal when the size of the network becomes larger.
- Reliability: reliable data transmission in term of packet loss is one of the main concerns to provide a high degree of efficiency in monitoring and control systems.

Therefore, employing energy-efficient communication techniques, taking into account multi-hop ability, scalability and reliability, is highly desired. As a direct result, the lifetime of the network will be improved.

## 1.2 Motivation and Research Issues

Most current energy minimisation approaches consider WSNs in terms of network layers: (1) the operating system, (2) the physical layer, (3) the MAC layer, (4) the network layer, (5) the application layer, and (6) the power harvesting layer. In this section we review related efforts in the minimisation of energy consumption at each layer.

At the operating system (OS) level, two major approaches have been used to optimise and manage the energy consumption of the sensor system under its control. At the OS kernel level, one technique for minimising the system energy consumption is processor scheduling with Dynamic Voltage Scaling (DVS). This technique may be deployed to allocate CPU time to tasks and manipulate the CPU power states (Sravan et al., 2007). Parallel thread processing techniques can also be used to reduce energy consumption of the processor. For example, with a cluster-based infrastructure WSN, cluster heads collect data and execute the necessity computation operations in parallel. As stated in (Min et al., 2001), energy can be saved using frequency and voltage scaling when there is great latency per computation; this latency results from partitioning of computations.

At the physical layer, energy is consumed when the radio channel sends or receives data. The radio channel has three modes of operation: idle, sleep and active. Thus, the key to effective energy management is to switch the radio off when the radio channel is idle. To consume less energy, it is important to minimise the time the radio is in the transmit and receive states and reduce the amount of switching between different modes (Raghunathan et al., 2002). Furthermore, a low-power listening approach may operate at the physical layer by periodically turning on the receiver to sample from incoming data. This duty-cycle approach reduces the idle listening overheads in the network (Halkes et al., 2005).

Efficient MAC protocols efficiently arbitrate the use of the shared channel while aiming to reduce packet collision, idle listening, protocol overhead, and overhearing. TDMA-based protocols effectively avoid packet collisions, but their deployment in multi-hop and ad hoc networks is very complex (Halkes et al., 2005). The PAMAS protocol offers a technique for reducing collisions where the nodes can calculate the finish time of another node's data transfer. It saves energy by turning nodes off during the data transfer duration of other nodes. In (Halkes et al., 2005), Halkes, Dam and Langendoen compare two MAC protocols (T-MAC, S-MAC) developed for wireless sensor networks. With the S-MAC protocol, nodes can send queued frames during the sleeping time. Accordingly, the time between frame transmissions and idle listening is reduced. Nodes, however, are required to send SYNC messages at the start of a frame for synchronisation. T-MAC adapts the duty cycle to the network traffic. It operates as S-MAC but also uses a time-out mechanism for determining the end of the active period. The adaptive duty cycle reduces traffic fluctuation in both time and space and allows longer sleeping times.

At the network layer, several approaches may be adopted to increase the network lifetime. Topology control and related routing mechanisms can be optimised for the purpose. Determining the best topology among nodes in order to provide a connected network to route packets to the destination is a significant operation in WSNs. The challenges in selecting a suitable topology include duty cycle control of redundant nodes, connectivity maintenance, self-configuration and redundancy identification in a localised and distributed fashion (Xu et al., 2003). Two significant methods for tackling these challenges are the Geographic Fidelity (GAF) and Cluster-based Energy Conservation (CEC) protocols. GAF uses the node's location information (as determined by a GPS) to configure redundant nodes and cluster them into small groups using localised and distributed algorithms. CEC has the same fundamental operation but it does not depend on location information. In (Xu et al., 2003), the two methods were compared by simulation. They found that CEC consumes much less energy than GAF (about half) if the nodes are stationary. However, GAF is more efficient than CEC in high mobility environments. In (Le et al., 2008), the authors suggested a new approach for reducing protocol overhead created by the CEC protocol and the energy consumption of GPS connected to sensors. In this approach, a base station informs the sensors about their cluster ID and cluster area by sending a sweeping beacon. If a node hears the beacon it can locate its cluster without the need for a GPS receiver. Various kinds of topology such as tree, mesh, clustered, ad-hoc and others can be employed. Authors in (Salhieh et al., 2001), examine the influence of different types of mesh topologies on the power dissipated.

Routing is a significant and costly task in WSNs as it plays a major role in determining the network lifetime. Al-Karaki and Kamal (Al-Karaki and Kamal, 2004) discussed types of networks, topologies and protocols and their influences on the energy cost. SPIN (Sensor Protocols for Information via Negotiation) (Heinzelman et al., 1999a) is a routing technique based on node advertisements where nodes only need to know their one-hop neighbours; however, it is not suitable for applications that need reliable data delivery. LEACH (Low-Energy Adaptive Clustering Hierarchy) (Heinzelman et al., 2000) is a clustered routing algorithm. In this method, the cluster heads are responsible for relaying data and controlling the cluster. Although LEACH is an effective technique for achieving prolonged network lifetime, scalability, and information security, LEACH does not guarantee an optimal route. Directed Diffusion technique is a data centric, localised repair, multi-path delivery for multiple sources, sinks and queries (Intanagonwiwat et al., 2000). Also, this method is able to find the optimal route.

everal technologies exist to extract energy from the environment, such as solar, thermal, kinetic energy, and vibration energy. Weddell, Harris and White (Weddell et al., 2008) explain the advantages of energy harvesting systems as the ability to recharge after depletion, as well as monitoring of energy consumption, which may be required for network management algorithms.

While efforts to reduce energy consumption have covered different aspects of WSNs, many important issues remain untouched:

- There is no a general approach for determining and optimising the energy consuming constituents of WSNs.
- Current approaches focus on one aspect and may load energy consumption in other aspects.
- Existing approaches miss quantitative measures of energy consumption of the entire network.
- Most of the current approaches are applicable for specific sensor networks with special properties.

In this thesis, we deeply tackle the first two issues as well as touching the third one. Briefly, we introduce a new energy-driven architecture by splitting the whole WSN system into a few main constituents. Then, energy-related parameters in each constituent are extracted. After reducing the number of parameters using the concept of machine learning, a new routing algorithm is designed. The result is an application-independent and constituent-based network model, such that existing approaches can be adapted to energy constituents of this architecture.

**1.3 Research Aims and Objectives**

The aim of this research is to propose an energy constituent-based model. Not only does modelling of constituents as single energy consumption units present many possible strategies for maximising the network's lifetime, it also has a few benefits when WSN is seen as a composition of these constituents. Figure 1-3 clearly shows how the energy of a node is consumed by tasks, operations, events, changes, demands and commands during its lifetime. This composition of constituents allows optimisation of the energy consumption of a node if desired, permits optimisation of a selected constituent for a specific application, and, more importantly, allows an overall optimisation of the energy consumption of the entire network by considering the play-off between constituents. Furthermore, the constituents can be adapted to suit the required application.

This research focuses on minimising and optimising energy consumption based on the energy consuming constituents as a general model for WSN deployment and development. The model deals with all common aspects of energy consumption in all types of WSNs. We believe designing wireless sensor networks with their energy constituents in mind will enable designers to balance the energy dissipation and optimise the energy consumption among all network constituents and sustain the network lifetime for the intended application.

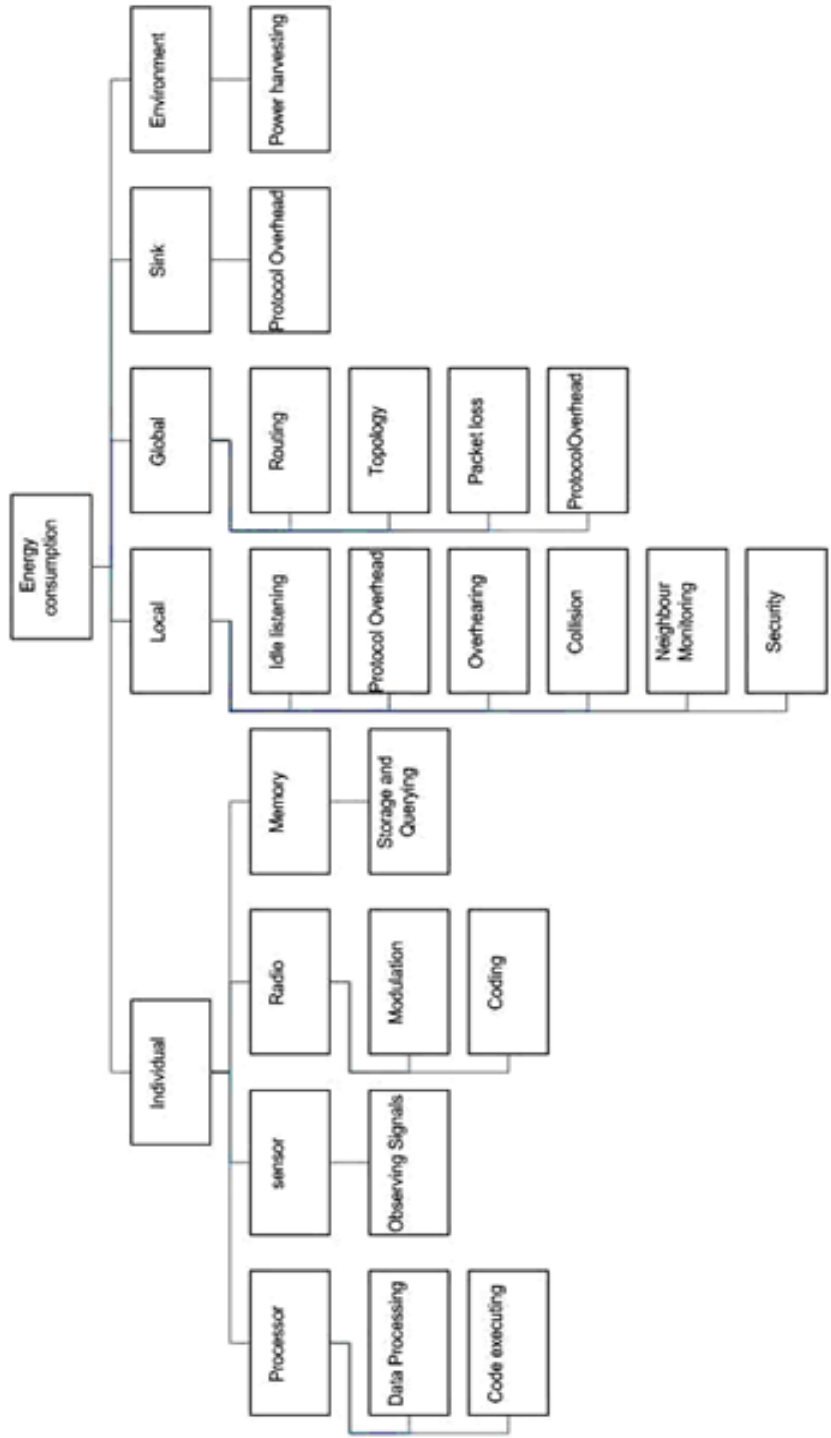

Figure 1-3. Energy Consuming Constituents

Our aim is to propose a single overall formulation of the energy consumption of the entire wireless sensor network. Another possible but more difficult formulation expresses the energy consumption model as a non-linear function of its constituents. This approach requires more extensive exploration, as we do not understand well enough the metrics associated with the energy of each constituent and we are unsure about the mathematical models that can describe such a non-linear relationship.

In this research we comprehensively model the components of each of the five energy constituents of the architecture. The aim is to provide an accurate account of all functional aspects of a constituent and their salient energy-related parameters. These parameters will allow us to evaluate the performance of WSNs, optimise their operation, and design more energy-efficient applications.

In the next stage of our research, we aim for an optimization of each constituent and a general optimisation with respect to a balance between the energy constituents. These optimisations will be confirmed by mathematical proof and simulation. Finally, we will use the outlines of the project to generate an algorithmic solution to minimise overall energy consumption and network performance.

## 1.4 Research questions and their contribution to knowledge

The energy minimisation challenge has been surveyed from different aspects but there are still unsolved problems that should be taken into account. The strategy of the present research (Figure 1-4) comprises three main stages: problem definition, developing new approaches and evaluation.
In addressing a significant and key problem, we developed the following research questions based on our preliminary study:

### 1.4.1 Characterisation

- Can the current approaches optimise WSNs in term of energy?
    - To the best of our knowledge, there is not an overall energy minimisation approach. It is expected that partial energy minimisation does not minimise the overall energy.

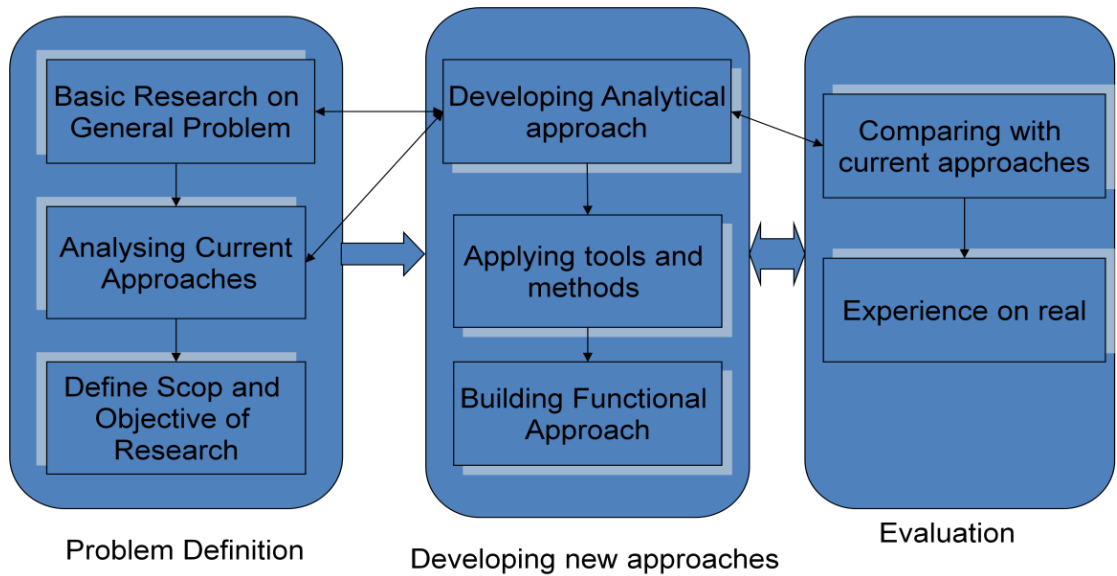

Figure 1-4. Research methodology

- Which kind of architecture can provide a fundamental concept for generating a mathematical model for overall energy consumption?
    - The desired architecture, which is based on the main sensor's operations (data generation and collaboration to deliver data) should cover all constituents of the energy consumption in WSNs.

### 1.4.2 Methods/Means

- Is the desired architecture appropriate for overall energy minimisation on all WSN applications?
    - The desired architecture should be adaptable to different kinds of WSN application in terms of the energy constituents.

### 1.4.3 Feasibility

- How can the model be used for existing and future wireless sensor networks?

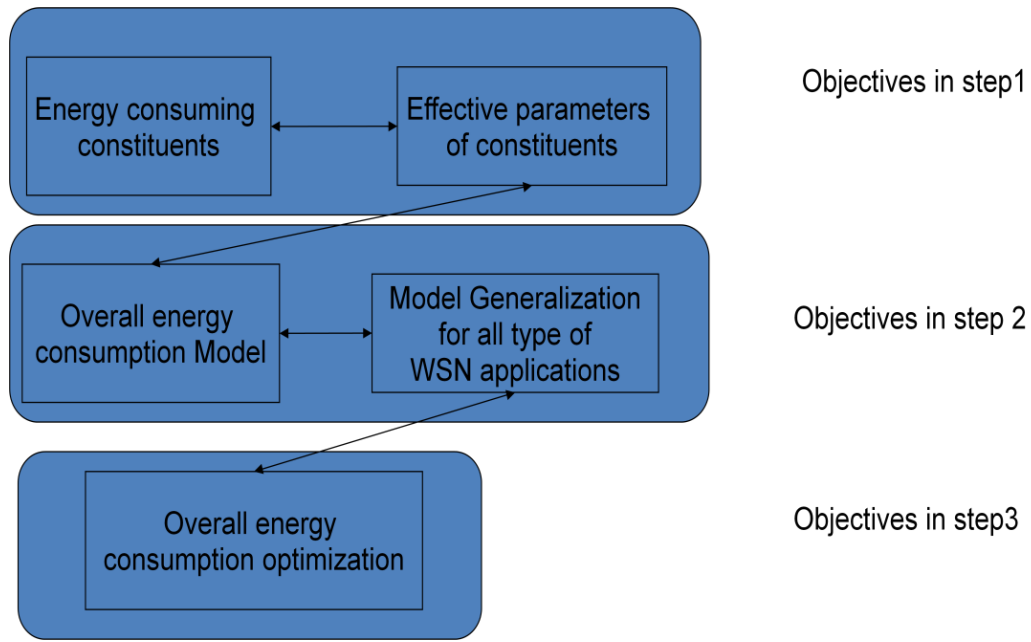

Figure 1-5. Research steps

- The overall energy consumption model should include almost all energy-consuming constituents in WSNs and applications. It can then be used to develop and improve energy consumption of a variety of sensor applications.

### 1.4.4 Selection

- What is the best and most practical method for validation of the architecture and model?
  - The best method will be to analyse an application based on the relationships between constituents, find the optimum values of parameters using the overall energy consumption model and then apply these parameters to the real application.

### 1.4.5 Generalisation

- Is the overall energy consumption model applicable to existing applications? Can it be used in the development of new applications?

- The model will be based on analysing energy consuming constituents and prevalent parameters. The optimum value of prevalent parameters can be used to develop new applications and improve existing applications.

**1.4.6 Risks**

- Is the problem of overall energy based on prevalent parameters modellable?
  - Different modelling techniques, such as machine learning-based methods, should be examined to obtain a suitable model with consideration of the modelling error.

## 1.5 Research objectives and scope

Addressing the above-mentioned research questions, Figure 1-5 shows the steps that will be followed in this thesis:

- Determine the effect of energy consuming constituents and their prevalent parameters on overall energy consumption in WSNs.
- Obtain a quantitative measurement and modelling of the overall energy consumption based on prevalent parameters.
- Propose a model which is applicable for all types of sensor network applications.
- Optimise overall energy consumption by optimising the model.
- The model should cover the challenging problems: scalability, reliability, and collaboration
- The overall model will offer the best approach to minimise the energy consumption by involving the prevalent parameters

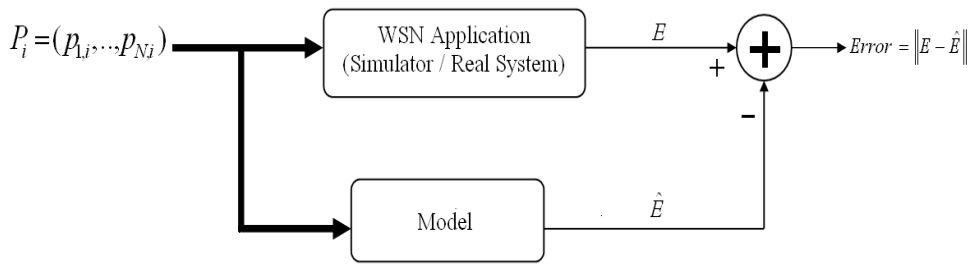

Figure 1-6. Model Evaluation

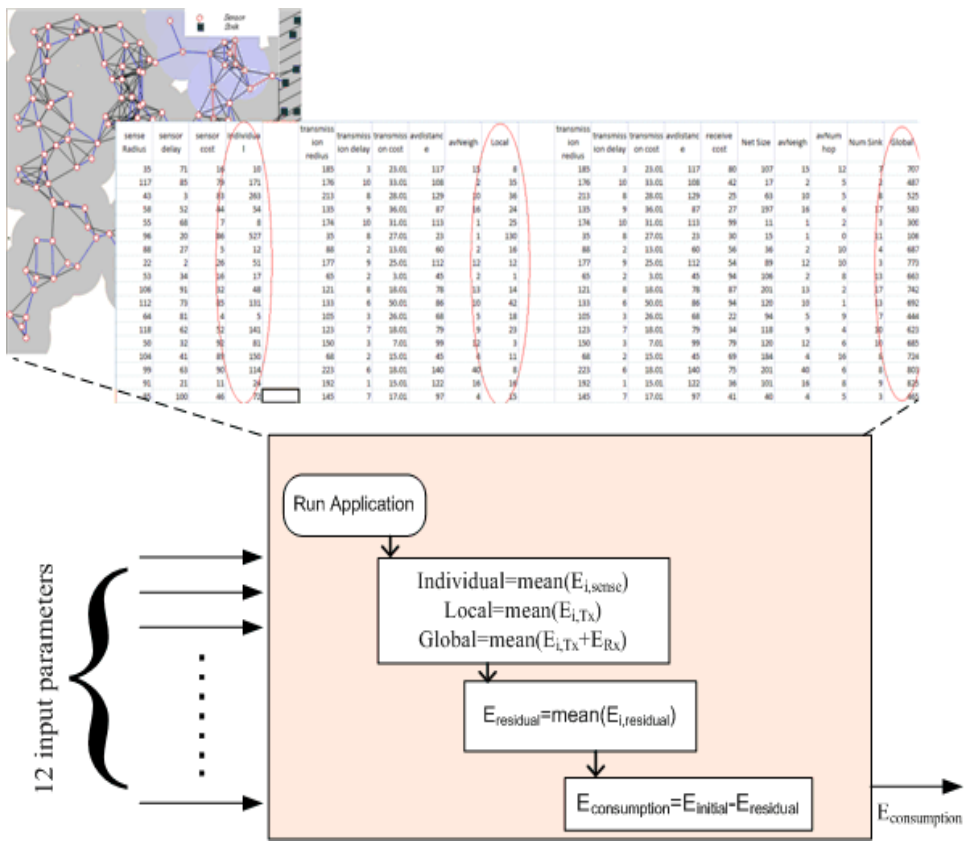

Figure 1-7. The procedure of capturing data from the event detector application
.

## 1.6 Research Contribution

Despite the widespread deployment of wireless sensors and sensor systems, a critical challenge has always been their limited power supply. The power supply of a sensor is limited by its battery and the lifetime of a wireless sensor application depends singularly on it. Energy minimisation has become one of the most challenging issues in sensor applications.

Current efforts in minimizing energy consumption have increased over the last few years with the expectation that by reducing energy consumption of a component of a sensor network, the overall energy consumption of the whole network would be reduced. Consequently, most efforts focused on some specific components of energy dissipation in WSNs. These components are, however, highly integrated within a WSN and their interplay cannot be easily taken into account as each energy consuming constituent is treated independently without regard for other constituents. As energy consuming constituents of a WSN are interrelated intimately, minimizing the energy consumption of one constituent may increase the energy consumption of other constituents and hence may not guarantee the minimization of the overall energy consumption of the entire network.

In our research, the ultimate aim is to produce an energy map architecture of a generic WSN application with essential and definable energy constituents and the relationship among these constituents so that one can explore strategies for minimizing the overall energy consumption of the entire application. The major contributions of this study are listed below:

- This research introduces a novel Architecture and its components as a single overall model and propose a feasible formulation to express the overall energy consumption of a generic wireless sensor network application in chapter 3.
- The fundamental aim is to model the energy consumption of the entire sensor network by taking into account of various constraints of energy consuming constituents of the network. To achieved this aim based on proposed architecture in chapter 3, energy consumption is modeled in terms of energy consuming constituents and their input parameters and tasks.
- The study investigates the dependency between extracted parameters and energy consumption in the network and consequently selecting the most important ones by taking advantage of statistical and machine learning tools in chapter 5.

- The research provides an energy-efficient dynamic topology management algorithm that aims to increase the overall lifetime of various mesh-topology wireless sensor networks by taking in to account the interconnection between energy consuming constituents and the most important parameters. This goal is achieved in chapter 6.

**1.7 Justification**

The proposed model (based on energy constituents' parameters) and algorithms in this research will be justified and evaluated using the following steps:
1. Extracting and examining the prevalent parameters in current approaches in the literature in order to achieve a better result.
2. The accuracy of the model will be tested by giving the same values of these prevalent parameters to both model and simulator/real WSN and measuring the square root error between their outcomes (Figures 1-6 and 1-7). The expectation is that the outcome of the model will be approximately similar to the output of the simulator/real WSN.
3. After justification of the model, it can be used to develop new approaches to minimise energy consumption by selecting a set of these parameters and tweaking them.

**1.8 Structure of the Thesis**

The author of this thesis is interested in energy efficiency techniques in wireless sensor networks. In a preliminary study of this topic, we completed an in-depth survey of the existing techniques in different layers of WSNs. Then a comprehensive model was proposed that involved splitting the WSN into a few energy consuming constituents. After extracting the parameters influencing energy consumption in each constituent, the relationships between them and the overall residual energy were studied. Due to the high number of these parameters, selecting and tweaking them is time-consuming and impractical, and therefore a subset of these parameters were favoured. The concepts of statistics and machine learning were employed to extract prevalent parameters, by first analysing the correlation between each parameter and residual energy and then removing correlated parameters which imply very small coefficients on a linear regression model between these correlated parameters and the residual energy. Towards the end, a new energy-efficient topology management algorithm was proposed. Unlike most of

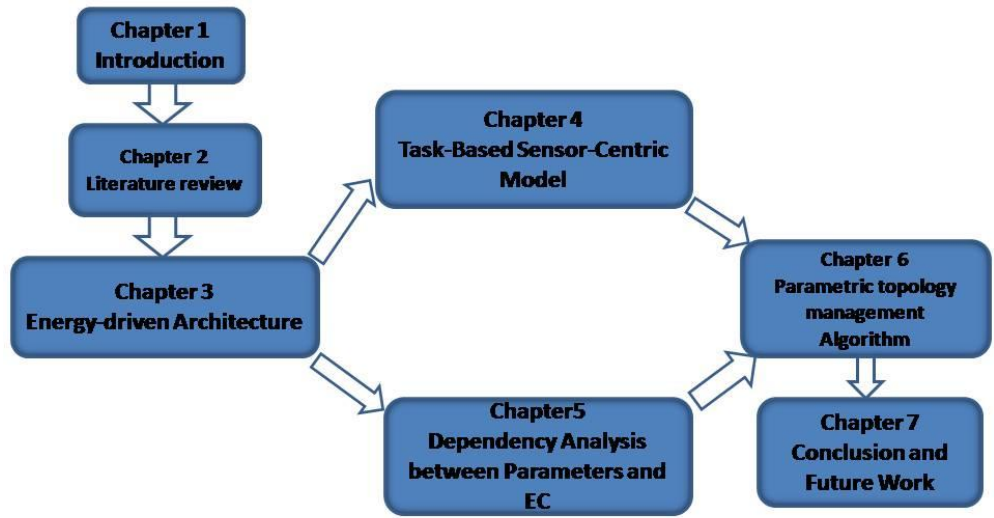

Figure 1-8. Thesis Structure

the research in this area, which uses a few parameters in only one layer and therefore ignores the interconnection between the selected layer and other parts of the network, the main purpose of this algorithm is to utilise parameters from different layers of the WSN; this is a promising way to reach a global algorithm that can be modified for different applications. The rest of this thesis is organised as follows (Figure 1-8):

Chapter 2 presents a review of the literature on the concepts and challenges in wireless sensor networks. We first introduce the characteristics and architecture of sensor networks and applications and then discuss techniques to minimise energy consumption. Future research orientations are discussed and a comparison of our study to other work is made at the end of this chapter.

Chapter 3 introduces the proposed Energy Driven Architecture and its components as a single overall model and proposes a feasible formulation to express the overall energy consumption of a generic wireless sensor network application in terms of its energy constituents. We then discuss the concept of each constituent.

Chapter 4 investigates all possible tasks of a sensor in its embedded network and proposes an energy management model. In this chapter we categorise tasks into five energy consuming constituents. The sensor's Energy Consumption (EC) is modelled on its energy consuming constituents and their input parameters and tasks.

In Chapter 5, statistical and machine learning tools are employed to reduce the number of parameters by analysing the dependency between these parameters and the target parameter (i.e., average energy consumption in the network) and consequently selecting the most important ones. The applied methods are correlation (Pearson, Spearman and nonlinear second and third degree correlation), Lasso regularisation and p-value. Later, random forest regression is applied to compare the accuracy of prediction for both original and reduced parameters in estimating the average energy consumption of the network.

Chapter 6 demonstrates an energy-efficient dynamic topology management algorithm that increases the overall lifetime of various mesh topologies of WSNs. The performance of the new topology management algorithm, which employs Dijkstra to find the energy-efficient lowest cost paths among nodes, is compared to similar topology management algorithms.

Chapter 7 summarises the ideas presented in this thesis, the major contributions of this research, and future research plans.

# Chapter 2. Background and literature review

Recently, wireless sensor networks (WSN) have become popular due to their exceptional capabilities. Applications for WSNs cover a substantial range of domains varying from military to farming applications; for example, precision agriculture, where a farmer can control temperature and humidity, or surveillance systems to detect and monitor enemies or threats. Other examples include observing the activities of birds, small animals and insects, tracking the effects on crops and livestock of various environmental conditions, monitoring earth's activities and planetary exploration, discovering forest fires, detecting floods, mapping environment bio-complexity and studying environmental pollution (Akyildiz et al., 2002a, Cerpa and Estrin, 2002) (Ibrahiem M. M. El Emary, 2013). WSNs can also be used to address numerous challenges in the field of health and medicine by monitoring and directing data to a base station; it can create an interface to observe conditions of disabled and integrated patients, monitor diagnostics and drug administration in hospitals, observe human physiological data and track doctors and patients inside a hospital (Young Han Nam et al., 29 Oct-1 Nov 1998). There are several examples of using WSNs in healthcare, on heart problems (K. W. Goh, 2005, Hsein-Ping and Do-Un, 2009), asthma (Chu et al., 2006), emergency response (Konrad et al., 2004) and stress monitoring (E. Jovanov, 2003).

Typically, a WSN application consists of a set of sensor nodes distributed in the studied area and a few sinks (i.e., base stations); all nodes cooperate with each other to create and pass generated data to the sink. The role of every node is to sense data, depending on the application, and then send it to the related sink via a single hop or through multiple hops. There are many parameters that should be considered when dealing with data dissemination, such as data reliability, congestion status and required delay, to name a few (Rahman et al., 2008).

Each application needs a different type of sensor network architecture and communication protocol; for example, military applications are designed based on a dense deployment of sensors supporting self-organising, rapid deployment, and fault tolerance (Akyildiz et al., 2002b). On the other hand, health applications need only a limited number of sensors connected to a patient with reliable data transmission. Through health monitoring applications, health industries are trying to change traditional healthcare approaches for the elderly and chronic illness by utilising low cost, ubiquitous and continuous healthcare monitoring; however, it is difficult to choose a suitable architecture and to fit one technology into the overall architecture. To design a suitable architecture, a number of important factors are taken into consideration. These factors include cost, size, power, mobility and processing (Hoang, 2007).

Another challenge in the implementation of healthcare monitoring systems is selecting a technology that fits to the architecture in order to offer a low cost service and to support mobile users. According to authors in (Yin et al., 2008), suitable technologies include a body sensor network, community server, and medical services. ZigBee, a communication protocol for WSNs, offers a wearable wireless body/personal area network and provides low cost, low power consumption and portability.

Because of the low cost and light weight of wireless sensors, they are a key device for monitoring systems. However, the short lifetime of these devices, supplying their power via batteries or other limited sources, means that they cannot offer a long lasting monitoring service. Thus, energy is a critical issue for sensor lifetime. Generally, sensors consume energy when they do individual operations such as data sensing and processing, or group-based operations such as running different communication protocols. There are also several methods for producing energy, but they cannot eliminate the need for energy management. In most situations, these techniques increase the complexity of systems and require new methods for energy management.

As a practical example, the critical issues in the energy consumption of healthcare monitoring systems are reviewed in the next section.

## 2.1 Healthcare Monitoring Networks

WSN has become a technology that promises to considerably improve the quality of healthcare across a wide range of configurations and for a diverse range of applications. For example, the potential of WSNs has been shown by early medical diagnosis via real-time tracking of patients in hospitals (Jeonggil et al., 2010, Octav et al., 2010), supply of emergency care in large disasters through automatic electronic triage (G. Virone, 2006), improvements in the quality of elderly life by means of providing smart environments (Nourchene et al., 2011) and in the study of human behaviour and chronic diseases (S. Kumar, September 8, 2011). For example, wireless biomedical sensors can be implanted to continuously and precisely monitor the level of glucose in diabetes patients ( December 2011). These sensors can also play an important role in the early detection of cancer (Akyildiz et al., 2002c) by noticing changes in blood flow in suspected locations. Generally, diagnoses are more likely to be made much earlier by using the sensors than without, via tracking and monitoring the patient in their everyday activities, processing this information and then relaying the data to a health check group.

The challenges of incorporating wireless sensor networks in healthcare applications range from computational capabilities to managing limited power. Healthcare monitoring networks involve three categories: body sensor networks (BSN), personal servers (PS) and medical servers (MS) (Yin et al., 2008). The local part (or BSN), including sensors connected to the patient, extracts raw data from the patient's body and sends it to a gateway connecting the sensor network to a local server (or PS). The local server gathers information from the gateway and sends it to a central server (or MS). The central server integrates the received data with other resources of the patient's medical record using the internet, then transmits this data to a medical centre via the internet for comment or even to inform an emergency service (Hoang, 2007). In this topology, all servers are powerful devices such as a laptop, PDA or a desktop computer. AutoSense systems (Hande and Cem, 2010), still in the early stages for population-scale health care studies, are another example of using WSNs in health care, in which measurements of personal psychosocial stress and alcohol consumption are monitored. A suit of deployable wireless sensors, producing a body-area wireless network, measure respiration rate, skin conductance, skin temperature, arterial blood pressure and blood alcohol concentration. Collected data, after validation and cleaning at the sensor, are transmitted to a smart phone as raw data, followed by real time computation of features representing the beginning of psychosocial stress and the occurrence of alcoholism. The processed data is then distributed to researchers investigating behavioural research questions about stress, addiction, and the relationship between the two. By collecting time-synchronised data about a subject's activities, social context and location, factors leading to stress are extracted and personalised guidance about stress reduction can be produced. To scale this system, various technical and algorithmic challenges (e.g., energy) need to be addressed. Energy is one of the main challenges due to high sampling rates of some on-body sensors, leading to significant energy consumption and consequently short lifetime; another bigger challenge is the issue of information privacy and its close relationship with the quality and value of information (Kumar, 2011).

There are a few issues limiting the usage of sensor networks in applications: the sensor itself suffers from many limitations such as insufficient energy sources, small memory and limited processing capability. Moreover, due to the deployment of sensors in large numbers, WSN applications deal with other issues such as efficient multi-hop routing, security, data reliability and scalability. Before starting our present work of analysing energy consumption, several points should be addressed: sensor hardware capabilities, network architecture and communication technology.

**2.1.1. Sensor hardware capabilities**

There are various types of sensors with specific uses in special environments. Some of the commercially available wireless sensor nodes for health monitoring – one of the widest applications of WSNs – include pulse oxygen saturation sensors (to evaluate the percentage of haemoglobin saturated with oxygen, and heart rate), blood pressure sensors, electrocardiograms (to detect heart abnormalities by measuring its electrical activity), electromyograms for evaluating muscle activities, temperature sensors, respiration sensors, blood flow sensors and blood oxygen level sensors (oximeters) for measuring cardiovascular exertion (distress), to name a few (Ramakrishnan, 2013). However, there are a few technical challenges in using WSNs in this domain (Ramakrishnan, 2013):

- Power: biosensors have a small range of resources to provide energy (e.g., a typical alkaline battery used in such sensors only produces about 50 Wh of energy); the lifetime of a biosensor is typically less than one month.
- Computation: due to lack of memory, the biosensors are not able to execute large-bit computation.
- Security and interference: the biosensor network must be secure enough to avoid illegal entities reporting false data to the control node or providing the wrong instructions to the other biosensors and possibly causing significant harm to the host.
- Material constraints: the size, shape and materials of biosensor must be safe and compatible with the body tissue.
- Mobility: the WSN of biosensors should support mobility through the development of multi-hop, multi-modal and ad-hoc sensor networks in order to provide location awareness.

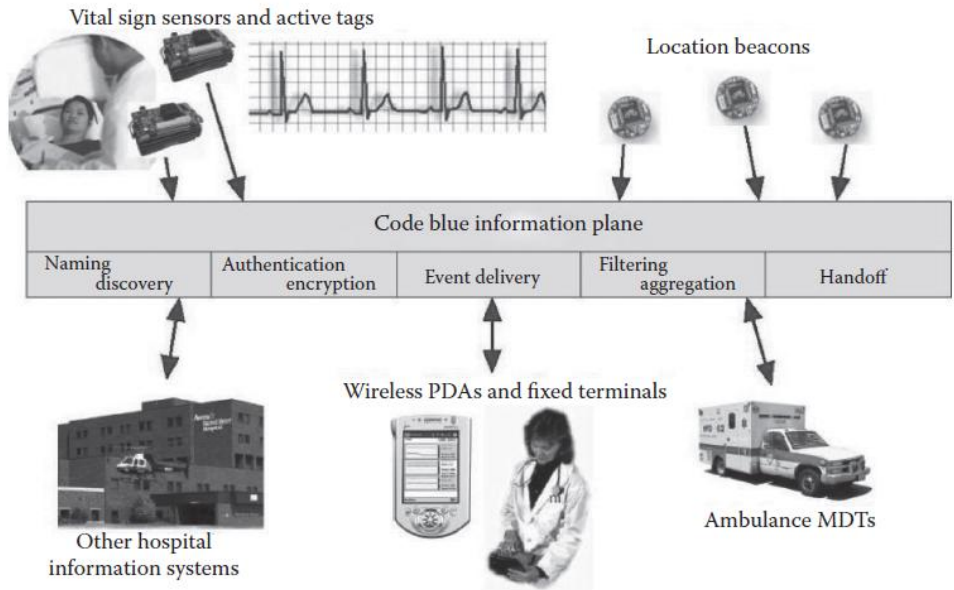

Figure 2-1. CodeBlue architecture for emergency response

- Robustness: in harsh environments, the failure rate of sensors is high, so routing protocols must be designed in such a way to minimise the effect of sensor failure on network performance.
- Continuous operation: a network requires data from all biosensors and heavily depends on continuous operation of the biosensor during its lifecycle, which may mean days, or sometimes weeks without operator intervention.

Currently the power of sensors can be provided by various energy sources such as batteries, vibration energy harvesters or solar cells. Selecting a power supply is completely determined by application and environment. Generally, the power supply is a major limitation of the system's lifetime. The authors of (Weddell et al., 2008) noted that solar cells, vibration energy harvesters, and LiSOCl2 batteries can deliver a long lifetime for sensors, and explored their performances over 10 years. Their results showed that the requirement for control of LiSOCl2 batteries makes them unusable for some kinds of applications in which they are not accessible after deployment. Solar cells and vibration energy harvesters are not cost efficient, especially for applications like health monitoring that require low costs. Also, their performance depends on the application's environment. In addition, a system relying on these two energy harvesting technologies could be unreliable (Berzosa et al., 2012) due to their inconsistency in producing energy, affecting the continuity and quality of the services provided. To conclude, the complexity of energy harvesting technologies and lifetime issues in networks highlights the fact that energy management and energy minimisation must play a significant role in WSN applications.

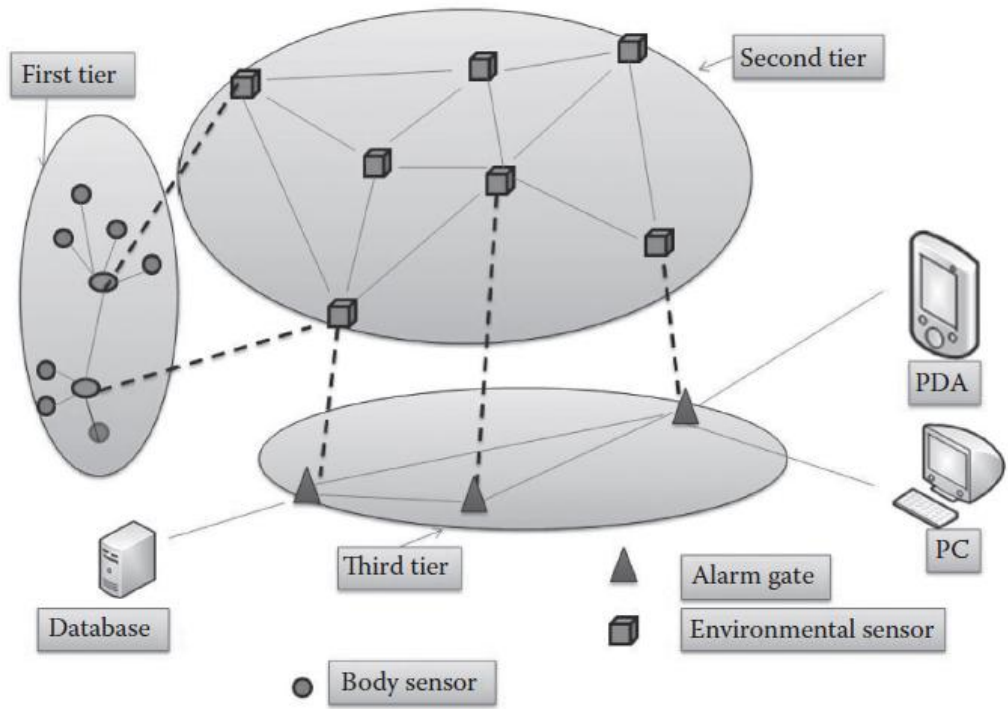

Figure 2-2. ALARM-NET architecture (P.Kumar, 2012)

**2.1.2. Architectures**

CodeBlue , one of the well known WSN health care projects, uses a large number of mica2 motes sensors ; after collecting data from the patient's body, these medical sensors transmit data to PDAs, mobiles, laptops and personal computers for further investigation. The general architecture of the CodeBlue is shown in Figure 2-1. The medical sensors send their data through a particular wireless channel; from the other side, hand-held devices (e.g., PDA and laptop) are locked to this channel providing a framework to deliver patient information to medical professionals. A TinyADMR routing element, using an adaptive demand-driven multicast routing (ADMR) protocol, is employed to facilitate node multicast routing, mobility and minimal path losses. CodeBlue also employs MoteTrack (Lorincz, 2006), a RF-based localisation, to detect the location of a patient or medical professional.

Alarm-Net (Wood, 2006) is another health monitoring system primarily designed to monitor a patient's conditions in the home environment. This system consists of a collection of body sensor networks and environmental sensor networks. As shown in Figure 2-2, three network tiers are used: in the first tier, sensor devices are deployed on the body of patient, which monitor and collect individual physiological signals; in the second tier, environmental sensors (e.g., temperature, dust, motion, and light) are located in the living space to accumulate data on the environmental conditions. The data from both network tiers are aggregated into the third tier where an internet protocol (IP)-based network, named as AlarmGate, is used to distribute data among hand-held devices (such as PDA, mobile) or desktop computers.

UbiMon, MobiCare and MEDiSN (Jeonggil et al., 2010) are other health monitoring systems using WSNs to monitor patients. Both UbiMon and MobiCare consist of a collection of wearable and implantable sensors joined with an ad hoc network; they both cover a major area in mobile patient monitoring systems by producing continuous and timely monitoring of a patient's physiological status. MEDiSN, however, was especially designed for use in hospitals and during disaster events, and is exclusively focused on reliable communications, data rate, routing and Quality of Service (QoS). It consists of multiple mobile battery-powered Physiological Monitors (PMs) that temporarily store measurements and transmit them later after encrypting and signing, as well as distinct Relay Points (RPs) (unlike CodeBlue in which PMs also relay data), to connect to one or more gateways through bidirectional wireless trees. Using hop-by-hop retransmissions, which take into account the effects of packet collisions and corruption, traffic flow in both directions is managed.

### 2.1.3. Communication technologies

In the domain of WSNs, there are two main communication protocols: 6LowPAN and ZigBee. 6LowPAN, released in 2007 by IEFT, is an open standard communication protocol on how to use IPv6 on top of low power, low data rate, low cost personal area networks; it works on top of physical and MAC layers, defining how IPv6 datagrams are transmitted using 802.15.4 frames by implementing compression/decompression of IPv6 headers. It also deals with the time varying link relationship among the nodes comprising the WSN. In addition, it supports implementation of routing protocols at either the link layer (mesh under routing) or network layer (route over routing) (J.N.M.Valdez, 2011).

ZigBee ( April 28, 2013), arguably as popular for a low-cost, low-power advanced communication protocol for small devices, builds on top of the physical layer and MAC defined in IEEE standard 802.15.4 for low-rate wireless personal area networks (LR-WPANs). It is widely used in Body Sensor Networks (BSNs). BSNs comprise a sensor or group of sensors attached to a patient and a coordinator for collecting raw data from the sensors. This data will be sent, analysed, and processed by control devices through the network. The ZigBee coordinator as a controlling device works with interrupt to reduce power consumption in the network – one of the key factors of healthcare monitoring – and gathers raw data. In addition, ZigBee is applied in a mesh network of routers to relay data from different patients to the Access Point (AP). The AP is connected to the internet to allow collaboration of the doctors, medical centre, and other data centres that gather patient records, so that decisions can be made. It is worth noting that the main difference between ZigBee and 6LowPAN is the IP interoperability of the latter. A ZigBee device requires an open 802.15.4/IP gateway to interact with an IP network while a 6LowPAN device communicates with other IP-enabled devices; which one is chosen highly depends on the application (C.Buratti, 2009).

Most sensor applications need to connect to internet, so ZigBee has to provide this feature. Authors in (Sveda and Trchalik, 2007) focused on designing software architecture between ZigBee and Internet by IEEE 1451. IEEE 1451 is a standard-base networking framework that includes a transducer information model called the Network Capable Application Processor (NCAP). NCAP is an object-oriented model that uses block classes, physical blocks, transducer blocks, function blocks, and network blocks. ZigBee Gateway and ZigBee Bridge are proposed to provide connectivity between ZigBee and the internet. Zigbee Gateway translates both addresses and commands between ZigBee and IP. ZigBee Bridge works over Ethernet or WiFi devices and is used to communicate with IP devices.

## 2.2. Energy consumption in WSN

WSN sensors, usually deployed in non-accessible environment, are powered using small batteries along with techniques for power harvesting; replacing batteries is not an option. Relying on a battery not only limits the sensor's lifetime but also makes efficient design and management of WSNs a real challenge. The limitation of energy supply, however, has inspired a lot of the research on WSNs at all layers of the protocol stack.

Network architectures, such as OSI and Internet, are basically functional models organised as layers where a layer provides services to the layer above (e.g. the application layer provides services to the end users). A network is often evaluated in terms of the quality of its service parameters, such as delay, throughput, jitter, availability, reliability and even security. However, when it comes to energy consumption (EC), one often encounters difficulty, as evaluation and optimization of the network as a comprehensive model that takes the EC into account hardly exists. Generally, researchers focus on the traditional network architecture and try to minimise a specific component of a single layer, with the hope that the overall EC of the network is reduced without regard for other components or layers. This is not an ideal situation, where one does not know how a single component fits within the overall energy picture of an entire wireless sensor network. Most current energy minimisation models focus on sending and receiving data (Wang et al., 2006a), while other parameters are neglected. In (Heinzelman et al., 2000) and (Heinzelman et al., 2002), the power consumption model focused on the cost of sending and receiving data and deduced the upper limit of the energy efficiency of single hop distance. This approach considers an intermediate node between source and destination so that the retransmission will save the energy. Other approaches evaluate the energy efficiency of wireless sensor networks by using the power consumption model mentioned in (Heinzelman et al., 2000) and (Heinzelman et al., 2002).

Since wireless networks have different specifications and challenges, the traditional network architecture cannot satisfy them. The cross layer idea was created to provide a flexible network architecture for wireless networks. The key idea in cross-layer design is to allow enhanced information sharing and dependence between the different layers of the protocol stack (Goldsmith and Wicker, 2002), (Shakkottai et al., 2003), (Raisinghani and Iyer, 2004). It is argued that by doing so, better performance gains can be obtained in wireless networks, and the resulting protocols are more suited to employment on wireless networks as compared to protocols designed in the strictly layered approach. Broad examples of cross-layer design include, say, design of two or more layers jointly, or passing of parameters between layers during run-time, etc.; but there is no criteria to determine which layers should be combined to give the best result for the overall EC (Mehmet C. Vuran and Akyildiz).

### 2.2.1. Physical Layer

Communication between wireless sensor nodes needs a radio connection as a physical layer in which energy is consumed when the radio sends or receives data. The physical layer involves modulating and coding data in the transmitter, and then in the receiver this layer must optimally decode the data. The radio channel has three modes: idle, sleep and active. Thus, the key to effective energy management is to switch the radio off when the radio channel is idle; to consume less energy, it is important to minimise the time and energy to switch between different modes and transmit and receive states (Raghunathan et al., 2002). Furthermore, a low-power listening approach may operate at the physical layer, in which the basic idea is to periodically turn on the receiver to sample the incoming data. This duty-cycle approach reduces the idle listening overhead in the network (Halkes et al., 2005). Moreover, the energy consumption of the radio channel for sending and receiving data is equal; consequently, energy efficient MAC protocols have to maximise the sleep time of sensors (Raghunathan et al., 2002). Due to real-time monitoring and interaction with different parts of a sensing node, the operating System (OS) is probably the best place to optimise and manage energy consumption of a WSN at the node level. Perhaps one of the best known techniques at the OS kernel level for minimising energy consumption in the anode is processing unit scheduling by Dynamic Voltage-Frequency Scaling (DVFS). This technique allocates CPU time to tasks and manipulates the CPU power states (Sravan et al., 2007). In other words, tasks are executed at different frequencies, where lower frequencies mean less power consumption, and the CPU is moved to the lowest power state when there is no task to execute.

Parallel thread processing techniques can be useful to reduce the energy usage of a node's processor; for instance, in a WSN with cluster-based infrastructure, cluster heads become responsible for collecting data and executing the necessity computation operations. As addressed in (Min et al., 2001), partitioning a computation, resulting in creating a greater allowable latency per computation, saves more energy through DVFS. Such partitioning makes a considerable improvement in energy dissipation by altering task scheduling algorithms, sequence of tasks execution, and communication scheduling among sensors. In (Tian et al., 2006) a task mapping technique followed by a scheduling solution for WSN applications was proposed to improve the partitioning technique.

Clustering is another technique to minimise energy consumption with a guarantee of deadline constraints. In (S. Park, 2012), the authors presented an energy-efficient fair clustering scheme that had a cluster head node at the centre of the cluster. Wei et al. in (D. Wei, November 2011) completed the work by proposing a procedure to choose the cluster head candidates; first, for each data collection round, the ratio of initial energy level to the average initial energy of the network is calculated; then, based on these values the cluster head candidates are selected. The node with more resources is picked for data transmission. Clustering, however, has a technical limitation: it can only be used in wireless sensor clusters where all sensors are equipped with DVS processors and have computation ability.

**2.2.2. Link Layer**

With regards to energy consumption, the link layer has received a remarkable amount of attention, mainly in energy-aware routing (Younis., 2005) where the aim is to minimise transmission power by multi-hop data transmission instead of direct sensor-link communication. Power consumption in this layer takes into account the consumed energy due to collisions between the radio transmissions of nodes, unnecessary active states due to keeping receivers in the active mode instead of switching to other modes, and the energy required to move from one mode to another mode in the radio circuit (Xu and Saadawi, 2001).

A sensor consumes a large amount of energy during data transmission through three major activities: transmission, reception, and being idle. One study (Langendoen., 2003) showed that the ratio of power consumption in a processor (including CPU, memory) compared to the radio for the sensor nodes alters from 1:12.5 when both processor and radio are in sleep mode, to 1:4.76 when both are in active mode. As the largest energy consumer in a sensor, radio should play an important role in managing energy consumption and extending sensor lifetime.

In (Dong et al., 2005), the authors addressed the problem of energy-efficient reliable wireless communication in the presence of unreliable or lossy wireless link layers in multi-hop wireless networks. Their main focus was on single path routing. Banerjee and Misra in (Misra and Banerjee, 2002) explored the effect of lossy links on energy efficient routing and solved the problem of finding the minimum energy paths in the hop-by-hop retransmission model. However, they all followed a conventional design principle in the network layer of wired networks: after the best path(s) between a source and destination is (are) calculated, all data flows from source and destination follow the selected path(s) until the path is updated after certain topology management update period. ExOR (Biswas and Morris, 2005) challenged this conventional design principle in the network layer. MORE was presented in (Chachulski et al., 2007) as a MAC-independent opportunistic routing protocol. It randomly mixes packets before forwarding them. This randomness ensures that routers that hear the same transmission do not forward the same packets. Thus, MORE needs no special scheduler to coordinate routers.

**2.2.3. MAC Layer**

The problem of how to efficiently employ the residual energy of sensors has been the main concern in designing and developing MAC protocols for WSNs (Kurtis Kredo and Mohapatra, 2007). In this layer, the major energy drift results from collision, control packet overhead, idle listening and overhearing, in which the former plays an undeniable role in designing and choosing energy-efficient MAC protocols in wireless networks. Among popular protocols, two are suitable for this case: time division multiple access (TDMA), and code division multiple access (CDMA).

As stated before, one of the approaches to save energy in the link layer is to switch the radio to sleep mode. To take advantage of this opportunity, the link layer requires a time-based medium sharing, e.g., TDMA, with accurate clock synchronisation to properly schedule state transitions; an alternative is to use two radios to separate channels for data and control messages. TDMA and similar approaches, however, are not suitable for many application in WSNs even though they stop medium contention and reduce energy consumption. Since scheduling time slots is NP hard problem, TDMA and time-based medium sharing approaches do not scale properly. Moreover, these approaches often adapt slowly to changes in the traffic flow and density due to the need for pre-scheduling control messages (Ibrahiem M. M. El Emary, 2013).

CDMA is a promising MAC protocol for most wireless sensor network applications in terms of avoiding collision and supporting bounded delay; however, implementing the original CDMA protocol requires significant changes in the design of sensors. For instance, this protocol needs a large memory to store the codes of all delayed sensors, which is in contrast to the small memory nature of sensors and therefore limits the scalability of CDMA. The transmission time of a message is also lengthened, resulting in an increase in energy consumption, due to the bit encoding part of CDMA. As a result of these limitations, in addition to the circuit complexity and cost of the radio circuit, the designer is required to use only a part of the CDMA protocol to allow a practical implementation of small inexpensive sensor devices, as well as to consume only a small portion of the sensors' energy (Ibrahiem M. M. El Emary, 2013).

**2.2.4. Network Layer**

The network layer consists of a few parts, each one involving different techniques to reduce energy consumption of the network and ultimately improve its lifetime; this section studies these strategies. Briefly, there are a few easy techniques to reduce communication load and therefore consume less energy: among them are decreasing the amount of transmitted data, reducing the number of reporting sensors, and shortening the communication range (Z. Tan, March 2011), to name a few. Since there are different types of nodes in a network and each one has its own energy requirement, assigning energy according to requirement makes it possible to avoid the wastage of residual energy. Non-uniform energy assignment achieves a balance between energy efficiency and energy balance simultaneously (Z. Tan, March 2011). Despite its benefit, monitoring the energy requirement of each node and assigning an appropriate task is very difficult. Generally, sensors have a high degree of cooperation in nature, and the authors of one study (Shaoqing and Jingnan, 2010) employed this behaviour to propose a data transmission policy called energy-efficient cooperative communication (EECC).

**2.2.4.1 Topology**

Determining the best topology among nodes in order to provide a connected network for routing packets to the destination is a significant operation in WSNs. There are several factors that are important in selecting a suitable topology, such as energy efficient deployment and maintenance during the network lifetime, so that the network achieves maximum connectivity with minimum energy consumption. Topology control protocols aim to establish resilient network topology at the same time as minimising the energy consumption in establishing and maintaining the topology. Xu et al. (Xu et al., 2003) mentioned a number of challenges, including duty cycle control of redundant nodes, connectivity maintenance, self-configuration and redundancy identification in a localised and distributed fashion. Two significant methods for tackling these challenges are Geographic Fidelity (GAF) and Cluster-based Energy Conservation (CEC) protocols. GAF uses a node's location information, determined by a GPS, to configure redundant nodes and configures them into small groups using localised and distributed algorithms. CEC has the same fundamental operation but does not depend on location information and radio propagation. The authors of (Anastasi et al., 2009) simulated these two methods in the same situation. The results show that CEC consumes half of the energy used in GAF protocols. In contrast, when the nodes move frequently, CEC turns off the nodes more often and consequently consumes more energy than GAF. Therefore, GAF is more efficient than CEC in high mobility environments. In (Le et al., 2008), a new approach was proposed to reduce protocol overheads created by the CEC protocol and the energy consumption of GPS-attached sensors. In this approach, an energy-rich node such as a base station informs the sensors about their cluster ID and cluster area by sending a sweeping beacon. Therefore nodes have information about which cluster they belong to and hence they do not need to carry a GPS. The authors of (Ramesh et al., 2012) compared balanced and progressive topologies for sensor networks. Both the balanced and progressive topologies provided energy efficiency, but the best choice depended on the network size.

Various kinds of topology, such as tree, mesh, clustered, ad-hoc, and others, provide a virtual backbone for routing in WSNs. Salhieh et al. (Salhieh et al., 2001) examined the influence of different types of mesh topologies (2D and 3D topologies with different numbers of neighbours) on the power dissipated. According to their results, "increasing the number of neighbours decreases the number of transmission and total power dissipated in the system." Their main point was that selecting a suitable topology is important as it can support more energy-efficient routing strategies.

In (Minhas et al., 2009), an online multipath topology management algorithm was presented; for a given topology management request, their technique maximises the lifetime of the network by fair distribution of source to sink traffic along a set of paths. In (Baccour et al., 2010), Fuzzy membership functions were applied to the distance between nodes and the nodes' residual energy to form an edge weight function in a multipath topology. The authors claimed a better lifetime for the multipath scheme over a single-path fuzzy topology management scheme and online maximum lifetime heuristic using extensive simulation on a variety of network scenarios. The authors of (Zhang et al., 2013) proposed a topology management algorithm called ESRAD using

Dijkstra to minimise energy consumption in WSNs. Under the assumption that energy consumption is proportional to the number of hops, ESRAD formulates energy consumption at both the node and edge and engages Dijkstra to find the shortest paths with the least energy consumption.

In (Musznicki et al., 2012), after categorising the most common WSN multicast procedures based on the geographic position of a target group, the authors presented an algorithm based on Dijkstra for discovering the shortest energy-efficient paths via nodes that provide the maximum geographical advance towards sinks. This algorithm is based on the assumption of availability of the position of the current node, nodes in its neighbourhood (in the radio range of the node) and the location of associated sinks. Bhattacharya et al. (Bhattacharya and Kumar, 2014) presented an algorithm that generates the minimum length multicast tree to send data from one node to multiple sinks in a WSN. Named the Toward Source Tree (TST) algorithm, it focuses on minimising the number of hops, one of the most important factors in wireless sensor networks, by producing an energy efficient multicast tree with a low complexity.

A comparison of energy consumption between chain, grid and random topologies was studied in (Qiong et al., 2013). The comparison revealed that grid topology had the highest energy consumption followed by random and chain topologies, in that order; chain topology also showed better packet delivery rate than the others. In fact, grid topology had the worst performance in both energy consumption and packet delivery rate. The authors concluded that the achieved results were a direct outcome of the routing protocols, as the main parameter in the comparison.

**2.2.4.2. Routing**

Since routing is a significant and costly task in WSNs, routing protocols should be energy efficient to increase the network lifetime. Al-Karaki and Kamal (Al-Karaki and Kamal, 2004) discussed types of networks, topologies and protocols and their influences on the energy cost. In homogenous sensor networks, all nodes are the same and the routing tasks are assigned equally among the nodes, while in heterogeneous networks the nodes have different capabilities. Nodes with high capability may be assigned more responsibility and overall energy consumption can be reduced by optimising arrangements.

Routing protocols operate on topologies such as tree, mesh, clustered, etc to deliver data to the destination. Different methods use different techniques to extend the lifetime of the sensor. SPIN (Sensor Protocols for Information via Negotiation) (Heinzelman et al., 1999b) is a routing technique based on node advertisements in which nodes only need to know their one-hop neighbours. The technique is, however, not suitable for applications that require reliable data delivery. LEACH (Low-Energy Adaptive Clustering Hierarchy) (Heinzelman et al., 2000) is a clustered routing algorithm where the cluster-heads are responsible for relaying data and controlling the cluster. Although LEACH is an effective technique for achieving prolonged network lifetime, scalability, and information security, LEACH does not guarantee optimum routes. Directed Diffusion technique is a data centric, localised repair, multi-path delivery for multiple sources, sinks and queries (Intanagonwiwat et al., 2000) aiming to find optimal paths.

There are a few energy-efficient topology management protocols that may be used to find the shortest paths among nodes and hops, mostly based on Dijkstra, with the condition of wireless link reliability; this condition implies lack of packet loss in wireless links (Li et al., 2005),(Wan et al., 2002). To remedy the unreliability of the wireless channels, multipath topology management (Heinzelman et al., 2000), (Garcia-Luna-Aceves et al., 2006), building reliable backbone (Wan et al., 2002), and an energy efficient reliable routing structure (Dong et al., 2005), (Misra and Banerjee, 2002) have been used. Usually the shortest or the lowest energy path is defined as the optimal path for relaying data. Thus, each node needs to be aware of its neighbours' capabilities, such as residual energy and memory capacity, in order to select the best neighbour to send data; however, Shah and Rabaey (Shah and Rabaey, 2002) argued that the lowest energy path is not the best choice. They suggested an energy-aware method where nodes select different directions based on the residual energy of their neighbours. Even though significant improvement in network survivability was achieved, the method requires frequent updates on path energy information in routing tables, resulting in an additional overhead in self-organised wireless sensor networks. In another paper (Fengyuan Ren, December, 2011 ), the authors pointed out that forwarding packets to the sink along the minimum energy path may reduce energy consumption but results in an unbalanced distribution of residual energy among the sensor nodes; therefore, they proposed an energy-balanced routing policy to tackle this problem. Hwang et al. (Ghaffari, 2014) proposed another protocol for energy-efficient routing for WSNs with holes, created due to uneven deployment.

**2.2.5. Congestion Control**

Congestion control algorithms used for wired networks are not appropriate for wireless networks, as packets need to be retransmitted and additional energy has to be consumed. The authors of (Scheuermann et al., 2008) simulated a TCP-like congestion control in a wireless network. As a result, they reported that the throughput drops rapidly when the traffic load increases beyond a certain optimal level due to congestion and packet collision. They proposed an alternative method called a hop-by-hop congestion control based on a backpressure mechanism. In this approach, the input flow is maintained below the output flow. This means that if the previous forwarded packet is overheard, the next packet may then be sent. According to their simulations, the approach is successful in increasing the network throughput and decreasing the delay and the retransmission load in different topologies compared to other existing protocols.

**2.2.6. Application Layer**

The centrepiece of this layer is an aggregator, which combines data arriving from different nodes, removes redundant data and compresses it before transmission to the intended destination, recalling that reducing the number of transmissions conserves energy. Generally, routing in WSNs considers data aggregation at some nodes. Similarly, for data aggregation, the routing protocol plays an essential role; in cluster-based protocols, cluster head nodes play the role of aggregator to compress data arriving, aggregate data and perform in-network processing (Al-Karaki and Kamal, 2004). Data aggregation techniques are tied to the method used to generate data in sensors and route packets through the network. Before forwarding to the sink, generated data from different sensors can be processed together. First, data from different nodes are fused together, then processed locally to remove redundant information and finally transmitted. Fusing data from different nodes, (a.k.a. data fusion), or in general aggregating data requires the WSNs to be time-synchronised.

A protocol at the routing level is required for proper data-gathering; this protocol is formulated to configure the network and collect information from the environment (P. Mohanty, 2010). In each round of data gathering, sensors (nodes) collect data from the environment and send them to the sink (Norman et al., 2010). In general, existing data-gathering protocols (P.

Samundiswary, 2010) are grouped into different categories depending on the network topology and routing protocols (P. Mohanty, 2010, Hussain., 2010). In a simple way, data from different sensors are aggregated (e.g., sum, average, min, max, count) (Jayanthy., 2010). In a more robust way, data fusion is used to combine several unreliable data measurements to produce a more accurate signal (i.e., enhancing the common signal and reducing the uncorrelated noise). The ultimate goal is to consume less energy while transmitting all the data to the sink in order to improve the lifetime of the network (Liang et al., 2009, S. K. Narang, March 2010).

**2.2.7. Energy Harvesting**

Several technologies exist to extract energy from the environment, such as solar, thermal, kinetic energy, and vibration energy, and the network lifetime may increase by using power harvesting technologies. Weddell, Harris and White (Weddell et al., 2008) explained the advantages of energy harvesting systems as the ability to recharge after depletion and to monitor energy consumption, which may be required for network management algorithms.

Energy harvesting technologies plays an important role in applications that are expected to operate for a long duration. There are various challenges in energy harvesting management. Kansal et al. (Aman Kansal et al., 2010) classified energy sources into four categories and corresponding challenges: uncontrolled/predictable, uncontrollable/unpredictable, fully controlled and partially controllable. They emphasised that energy management in energy harvesting systems is fundamentally different from battery operated systems because of the unpredictable available power. They showed that the power availability varies in time and for different nodes in the network. This presents some difficulties to a node when it has to make decisions based on knowledge of the residual energy of the network. Additionally, different nodes may have different harvesting opportunities, so it is important to assign the workload according to the energy availability at the harvesting nodes. To solve these problems, they proposed an analytical model for energy harvesting and performance. Moreover, Kansal et al. at (Aman Kansal et al., 2010) suggested an approach to balance the harvesting energy and the load in a node. They explained the requirement for collaboration between power management applications when the harvesting source cannot support the consumption level of the node's load.

There is a significant interest in energy harvesting for different wireless sensor applications to improve their sustainable lifetimes, but there is also a balanced need to guarantee performance and exploit the available energy efficiently. Most of the studies in the field of wireless sensors are based on residual battery status, while in harvesting systems the problem still is the estimation of the environmental energy availability at nodes. Although Kansal et al. (Aman Kansal et al., 2010) proposed an environmental energy availability method for power management, their method is based on a predictable energy resource and cannot be used with an unpredictable resource.

# Chapter 3. A Generic Energy-Driven Architecture in Wireless Sensor Networks

In this chapter, Energy Driven Architecture (EDA) is proposed as a robust architecture taking into account all principal energy constituents of wireless sensor network applications, published in (Kamyabpour and Hoang, 2010). By building a single overall model, a feasible formulation is then proposed to express the overall energy consumption of a generic wireless sensor network application in terms of its energy constituents. The formulation offers a concrete expression for evaluating the performance of a wireless sensor network application, optimising its constituent's operations, and designing more energy-efficient applications. Extensive simulations are used to demonstrate the feasibility of our model and energy formulation.

## 3.1 Problem statement

Energy consumption is easily one of the most fundamental and crucial factors determining the success of the deployment of sensors and wireless sensor networks (WSNs) due to many severe constraints, such as the size of sensors, the unavailability of a power source, and inaccessibility of the location, which prevents further handling of sensor devices once they are deployed. Efforts have been made to minimise the energy consumption of wireless sensor networks and lengthen their useful lifetime using various approaches at different levels. Some approaches aim to minimise the energy consumption of the sensor itself at its operating level (Min et al., 2001), some aim to minimise the energy spent in the input/output operations at the data transmission level (Alzoubi et al., 2002), and others target the formulation of sensor networks in terms of their topology and related routing mechanisms (Shah and Rabaey, 2002). The common goal here is to reduce the energy consumption of some components of the application as much as possible by reducing the tasks that have to be performed by the sensors and the associated networks, yet still fulfill the goal of the intended application. In addition to the minimisation effort, some approaches have tried to replenish the energy capacity of the sensors by building into them components and mechanisms for harvesting additional energy from available energy sources within their environments, such as solar, thermal, or wind power sources (Raghunathan and Chou, 2006).

Yet another approach is to scan systematically through the levels of the OSI network reference model and minimise the energy consumption at some level (if feasible) with the hope that this will reduce the overall energy consumption of the entire network and the application (Joaqu et al., 2007).The main problem with this approach is that it may succeed in reducing the energy consumption in one component of the overall WSN application, but this gain is often negated by an increase in the energy consumed by other components of the application. There has been very little understanding of the overall energy consumption map of the entire application, the major components of this energy map and the interplay between the components.

This chapter discusses our approach to tackle the problem from a different angle by focusing on energy constituents of an entire sensor network application. An energy constituent represents a major energy-consuming entity that may be attributed to a group of functional tasks. Eventually, these tasks have to be mapped to energy consumed actions that have to be performed by sensors and other components such as sensors' antennas, transceivers and central processing units.

The ultimate aim behind this approach is to produce an energy map architecture of a generic WSN application with essential and definable energy constituents and the relationships between these constituents so that one can explore strategies for minimising the overall energy consumption of the entire application. The EDA architecture is the result of our efforts in this direction. Based on this architecture, this chapter proposes a formulation of the energy consumption of an entire application in terms of mathematical expressions that enable one to analyse and optimise the energy consumption function. The architecture focuses on energy constituents rather than network layers or physical components. Importantly, it allows the identification and mapping of energy-consuming entities in a WSN application to energy constituents of the architecture. Specifically, in this chapter we not only identify energy constituents in WSN applications but also identify individual components and their contributions to each of the constituents of the EDA architecture. The energy consumption of each constituent is formulated in terms of its components. Furthermore, we identify and take into account in the mathematical expressions salient parameters (or factors) that are believed to play a significant role in an energy component. Preliminary simulation results are also presented to demonstrate the feasibility of the model for further study and evaluation.

## 3.2 Overall Energy Consumption Formulation

We suppose a continuous time between t1 and t2 for the energy consumption measurement. Residual energy in time t is defined by omitting consumed energy in Δt from the initial battery power in t-Δt. Thus, the energy consumption will be determined in Δt as:

$$\begin{cases} E_{residual,i}(t_2) = E_{residual,i}(t_1) - E_{consumed,i}(\Delta t) \\ E_{consumed,i}(\Delta t) = \frac{\partial E_{residual,i}(t)}{\partial t} \Delta t \\ \Delta t = t_2 - t_1 \end{cases} \quad (3-1)$$

Realistically, there is a nonlinear relationship between the overall energy consumption of the system and its constituents depending on the application and the overall design. However, this nonlinear formulation requires more extensive exploration as there is no deep understanding of metrics associated with the energy of each constituent; also, there are no solid mathematical models that can handle such a non-linear relationship. Therefore, a simpler linear approach is adopted to model the overall energy consumption and explore the implications; future work will explore nonlinear approaches. In the following formula the overall energy is expressed as a linear combination of the EDA's constituents. Interplay among the components can be taken into account in terms of their weights as some function of the design of the WSN and the application.

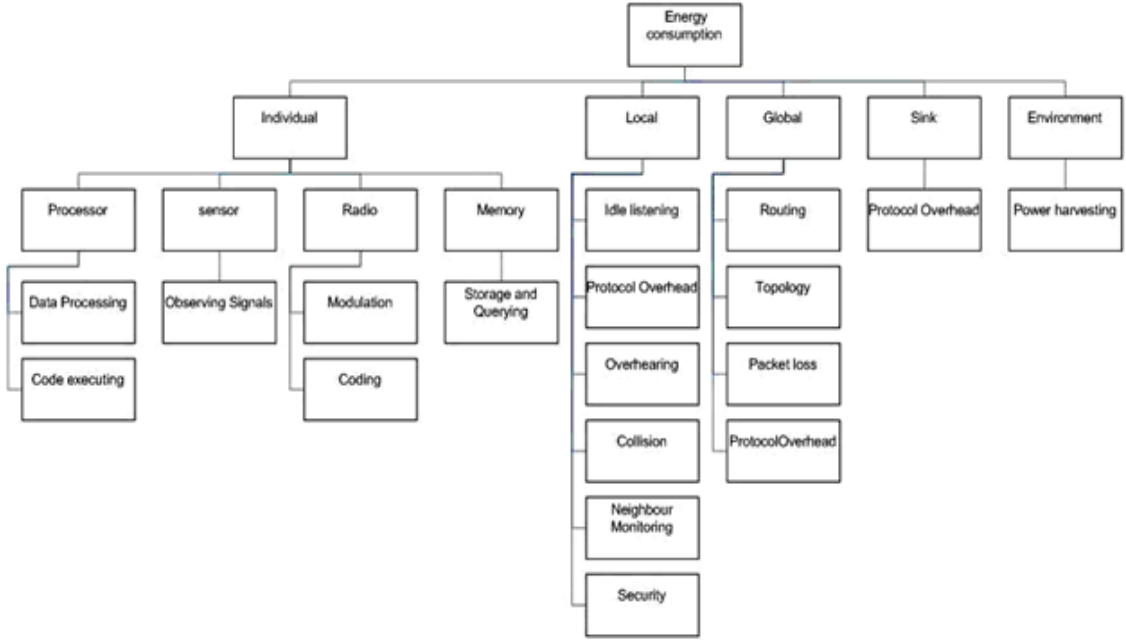

Figure 3-1. Energy consuming constituents

The total energy consumption of node $i$ in the interval $\Delta t$ based on constituent of EDA can be formulated as follows:

$$\begin{cases} E_{consumed,i}(\Delta t) = \lambda_1 E_{indiviual,i}(\Delta t) + \lambda_2 E_{local,i}(\Delta t) + \\ \qquad\qquad \lambda_3 E_{global,i}(\Delta t) + \lambda_4 E_{battery,i}(\Delta t) + \lambda_5 E_{sink,i}(\Delta t) \\ subject\ to: \\ 1.\ E_{local,i}(\Delta t) > 0 \\ 2.\ E_{global,i}(\Delta t) > 0 \\ 3.\ \lambda_1 E_{indiviual,i}(\Delta t) + \lambda_2 E_{local,i}(\Delta t) + \lambda_3 E_{global,i}(\Delta t) + \lambda_5 E_{sink,i}(\Delta t) < \lambda_4 E_{battery,i}(\Delta t) \end{cases} \quad (3-2)$$

Figure 3-1 shows each constraint in terms of their energy consuming tasks in the network. The first constraint expresses the condition of the necessity to establish a collaboration connection. The second constraint shows the necessary and sufficient condition of accessibility of the node in the network. The third constraint means a node should have enough energy to do network tasks otherwise it is not active and should be removed from the network calculations. Each constituent is expressed in terms of key parameters (or factors). These key factors are determined based on application requirements. On the other hand, these parameters may influence more than a single

constituent; hence energy constituents may partially overlap. Consequently, the interplay among energy constituents must be taken into account in evaluating the overall energy consumption of the entire setup. For example, the number of neighbours determined by topology in the global constituent has a direct influence on energy consumption of the local constituent. We will elaborate on the model for each of the constituents in the following sections.

## 3.3 Individual Constituents

The individual constituent can be a state-based constituent. Figure 3-2 shows a typical example of an individual constituent's energy consuming states. Each unit has different energy consumption levels in different states. In addition, this constituent involves two different types of transitions: transitions between units and transitions between states of a single unit. The overall energy consumption in individual constituents is expressed as follows:

$$\begin{cases} E_{individual\ ,i}(\Delta t) = \sum_{u=1}^{N_u} \sum_{s \epsilon S} \sum_{w \epsilon W} I(e_{u,s}, e_{u,w}, t_{u,s}) & (3-3) \\ where\ \sum_{s \epsilon S} e_{u,s} > \sum_{w \epsilon W} e_{u,w} \end{cases}$$

where $t_{u,s}$ is the duration of the activity in each state. Since most energy minimisation methodologies use idle and sleep states to avoid wasting energy in idle states, the above constraint states that the total energy consumed for switching among states $e_{u,w}$ should be smaller than the total energy consumption of states $e_{u,s}$. Energy consumption in an active state for each unit depends on several factors as follows:

- Energy consumption of the processor unit in an active state depends on the number of processed bits $b_{proc.}$ and its operating voltage and frequency, as

$$e_{1,active}(\Delta t) = F_1(f, b_{proc.}) \qquad (3-4)$$

In most modern processors, energy consumption of the processor is proportional to the voltage and the frequency of the operation, as (N.B.Rizvandi et al., 2011)

$$p \propto cv^2 f \qquad (3-5)$$

Since the frequency and the voltage can be related, frequency is considered as an prevalent parameter in this unit.

- Energy consumption of a sensor unit in an active state depends on the sensor radius $r_{sense}$, the data generation rate $g_{sense}$, and the number of generated bits $b_{sense}$, as

$$e_{2,active}(\Delta t) = F_2(r_{sense}, g_{sense}, b_{sense}) \qquad (3-6)$$

- Energy consumption of a memory unit in an active state depends on the number of stored bits $b_{store}$, the number of memory read $e(rd)$ and write $e(wt)$, and the duration of storage $t_{store}$, as

$$e_{3,active}(\Delta t) = F_3(b_{store}, e(rd), e(wt), t_{store}) \qquad (3-7)$$

- Energy consumption of the transceiver unit for digital signal processing in an active state depends on the number of received $b_{Rx}$ and transmitted bits $b_{Tx}$, and the amount of needed energy for coding $e(code)$ and decoding packets $e(decode)$:

$$e_{4,active}(\Delta t) = F_4(b_{Rx}, b_{Tx}, e(code), e(decode)) \qquad (3-8)$$

The energy wastage in idle and sleep states can be measured according to the base amount of energy consumption in these states, which depends on unit type and the duration of staying in the state (Kamyabpour and Hoang, 2010). Moreover, switching among the unit's states also consumes a considerable amount of energy, which is measured differently for different types of unit.

Explicitly, Figure 3-3 shows an example of a data generation sequence in an individual constituent. Data generation time (sensing time), process time, storage time, and data transmission time may all contribute to the overall energy consumption of the individual constituent. These parameters can determine the number of task transmissions between units. For example, if the data generation time is smaller than the process time, the number of memory read and writes will increase because the data should be stored until the processor completes the task. Also, if the process time is smaller than the transmission delay then the number of memory read and write will increase. Limited resources of a sensor, such as memory units, should be used carefully. For instance, if a sensor does not have enough memory it cannot process received packets. Thus, we need to optimise the parameters of each unit with respect to the parameters of other units. Therefore, the active time in each constituent is one of the most important factors in the energy consumption of other units.

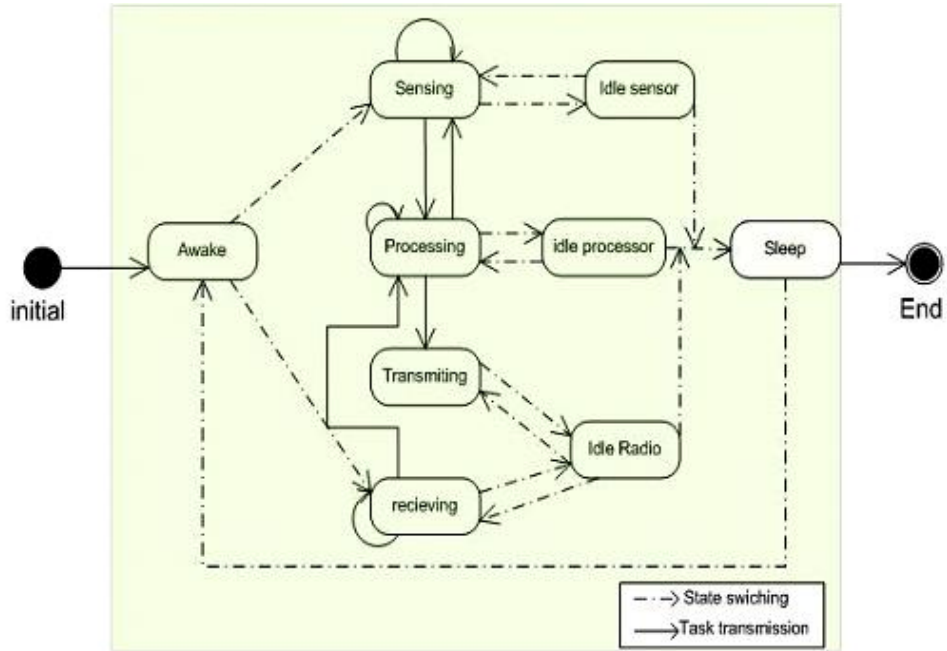

Figure 3-2. General State Diagram of an individual constituent.

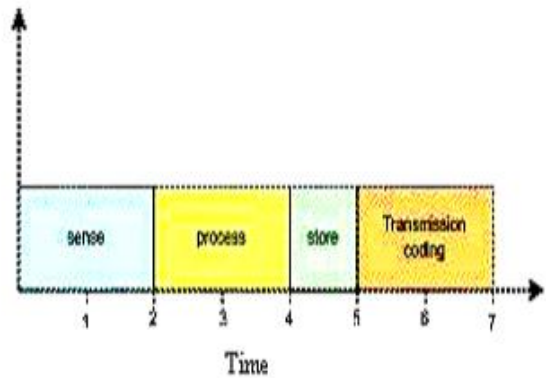

Figure 3-3. An example of the data generation sequence in an individual constituent

## 3.4 Group Energy Consumption

Generally, transmission is a key task in communication among nodes. Energy consumption for packet transporting in the network is proportional to the distance. The distance to neighbours can increase or decrease the energy consumed by a radio channel to transmit a single data bit. Heinzelman et al. (Heinzelman et al., 2000) derived the energy consumed to transmit and receive of k-bit message for a microsensor. The required energy for the transmit amplifier to send a bit is shown as $e_{amp}$. Hence, in local and global constituents, the energy consumption for transmitting $k$ bits to a node at distance $d$ from the transmitting node is defined as follows:

$$E_{Tx}(d) = e_{amp} d^2 k \qquad (3-9)$$

and energy consumption of receiving $k$ bits from a node is proportional to the receiver electronics energy per bit, $e_{elec}$, which is defined as follows (Heinzelman et al., 2000):

$$E_{Rx} = e_{elec} k \qquad (3-10)$$

These equations are general forms of energy consumption for communication. The important factors, which increase or decrease the energy consumption of transmission and receiving operations, should be considered by network designers. Determining the number and the distance to neighbours, transmission rate, receive rate, optimum size of data and message packets are all important in determining the energy consumption of the radio channel. Each factor is thus considered in a component of a constituent of the EDA architecture. Although the transmit amplifier is shared among group constituents, its energy consumption is determined based on its different roles in different constituents. The following is a discussion of each constituent.

### 3.4.1 Local Constituent

The local communication is concerned with initiating and maintaining all communications between a sensor node and its immediate neighbours so that they can co-exist to perform their roles within a WSN as dictated by the objective of the application. The following equation shows the local energy consumption of a node in interval time $\Delta t$:

$$\begin{cases} E_{local,i}(\Delta t) = \sum_{j \in neighbour_i} L\left(e_{ij}(mon), e_{ij}(sec), e_{ij}(idle), e_{ij}(local), e_{ij}(coll), e_{ij}(ohear)\right) \\ subject\ to: \\ 1.\ neighbour_i \geq 1 \\ 2.\ e_{ij}(local) \leq e_{ij}(idle) + e_{ij}(coll) + e_{ij}(ohear) \end{cases} \quad (3-11)$$

The first constraint shows that the node has to have at least one neighbour to be able to relay data and survive in the network. The second constraint is the condition of having optimum energy consumption in the local, which is energy consumed by control packets of the local protocol; the aim is to manage effective access to the shared media in order to avoid collision, idle listening and overhearing. The energy consumed by control packets in local should not be bigger than the sum of energy consumption of these costly problems in the network when there is no management of the shared media.

- Neighbour monitoring is used for gathering information on a neighbour's available resources, such as residual energy and memory space. Therefore energy consumption depends on the distance between neighbours and number of exchange bits. If $d_{ij}$ is the distance between node $i$ and its neighbour $j$ and $b_{mon}$ is the number of exchange bits between them, $r_{Tx}$ is the transmission radius and the number of neighbours is proportional to $r_{Tx}$, the energy consumed by monitoring $e_{ij}(mon)$ is given by

$$e_{ij}(mon) = F_5(d_{ij}, b_{mon}, r_{Tx}) \quad (3-12)$$

- Security management is for preventing malicious nodes from destroying the connectivity of the network and tampering with the data. Energy consumed by security $e_{ij}(sec)$ depends on the distance between neighbours $d_{ij}$ and the number of exchange bits, where $b_{sec}$ is the number of exchange bits between node $i$ and its neighbour $j$:

$$e_{ij}(sec) = F_6(d_{ij}, b_{sec}, r_{Tx}) \quad (3-13)$$

- Various local communication protocols have to be performed to maintain the node's relationship with its neighbours. This type of protocol overhead must be taken into account in terms of energy consumption. Energy consumed by local communication $e_{ij}(local)$ depends on the distance between neighbours and the number of exchange bits, where $b_{local}$ is the number of exchange bits between node $i$ and its neighbour $j$:

$$e_{ij}(local) = F_7(d_{ij}, b_{local}) \qquad (3-14)$$

- If a node does not receive an acknowledgment for the transmitted packet, it has to retransmit the packet. This situation happens when neighbours transmit packets on the shared medium at the same time. In this case, some parameters come into consideration: distance between neighbours, number of retransmitted bits, number of neighbours, and data transmission rate. The energy consumed by collision $e_{ij}(coll)$ is given below, where $d_{ij}$ is the distance between node $i$ and its neighbour $j$ and $b_{reTx}$ is number of retransmitted bits between them. $n_i$ is the number of neighbours of node $i$, $g_{Tx}$ is transmission rate of node $i$, and $r_{Tx}$ is the transmission radius. The network density, $net_{dens}$, may increase or decrease the probability of collision.

$$e_{ij}(coll) = F_{87}(d_{ij}, b_{reTx}, n_i, g_{Tx}, r_{Tx}, net_{dens}) \qquad (3-15)$$

- A node receives packets that are sent to the shared medium. Even when the node is not the destination, it still has to examine the packet to figure out what to do. Energy consumed by overhearing $e_i(ohear)$ depends on the number of overheard bits in node $i$ ($b_{ohear}$) and network density ($net_{dens}$), which may increase or decrease the probability of collision of overheard packets:

$$e_i(ohear) = F_9(b_{ohear}, r_{Tx}, net_{dens}) \qquad (3-16)$$

### 3.4.2 Global Constituent

The global constituent is concerned with maintenance of the whole network, selection of a suitable topology and the employed routing strategy. This may include energy wastage from packet retransmissions due to congestion and packet errors. The global constituent is defined as a function of energy consumption for topology management, packet routing, packet loss, and protocol overheads.

$$\begin{cases} E_{global,i}(\Delta t) = G\big(e_i(topo), e_i(rout), e_i(global), e_i(pktls)\big) \\ subject\ to: \\ \quad 1.\ e_i(rout) \geq 0 \\ \quad 2.\ e_i(rout) > e_i(topo) + e_i(global) + e_i(pktls) \end{cases} \qquad (3-17)$$

The first constraint shows that there is at least a path from node $i$ to destination within the network so that it participates in the global communication. The next constraint shows that the energy consumed for control packets and the retransmitted packet should be smaller than the routed data packets from an effective energy consumption point of view, otherwise this constituent wastes the node's energy.

- The energy consumption for establishing a relevant topology through the nodes based on the application's objective, $e_i(topo)$, can be calculated as:

$$e_i(topo) = F_{10}(a_i, b_{topo}, d_{iA}, n(\Delta t)) \qquad (3-18)$$

where $a_i$ is the number of nodes accessible nodes for node $i$, $d_{iA}$ is the distance between node $i$ and an accessible node, $b_{topo}$ represents the number of exchange bits for topology management and $n(\Delta t)$ determines the number of active nodes in the network in interval time $\Delta t$.

- The energy consumption for determining and maintaining hops and transporting packets to the destination, $e_i(rout)$, is a function of a few parameters:

$$e_i(rout) = F_{11}(h_{iD}, b_{rout}, d_{iD}, n(\Delta t)) \qquad (3-19)$$

The number of relaying hops can be expressed as a cost component in term of energy dissipation. It should be determined and minimised by a suitable routing method. The cost for maintaining the network connectivity should also be accounted for if hops fail during the network lifetime. $n(\Delta t)$ determines the number of active nodes in the network in interval time $\Delta t$. This may be useful for selecting the best routing method during the network lifetime. Therefore the routing method can be calculated dynamically according to the current network situation. $d_{iD}$, $h_{iD}$ are distance and number of hops between node $i$ and the destination, respectively.

- $e_i(global)$ represents the energy consumption due to protocol overheads. It is calculated based on the transporting cost of control packets for maintaining the overall network topology and configuration. $d_i$ is the distance between node $i$ and its neighbour and $b_{global}$ is the number of exchange bits between neighbours.

$$e_i(global) = F_{12}(d_i, b_{global}) \qquad (3-19)$$

- $e_i(pktls)$ represents the energy consumption due to packet loss. Selecting inappropriate topologies and routing methods may cause congestion and packet loss in the network. In this case, extra energy consumption has to be added if a node is required to retransmit a packet. $d_i$ is the distance between node $i$ and its neighbour and $b_{pktls}$ is an indicator of packet loss between neighbours.

$$e_i(pktls) = F_{12}(d_i, b_{pktls}) \qquad (3-20)$$

## 3.5 Environment Constituent

In cases where nodes are capable of extracting or harvesting energy from the environment, we propose to take into account this positive energy component in determining the lifetime of the WSN. The environment constituent as a positive energy component can be formulated as follows:

$$E_{battery,i}(\Delta t) = -H_i(t) \qquad (3-21)$$

where $H_i(t)$ is the amount of harvested energy at time $t$ by node $i$.

## 3.6 Sink Constituent

Energy consumption of nodes from a sink constituent viewpoint can be formulated as:

$$E_{snk,i}(\Delta t) = K(e_i(snk)) \qquad (3-22)$$

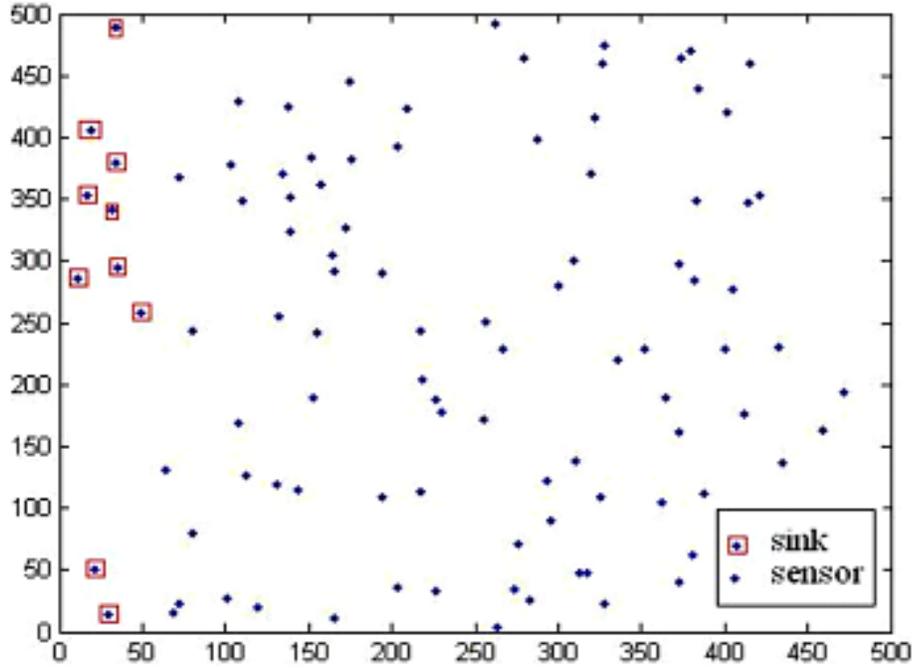

Figure 3-4. Randomly deployed sensors and sinks in the application

where $e_i(snk)$ shows energy consumed by each node to communicate with the sink and perform the sink's commands.

$$e_i(snk) = F_{14}(b_{snk}) \qquad (3-23)$$

The above equation means that the energy consumption of node $i$ for a sink constituent depends on the number of received bits from the sink.

## 3.7 Experimental Results

This section describes a range of simulated experiments conducted to evaluate the residual energy in the network with respect to different constituents of the EDA. Because events in the network occur at millisecond intervals and the initial power of the sensors is limited, the network is usually one to two minutes. Therefore the residual energy of the wireless sensor application was evaluated within an interval of sixty seconds. In particular, we focused on the individual, the local, and the global constituents. To gain a better understanding of the energy consumption of these constituents and their main parameters, the focus at this stage was on several parameters that are believed to play significant roles in the overall energy consumption. For the individual constituent, the sensor's sensing radius was selected as it determines the coverage of the sensor field. For the local constituent, we selected the transmission radius of a sensor, as it concerns the number of neighbours. For the global constituent, the routing scheme was chosen as it affects data transport from sensors to sinks. We investigated the influence of these constituents by measuring the residual energy and energy consumption of the network. Sensor sensing radius, sensor transmission radius, and routing scheme were considered as variables in our simulated experiments, while all other parameters were fixed; then, the variations in residual energy were compared for the different constituents' parameters to obtain the best result for an energy consuming component of a constituent.

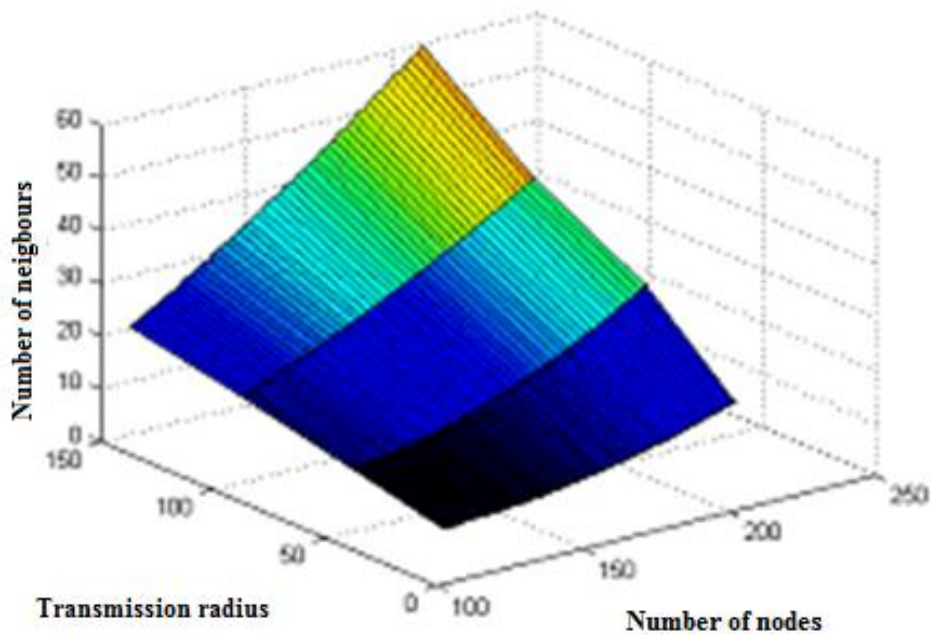

Figure 3-5. Average number of neighbours for different transmission radius
and network size

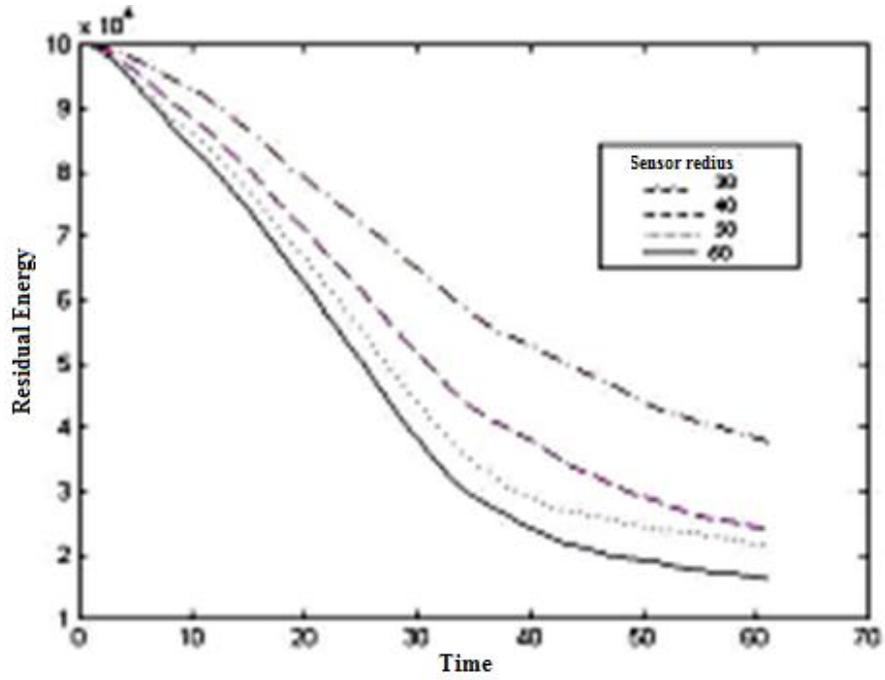

Figure 3-6. Residual energy for different sensor radius ($r_{Tx} = 130$, Selective)

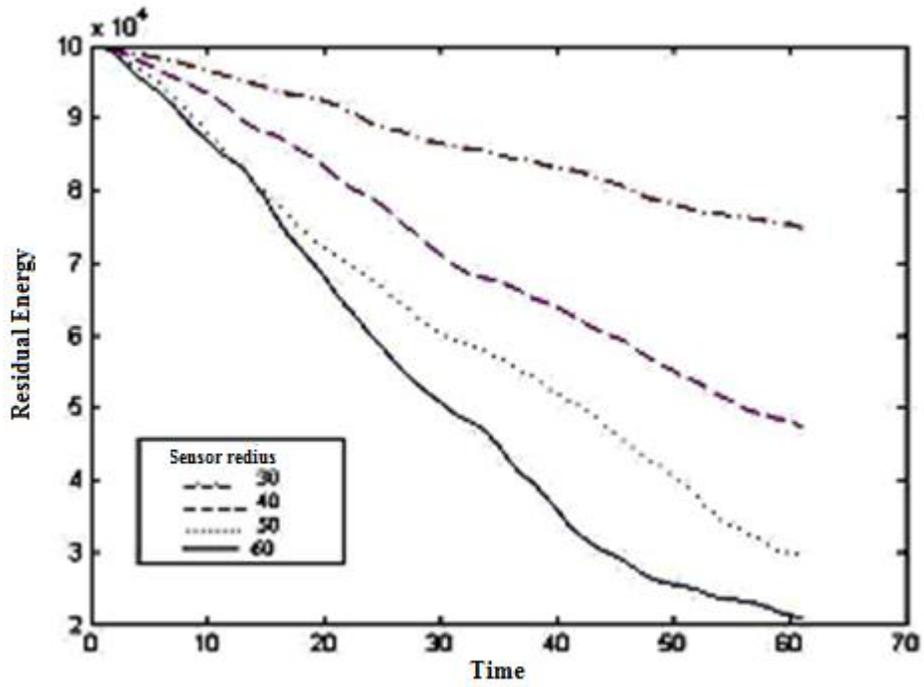

Figure 3-7. Residual energy for different sensor radius ($r_{Tx} = 130$, Random)

In our simulation, 100 sensors were deployed in a 500*500 pixel area (Figure 3-4) that generates data from environment events at random times and places in the area. In response to the type of sensing applications, we addressed generic data collected from environment as the central idea is not about the type of sensing but the relationship between the tasks to be performed by sensors and sensor networks and the total energy consumption. Sensing applications could equally environment temperature, pollution, or others. We considered the prevalent parameters of energy consumption of process, memory and radio units as constant in the individual constituent of all sensors. Also the duration of the experiments was assumed as constant 60 seconds. As the model is a task-oriented, 60 seconds of simulation time is adequate to account for all the tasks that a sensor can performed. Longer simulation times will not to alter the results of our task based model. A sensor generates data from the environment .We consider cost of sensing as a constant and it is clear that the frequency of sensing, the amount of generated data or sensing radius will increase or decrease the total energy consumption for sensing process. Relationships between the overall energy consumption and relevant tasks remain the same, except scaling factors. For our experiments, sensor radius was considered as a prevalent parameter of the sensor unit; other parameters of the individual constituent such as the sensing rate and the costs for different states of various units were kept constant for selective study. The influence of different sensor radius was measured on the overall residual energy of the network. Also, the considered variation of sensor radius parameter was the same for all sensors in the network.

For the local constituent parameters, the number of energy-consuming bits required to maintain an individual sensor's local environment and network density were kept constant for the duration of the experiments, but sensor transmission radius was varied. Neighbour selection usually is application-dependent, and a node placed in the area covered by another node may be chosen as a neighbour of that node. In our application the number of neighbours was changed based on the variation in transmission rate. Figure 3-5 shows how the number of neighbours varies according to transmission radius and network density. Clearly, the maximum number of nodes (200) with the maximum value for transmission radius (150) results in the highest average number of neigbours in the network. In addition, to be realistic, the cost of distance (Eqn. 3-9) was considered in transporting packets through to the network.

For the global constituent, we considered the routing method as the variable of interest. As topology and routing are costly and significant energy consuming components of the global constituent, they play the main roles in determining the residual energy of the network. Increasing the transmission radius increases the number of possible connections of each node and decreases the number of hops from nodes to their sinks. Nodes can establish connections with all nodes located in area reached by the node's transmission radius, and the type of connection among nodes is determined based on the geographic positions of nodes and sinks. We defined two types of connection: sender and receiver. A sender connection of a node is a connection used to send data. A receiver connection is one that the node uses only to receive data. Therefore nodes have knowledge of the position of their neighbours and the position of the sinks; they choose a sender connection with nodes that are closer to the sinks. In our experiments, two different routing methods were considered: Selective and Random. The selective routing method is based on the residual energy and busy degree of nodes (Kamyabpour and Hoang, 2010). In the random method, nodes randomly select a sender connection to send data to the sinks. In this application, nodes did not have the capability to extract energy from their environment, and had only their initial power. They consumed their initial power by surviving in the network. For this reason we did not consider an energy contribution from the environment constituent. A sink may be one or a group of powerful nodes and can be applied at any place in the application's area; we deployed a group of nodes as a sink in a special section of the area (the left side of Figure 3-4); in this way, it was easy to initialise, maintain, manage, and control network connections and paths to the sinks. Sinks do not manage or control sensors in the network; their only role is to collect received packets. Therefore, the energy consumption of the sink constituent is not considered.

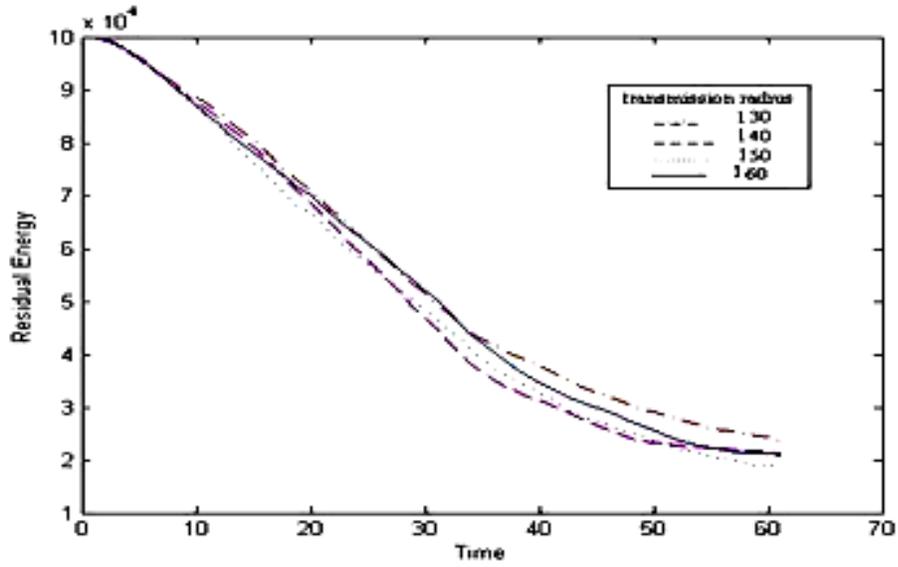

Figure 3-8. Residual energy with respect to different transmission radius ($r_{sense} = 30$, Selective)

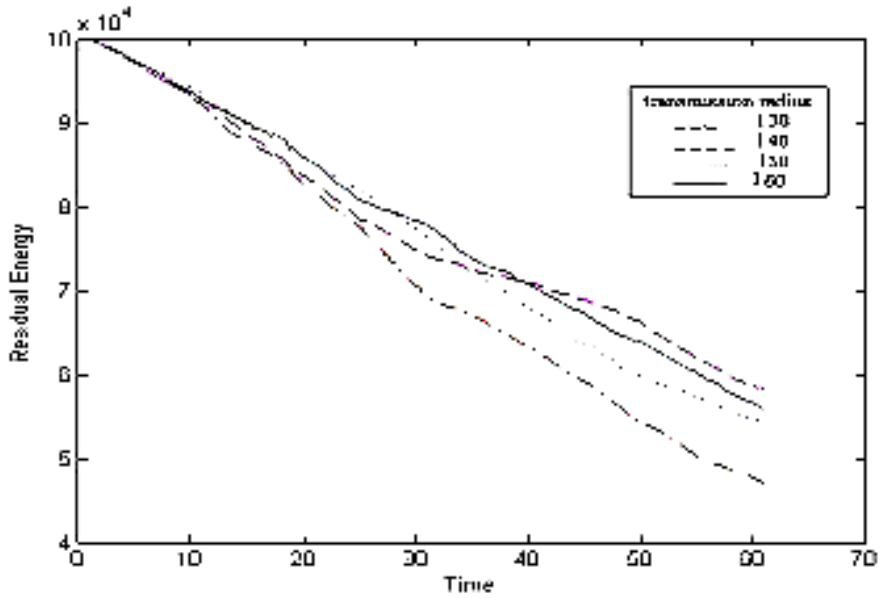

Figure 3-9. Residual energy with respect to different transmission radius ($r_{sense} = 30$, Random)

Figure 3-6 shows the variation in residual energy for four different sensor radiuses with constant transmission radius and selective routing method. The data show that the maximum and minimum residual energy, respectively, belong to the sensing radius 30 and 60 pixels . Because the shared sensing area increases if we increase the sensing radius, data redundancy in the network increases and consequently the energy consumption of routing increases. Thus there is an increase in energy consumption of the local constituent and a dramatic drop in network energy. As a result, with respect to the selective routing method and constant transmission radius, optimum energy consumption is obtained with the possible smallest sensing radius that covers the application environment. We repeated this test for the random routing method (Figure 3-7). The result was similar to the selective method test: the smallest sensing radius caused the biggest residual energy. Hence, an increase in sensing radius causes $e_{sense,active}$ to increase due to a longer duration of the active state of the sensing unit, and also $e_{local,i}$ rises because of increasing $e(rout)$; thus, in both tests it is expected that increasing the sensing parameter decreases the network's residual energy.

In Figure 3-8, transmission radius is considered as a prevalent parameter of the local constituent. Because the positions of nodes are not changed during the tests, and nodes should have at least one connection with another node in the network, we had to find a base value for the transmission radius. In this network, the limit of transmission radius was $r_{Tx} \geq 130$ in order to have a connected network at the initial time. The sensing radius was considered constant ($r_{Tx} = 40$) during the test and the applied routing method was selective. As can be seen, the variation in residual energy with transmission radius while keeping a constant sensing radius ($r_{sense} = 40$) different to the variation shown in Figure 3-6. Figure 3-9 is the same test using the random routing method. Comparing Figures 3-8 and 3-9 ($r_{sense} = 40$), the network has different residual energy with respect to the variations of the transmission radius during the test. Generally, increasing the transmission radius results in an increase in the number of neighbours ($neighbore_i$) and defines new neighbours at a longer distance. Collaboration with these new neighbours is costly; as a consequence, $E_{local,i}$ increases.

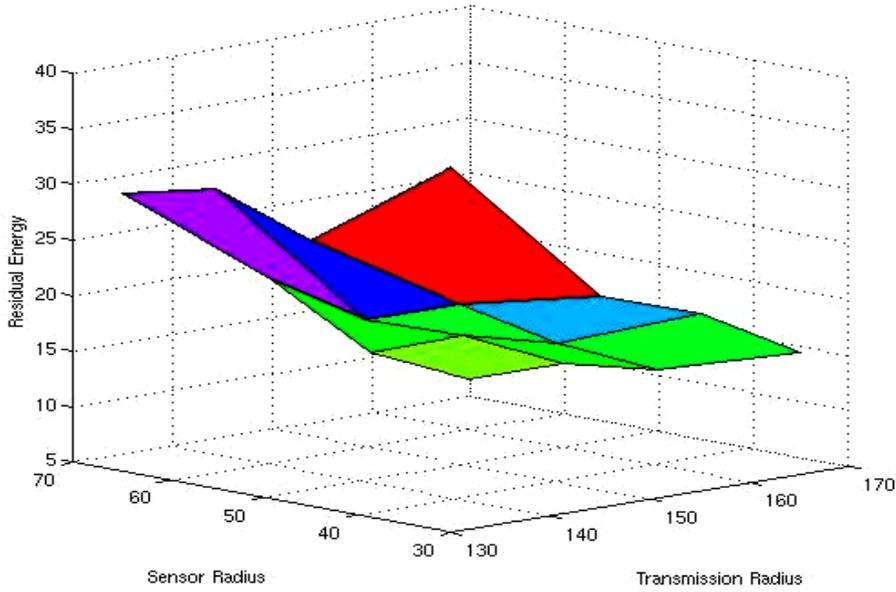

Figure 3-10. Optimum sensor radius and transmission radius for Selective routing method

On the other hand, increasing the transmission radius creates new paths with a smaller number of hops, and accordingly decreases energy dissipation in the network. Thus, while the cost increases with increasing distance, on the other hand there is a decreasing number of hops, and it is difficult to determine which has more weight in consuming energy. The behaviour of the network for the two routing methods shows how these parameters (number of hops, distance) have different influences on residual energy. With the selective method, the decrease in number of hops and related paths keep the network connected and therefore the application performs for longer; in contrast, with random routing, the network disconnects from the sink earlier as the load of the network is not controlled on short paths. This means that the nodes consume more energy because of the cost of distance between these nodes and sinks. The difference between the residual energy of the random and selective methods is due to counting the energy of inaccessible nodes in the overall residual energy. This situation in our model is controlled by considering constraints 1 and 2 in the overall energy consumption (Eqn. 3-2), and constraint 1 of the local (Eqn. 3-11) and global (Eqn. 3-17) constituents. However, in these tests the energy consumption of inactive nodes is counted in the overall energy consumption because the aim is to show how the constituent's parameters effect the overall energy consumption, not to compare routing methods.

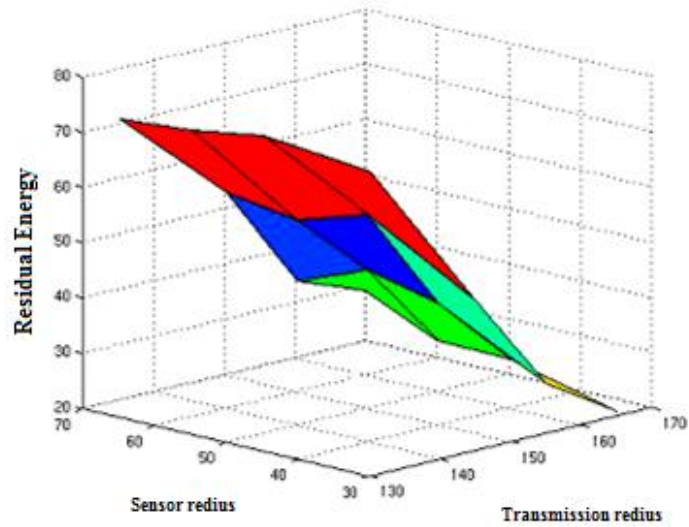

Figure 3-11.Optimum sensor radius and transmission radius for Random routing method

Figures 3-10 and 3-11 show the change in the overall residual energy of the network based on variation of parameters of the individual and local constituents, for the random and selective routing methods, respectively. These figures show that the largest residual energy (30 pixels) occurs with the smallest transmission (130 pixels) and sensing radius (30 pixels) for the selective method. This can be explained as follows: the increasing sensing radius results in increasing $e_{sense,active}$ (due to increase in covered sensing area) and $e(rout)$ (due to the increase in data

redundancy) and therefore raises $E_{individual,i}$ and $E_{global,i}$, respectively. Moreover, increasing transmission radius rises $E_{Local,i}$ and $E_{global,i}$ indirectly by increasing $neighbour_i$, $d_{ij}$ and $h_{iD}$, in that order; however, a larger transmission radius results in higher $d_{iA}$ and consequently more $E_{global,i}$. As a general rule, the selective routing method spends more global energy than the random routing method $(E_{global,i}(selective) > E_{global,i}(random))$ due to maintenance connectivity, choice based on residual energy and busy degree, so $e(rout)$, as the energy consuming component defined for the global constituent, is bigger than zero for a longer time with the selective method than with the random method. This hypothesis explains the variation in residual energy with sensing and transmission radius and routing methods shown in Figure 3-10 and 3-11. The connection between consumed and residual energies can be defined as follows:

$$E_{consumed,i}(\Delta t) = E_{initial,i}(t - \Delta t) - E_{residual,i}(t) \qquad (3-25)$$

where $E_{consumed,i}(\Delta t)$ is proportional to individual, local, and global constituents, and therefore their parameters have a direct effect on the overall energy consumption. In conclusion, to manipulate the overall energy consumption, the interactions, overlaps and influences of all constituents should be taken into account.

## 3.8 Summary and Remarks

In this chapter, a new approach for minimising the total energy consumption of wireless sensor network applications was presented based on the Hierarchy Energy Driven Architecture. In particular, we identified components of each constituent of the EDA. A model was extracted for each of the constituents from a sensor-centric viewpoint; a sensor node within a WSN spends its energy on three constituents: the individual constituent (the existence of the sensor itself), the local constituent (the sensor as a member of its local community), and the global constituent (as a member of the sensor network). Then a formulation for the total energy cost function was proposed in terms of their constituents. Simulation results for lifetime and residual energy of a sample network with different sensor radius, transmission radius and random and selective routing methods demonstrated that our model and formulation can be used to optimise overall energy consumption, and determine the contribution of each constituents and their relative significance. The implication is that optimising the energy of the general model with respect to all constituent parameters will enable one to engineer a balance of energy dissipation among constituents, optimise the energy consumption among them and sustain the network lifetime for the intended application. In the next chapter we will model overall energy consumption in terms of tasks and parameters defined by the EDA model.

# Chapter 4. Task-based sensor-centric model for overall energy consumption

In existing energy models, hardware is considered, but environment and network parameters do not receive adequate attention. Energy consumption (EC) components of traditional network architecture are often considered individually and separately, and their influences on each other are not considered in these approaches. In this chapter we consider all possible tasks of a sensor in its embedded network and propose an energy management model. We categorise these tasks into five energy consuming constituents. The sensor's energy consumption is then modelled on its energy consuming constituents and their input parameters and tasks. The sensor's EC can thus be reduced by managing and executing efficiently the tasks of its constituents. The proposed approach can be effective for power management, and also can be used to guide the design of energy efficient wireless sensor networks through network parameterisation and optimisation.

## 4.1 Problem statement

Our approach in this chapter can be considered as a sensor-centric approach that takes into account a sensor's constituents and its energy-consuming activities (or tasks) in performing its role within the sensor network and the associated application. As a result, the architecture has a modular structure, yet embraces cross-layer ideas. The proposed EC model is used for overall EC minimisation and power management for the sensors' resources. We will show how this model helps a sensor to manage power usage and lengthen its life in the network.

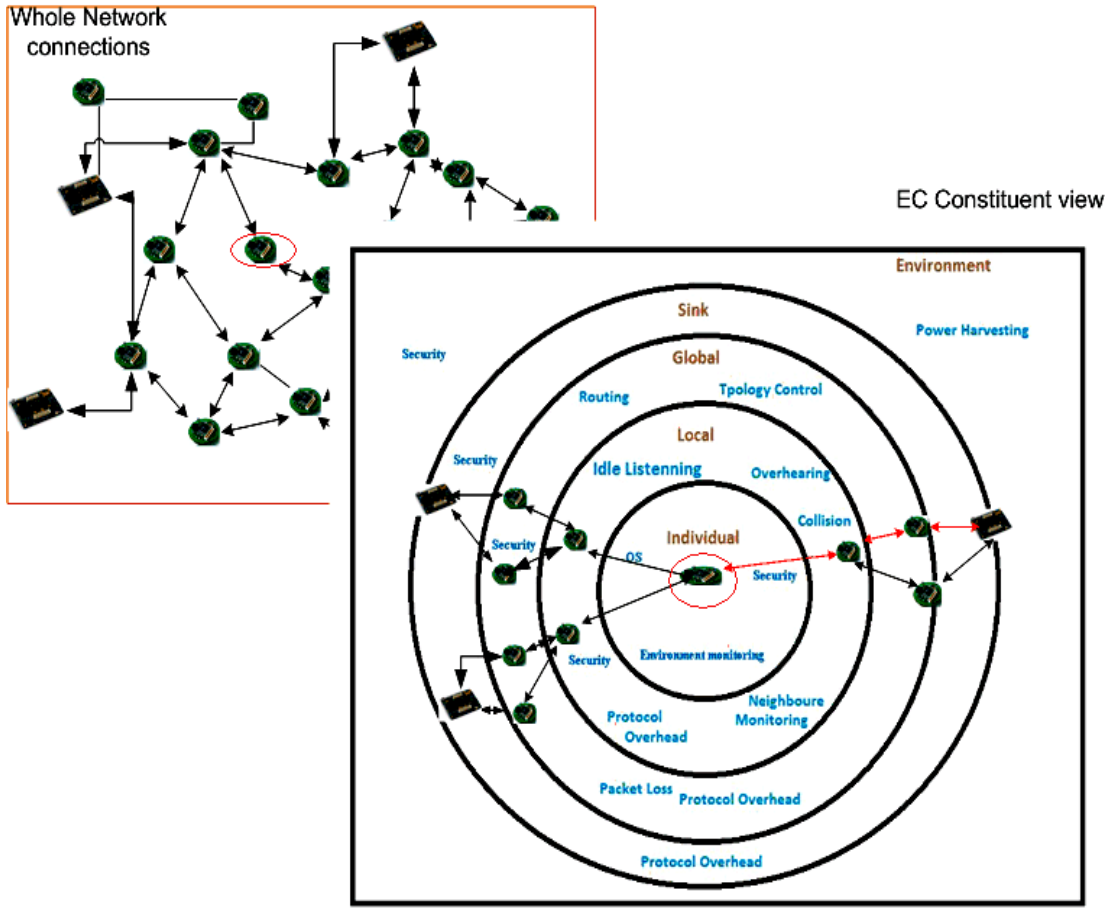

Figure 4-1. Sensor-centric view of a Wireless Sensor Network

We assume five energy consuming constituents: individual, local, global, environment, and sink (Figure 4-1). Starting from an individual sensor, the individual constituent represents all the activities the sensor has to do to survive and perform its sensing function. The local constituent represents all the activities the sensor has to perform to build a relationship with its neighbour. The global constituent represents all the activities the sensor has to perform to establish possible transport paths and carry data from itself to the destination (sink). The sink constituent represents all the activities the sensor has to perform as directed by the sink. The final constituent, the environment, represents activities the sensor may perform to harvest energy available from the environment.

EC minimisation based on these constituents involves the identification of sensor workload attributable to each constituent, improvement of resource utilisation through selection and load balancing among constituents, and reduction of power usage. In principle, the constituent power can be metered by tracking each hardware resource used by a constituent task and converting the resource usage to power usage based on a power model for the resource. This approach does not require any additional instrumentation of the application workload or operating system within the constituents. The constituent-based approach can naturally adapt to changes in applications and even hardware configurations. While prior studies have proposed mechanisms to design energy-efficient individual network protocols or network layers, they are not capable of optimising the overall EC of a sensor within the application.

Generally, a sensor must process and execute assigned tasks while it has enough power. This constitutes a basis for our model, which covers all possible energy consuming constituents. The sensor battery lifetime depends on how the functional tasks are distributed and executed among its individual, local, global, environment, and sink constituents. To execute a task, the sensor needs to exchange a number of packets. A sequence of data and control packets to complete a task is called a Packet Flow (PF). Sensors can manage their power by defining priorities for tasks with the help of an internal EC model. Moreover, power usage may be minimised by the developer assigning optimum values to effective network parameters with the help of an external EC model. In this chapter we focus on the internal EC structure by modelling incoming tasks so that a sensor can prioritise them in a way that minimises the EC.

## 4.2 Tasks-based Energy Driven Model (EDM)

The current EC models are specified for sensor network factors like radio (Heinzelman et al., 2000), data (Rabbat and Nowak, 2004), and hop (Wang et al., *2006b*); however, there are some other significant factors, such as the number of packets a node creates, processes, transmits, receives, senses and etc. Moreover, wireless network characteristics are quite different from wire line systems. The wireless channel characteristics generally affect all OSI layers. Manipulating a layer locally has direct influence on the energy consumption of other layers in WSNs. Optimising each layer individually to fix the problem leads to unsatisfactory results. It has been argued (Heinzelman, 2000) that it is hard to achieve design goals such as energy efficiency using a traditional layered approach. Hence, the cross layer approach was created to enhance performance of the system by jointly optimising multiple protocol layers (Schaar and Shankar, 2005). However, it is argued that cross layer designs with tight coupling between layers become hard to review and redesign. Changing one subsystem implies changes in other parts, as everything is interconnected. Moreover, cross layer designs without solid architectural guidelines inevitably have reduced flexibility, interoperability and maintainability. In addition, systems may become unpredictable, such that it is hard to foresee the impact of modifications.

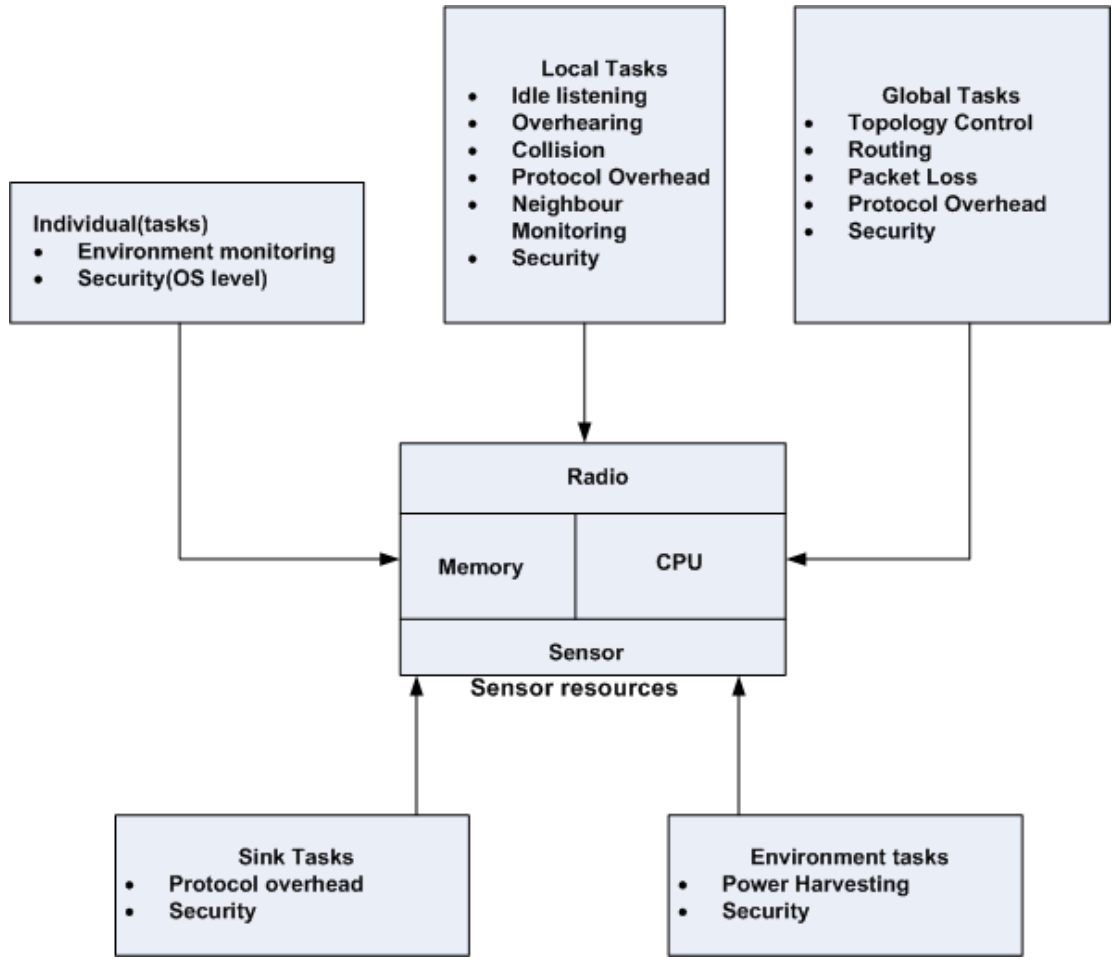

Figure 4-2. System design in term of constituents' tasks.

By creating a new modular view involving energy consuming constituents (Figure 4-1), an approach is proposed for modelling overall energy consumption (EC) in terms of prevalent parameters and energy consuming constituents. We consider five energy consuming constituents based on their tasks, as shown in Figure 4-2. The individual constituent is defined as all essential and basic operations or tasks required for the sensor to just exist, i.e., monitoring environmental events, executing the OS and providing security at the OS level. The local constituent deals with initiating and maintaining all communications with a node's immediate neighbours, i.e., monitoring neighbours and providing a secure communication with neighbours at the local level. In addition, the local constituent may include overhearing, idle listening and collision, if they occur. The global constituent is concerned with maintenance of the whole network, selection of a suitable topology and an energy efficient routing strategy based on the application's objective. This may include energy wastage from packet retransmissions due to congestion and packet errors. The global constituent is defined as a function of energy consumption (EC) for topology management, packet routing, packet loss, and protocol overheads. The sink constituent includes the roles of manager, controller or leaders in WSNs. The sink tasks include directing, balancing, and minimising EC of the whole network, and the collection of generated data by the network's nodes. The environment tasks are energy-harvesting operations, in the case where nodes have the capability to extract energy from the environment. Execution of these tasks requires sensor resources, CPU, memory, radio, and sensing units.

Table 4-1. Individual parameters

| Parameter | Description | Boundary |
|---|---|---|
| $r_{sense}$ | Sensing radius points to the covered area of the sensor: this will have different meaning in different applications, e.g., a temperature application and a radar application. | $r_{sense} > 0$ |
| $g_{sense}$ | Sensing delay | $g_{sense} \geq 0$ |
| $b_{sense}$ | Number of packets created by the sensor itself that include environment data. | $b_{sense} \geq 0$ |
| $b_{store}$ | Numbers of packets stored in the memory. | $b_{store} \geq 0$ |
| $b_{OS}$ | Number of OS instructions | $b_{OS} \geq 0$ |
| $b_{sec}$ | Security at Individual level | $b_{sec} \geq 0$ |

A knowledge of costly functions can guide a sensor to run tasks based on its residual power and the importance of tasks. Establishing a balance between the EC of constituents can also guide a sensor to minimise its energy usage. Thus, from a sensor viewpoint, the challenge is selecting and executing significant tasks efficiently in the optimum order to minimise its EC. In addition, moving tasks from a constituent with high level EC to a low level EC constituent can minimise its energy consumption, e.g., data aggregation that reduces global tasks and increases individual tasks. Moreover, we may split a high EC task into low level tasks that are suitable for low EC constituents. Thus, sensors can act intelligently to manage task execution in an efficient way based on the network energy consumption model.

Generally when a sensor runs a typical task, the energy will be consumed by the CPU, memory, radio and sensor units:

$$e_t = e_{cpu} + e_{mem} + e_R + e_{sens} \qquad (4-1)$$

For each task, a sensor runs the basic operations. We assume that a sensor is a server that should execute incoming tasks. We use the approach proposed in (Aman Kansal et al., 2010) to model energy consumption from a hardware perspective:

$$e_{cpu} = p_{cpu} b_{cpu} \qquad (4-2)$$

$$e_{mem} = p_{mem} b_{mem} \qquad (4-3)$$

$$e_{Radio} = e_{Tx} + e_{Rx} \qquad (4-4)$$

$$e_{Rx} = p_{Rx} b_{Rx} \qquad (4-5)$$

$$e_{Tx} = p_{Tx} b_{Tx} \qquad (4-6)$$

$$e_{sens} = p_{sens} b_{sens} \qquad (4-7)$$

Where $b_{cpu}, b_{mem}, b_{Rx}, b_{Tx}, b_{sens}$ are the number of packets processed in the CPU, stored in memory, received, transmitted by radio, and sensed, respectively.

Every task that sensor does in its lifetime is assigned to a constituent. The overall EC of a typical sensor can be calculated by adding the power usage of the individual, local, global, environment, and sink tasks (Figure 4-1):

$$E_{Overall} = E_{Individual} + E_{local} + E_{global} + E_{environment} + E_{Snk} \qquad (4-8)$$

Since each constituent includes a number of tasks, and tasks include a Packet Flow (PF), the EC of a sensor is as follows:

$$E_{Overall} = \alpha_0 + \underbrace{(\lambda_1 p_{cpu} + \lambda_2 p_{mem} + \lambda_3 p_{Rx} + \lambda_4 p_{Tx} + \lambda_5 p_{sens})}_{\alpha_1} b_{Individual} + \quad (4-9)$$

$$\underbrace{(\lambda_6 p_{cpu} + \lambda_7 p_{mem} + \lambda_8 p_{Rx} + \lambda_9 p_{Tx} + \lambda_{10} p_{sens})}_{\alpha_2} b_{local} +$$

$$\underbrace{(\lambda_{11} p_{cpu} + \lambda_{12} p_{mem} + \lambda_{13} p_{Rx} + \lambda_{14} p_{Tx} + \lambda_{15} p_{sens})}_{\alpha_3} b_{global} +$$

$$\underbrace{(\lambda_{16} p_{cpu} + \lambda_{17} p_{mem} + \lambda_{18} p_{Rx} + \lambda_{19} p_{Tx} + \lambda_{20} p_{sens})}_{\alpha_4} b_{environment} +$$

$$\underbrace{(\lambda_{21} p_{cpu} + \lambda_{22} p_{mem} + \lambda_{23} p_{Rx} + \lambda_{24} p_{Tx} + \lambda_{25} p_{sens})}_{\alpha_5} b_{snk}$$

or in a shorter way,

$$E_{Overall} = \alpha_0 + \alpha_1 b_{Individual} + \alpha_2 b_{local} + \alpha_3 b_{global}$$
$$+ \alpha_4 b_{environment} + \alpha_5 b_{snk} \quad (4-10)$$

In the following sections, we explain each constituent in terms of the prevalent parameters in the sensor's EC model.

### 4.2.1 Individual

$b_{Individual}$ consists of the PF of individual tasks (Table 4-1), i.e., sensing, executing OS and installed applications and also providing security for the sensor individually. Therefore, PF in the individual constituent is as follows:

$$b_{Individual} = b_{sens} + b_{OS} + b_{sec} \quad (4-11)$$

Table 4-2. Local parameters

| Parameter | Description | Boundary |
|---|---|---|
| $n$ | Number of neighbours | $n \geq 1$ |
| $e_i(idle)$ | Idle power consumption | |
| $d_{ij}$ | Distance to the neighbour | $0 < d_{ij} \leq r_{Tx}$ |
| $b_{mon}$ | Packet overhead for monitoring: depends on the application and its topology. | $b_{mon} \geq 0$ |
| $r_{Tx}$ | Transmission radius | $r_{Tx} \geq 0$ |
| $b_{sec}$ | Local security packet overhead: depends on application. | $b_{sec} \geq 0$ |
| $b_{local}$ | Packet overhead to avoid collision problem policy. | $b_{local} \geq 0$ |
| $b_{reTx}$ | Number of retransmission packets: depends on probability of collision and number of neighbours | $b_{reTx} \geq$ |

The number of sensed and produced packets by a sensor depends on its covered area, $r_{sens}$, and sensing delay, $g_{sens}$, therefore,

$$b_{sens} = P(\text{Sense}|r_{sens}, g_{sens})b_{Individual} \qquad (4-12)$$

According to these equations,

$$b_{Individual} = \frac{b_{OS} + b_{sec}}{1 - P(\text{Sense}|r_{sens}, g_{sens})} \qquad (4-13)$$

### 4.2.2 Local

$b_{local}$ includes packet flow for neighbour monitoring to gather information on available resources, such as their residual energy, memory space, security management (to prevent malicious nodes from destroying connectivity of the network and tampering with the data), idle listening packets, overhearing packets and retransmission packets (due to collision and the tasks to prevent them (Table 4-2)). Therefore, the local constituent's packet flows can be given as

$$b_{local} = b_{coll} + b_{idle} + b_{ohear} + b_{sec} + b_{mon} + b_{ohead} \quad (4-14)$$

where

$$b_{coll} = P(\text{coll}|n, g_{Tx}, net_{dens})b_{local} \quad (4-15)$$
$$b_{ohear} = P(\text{ohear}|n, net_{dens}, r_{Tx})b_{local} \quad (4-16)$$
$$b_{idle} = P(idle|n)b_{local} \quad (4-17)$$

Table 4-3. Global parameters

| Parameter | Description | Boundary |
|---|---|---|
| $n$ | Number of neighbours | $n \geq 1$ |
| $g_{Tx}$ | Transmission interval | Application dependent |
| $net_{dens}$ | Network density | $net_{dens} \geq 2$ |
| $b_{ohear}$ | Overheard packets | $b_{ohear} \geq 0$ |
| $a_i$ | Number of sensors in covered area | $a_i > n_i$ |
| $d_{iA}$ | Distance between sensors and other sensors inside the covered area | $d_{iA} \geq 0$ |
| $b_{topo}$ | Packet overhead for topology | $b_{topo} \geq 0$ |
| $N(t)$ | Number of nodes in time $t$ | $N(t) \geq 2 + n_{snk}$ |
| $d_{iD}$ | Distance between source and destination. | $d_{iD} > 0$ |
| $h_{iD}$ | Number of hops | $0 \leq h_{iD} \leq net_{dens}$ |
| $b_{rout}$ | Number of routing packets | $b_{rout} \geq 0$ |
| $b_{global}$ | Number of packet to avoid packet loss | $b_{global} \geq 0$ |
| $d_i$ | Distance between node $i$ to nearest sink | $d_i > 0$ |
| $b_{pktls}$ | Number of packet losses | $b_{pktls} \geq 0$ |
| $Snk$ | Number of sinks | $Snk > 0$ |
| $b_{sec}$ | PF security in global level | $b_{sec} \geq 0$ |

Where $n$ is the number of neighbours, $net_{dens}$ is total number of nodes in the network, $g_{Tx}$ is transmission delay, and $r_{Tx}$ is transmission radius. Therefore, Eqn. 4-14 can be rewritten as

$$b_{local} = \frac{b_{sec} + b_{mon} + b_{ohead}}{1 - \left(P(\text{coll}|n, g_{Tx}, net_{dens}) + P(\text{ohear}|n, net_{dens}, r_{Tx}) + P(\text{idle}|n)\right)} \quad (4-18)$$

### 4.2.3 Global

The global constituent consists of a number of tasks: topology control, routing, retransmission due to packet loss, and re-performing tasks to prevent packet loss.

$$b_{global} = b_{pktls} + b_{sec} + b_{topo} + b_{rout} + b_{ohead} \quad (4-19)$$

The possibility of packet loss in the network depends on the prevalent parameters, such as $d$, the distance between node and destination, and $net_{dens}$, the number of nodes in the network (Table 4-3):

$$b_{pktl} = P(pktls|D, net_{dens})b_{global} \quad (4-20)$$

Therefore,

$$b_{global} = \frac{b_{sec} + b_{topo} + b_{rout} + b_{ohead}}{1 - P(pktls|D, net_{dens})} \quad (4-21)$$

### 4.2.4 Environment

The environment constituent includes providing security and power harvesting management if a node has the ability to harvest energy from the environment (Table 4-4):

$$b_{environment} = b_{sec} + b_{ph} \quad (4-22)$$

### 4.2.5 Sink

The sink constituent includes providing security for sink communication and performing sink directions if it is applicable to the application (Table 4-5):

$$b_{snk} = b_{sec} + b_{ohead} \quad (4-23)$$

So far, we have described the constituents and their relationships with overall energy consumption (Eqn. 4-10); however, the coefficients of these constituents are still unknown. In the next section, multiple linear regression is employed to estimate values of these coefficients (i.e., $\alpha_1, \alpha_2, \alpha_3, \alpha_4, \alpha_5$ ).

Table 4-4. Environment parameters

| Parameter | Description | Boundary |
|---|---|---|
| $H_i$ | Harvested energy (Watt) | $H_i \geq 0$ |
| $b_{ph}$ | Overhead produced due to harvesting power. | $b_{ph} \geq 0$ |

Table 4-5. Sink Parameters

| Parameter | Description | Boundary |
|---|---|---|
| $b_{ohead}$ | Network management policy | $b_{ohead} \geq 0$ |
| $b_{sec}$ | PF security in sink level | $b_{sec} \geq 0$ |

## 4.3 Estimating model coefficients and evaluation

Taking multiple observations of the observable quantities allows estimation of the model parameters using learning techniques such as linear regression. We used linear regression with minimisation of least square error between observations and predictions (i.e., L2-norm), due to its closed-form calculation. We generate the model in four steps: profiling, model regression, evaluation criteria and refine the model. in the next subsection we explain each step in more detail.

### 4.3.1 Profiling

We carried out several experiments by loading sensors with a set of packet flows for the constituents, and therefore calculated the energy consumption of each constituent in Eqn. 4-10. Meanwhile, the overall energy consumption of the network was detected. Values were collected in time periods $\Delta t$, and each experiment was repeated for several time slices, as the tasks in each constituent and therefore their energy usage changes over time. For example, a sensor may become a head cluster for while and later act as an accelerator to monitor the environment. Repeating experiments and making a model based on different scenarios helps to provide knowledge of the cost of constituents' tasks. Such a model gives a sensor the ability to decide which tasks should be run in order to be energy efficient.

Due to the temporal changes, it is expected that several runs of an experiment – with the same packet flows for constituents – may result in slightly different overall energy consumption. Therefore, the average of several runs of an experiment was considered as the overall energy consumption of the experiment.

### 4.3.2 Model generation

This section explains how to model the relationship between the prevalent parameters and the overall energy consumption of the WSN. The problem of modelling based on linear regression involves choosing suitable coefficients of the modelling such that the model's output accurately approximates a real system's response. Consider one degree linear algebraic equations for $M$ number of experiments of a WSN application for five parameters ($M \gg 5$) (Rizvandi et al., 2012):

$$\overbrace{\begin{bmatrix} \widetilde{E_1} \\ E_2 \\ \vdots \\ E_M \end{bmatrix}}^{E} = \overbrace{\begin{bmatrix} 1 & b_{Ind}^{(1)} & b_{local}^{(1)} & b_{global}^{(1)} & b_{snk}^{(1)} & b_{env}^{(1)} \\ 1 & b_{Ind}^{(2)} & b_{local}^{(2)} & b_{global}^{(2)} & b_{snk}^{(2)} & b_{env}^{(2)} \\ & & & \vdots & & \\ 1 & b_{Ind}^{(M)} & b_{local}^{(M)} & b_{global}^{(M)} & b_{snk}^{(M)} & b_{env}^{(M)} \end{bmatrix}}^{B} \overbrace{\begin{bmatrix} \alpha_0 \\ \alpha_1 \\ \alpha_2 \\ \alpha_3 \\ \alpha_4 \\ \alpha_5 \end{bmatrix}}^{A} \Rightarrow A = ? \quad (4-24)$$

where $E_k$ is the value of overall energy consumption in the $k^{th}$ experiment and $\left(b_{Ind}^{(k)}, b_{local}^{(k)}, b_{global}^{(k)}, b_{snk}^{(k)}, p_{env}^{(k)}\right)$ are the values of parameters in Eqn.4-10 for the same experiment, respectively. Using the above formulation, the approximation problem is converted to estimating the values of model parameters, i.e. $\widehat{\alpha_0}, \widehat{\alpha_1}, \widehat{\alpha_2}, \widehat{\alpha_3}, \widehat{\alpha_4}, \widehat{\alpha_5}$, to optimise a cost function between the approximation and real values of overall energy consumption. Then, an approximated energy consumption $\left(\widehat{E^{(*)}}\right)$ of the application for a new unseen experiment is predicted:

$$\widehat{E^{(*)}} = \widehat{\alpha_0} + \widehat{\alpha_1}b_{Ind}^{(*)} + \widehat{\alpha_2}b_{local}^{(*)} + \widehat{\alpha_3}b_{global}^{(*)} + \widehat{\alpha_4}b_{snk}^{(*)} + \widehat{\alpha_5}b_{env}^{(*)} \qquad (4-25)$$

It can be mathematically proved that the model parameters can be calculated by minimising least square error between real and approximated values:

$$\boldsymbol{A} = (\boldsymbol{B}^T\boldsymbol{B})^{-1}\boldsymbol{B}^T\boldsymbol{E} \qquad (4-26)$$

### 4.3.3. Evaluation Criteria

We evaluate the accuracy of the fitted models, generated from regression based on a number of metrics (Rizvandi et al., 2012): Mean Absolute Percentage Error (MAPE), PRED(25), Root Mean Squared Error (RMSE) and R2 Prediction Accuracy. We describe the metrics in the following subsections:

- **Mean Absolute Percentage Error(MAPE)**

    The Mean Absolute Percentage Error for the prediction model is given by the following formula:

    $$MAPE = \frac{\sum_{i=1}^{M}\frac{\left|E^{(i)} - \hat{E}^{(i)}\right|}{E^{(i)}}}{M}$$

    where $E^{(i)}$ is the actual overall energy consumption of the network, $\widehat{E^{(i)}}$ is the predicted output and $M$ is the number of observations in the dataset for which the prediction is made. A lower value of MAPE implies a better fit of the prediction model, i.e., indicating superior prediction accuracy.

- **PRED(25)**

    The measure PRED(25) is defined as the percentage of observations whose prediction accuracy falls within 25% of the actual value. A more formal definition of PRED(25) is as follows:

    $$PRED(25) = \frac{\#\ of\ observations\ with\ relative\ error\ less\ than\ 25\%}{\#\ of\ total\ observations}$$

    It is intuitive that a PRED(25) value closer to 1.0 indicates a better fit of the prediction model.

- **Root Mean Squared Error (RMSE)**

    The metric Root Mean Square Error (RMSE) is defined by the following formula:

    $$RMSE = \sqrt{\frac{\sum_{i=1}^{M}(E^{(i)} - \widehat{E^{(i)}})^2}{M}}$$

    A smaller RMSE value indicates a more effective prediction scheme.

- $R^2$ **Prediction Accuracy**

    The $R^2$ Prediction Accuracy(Islam et al., 2012) is a measure of the goodness-of-fit of the prediction model. The formula of $R^2$ Prediction Accuracy is:

    $$R^2 = 1 - \frac{\sum_{i=1}^{M}(E^{(i)} - \widehat{E^{(i)}})^2}{\sum_{i=1}^{M}(\widehat{E^{(i)}} - \sum_{r=1}^{M}\frac{E^{(r)}}{M})}$$

    Note that the $R^2$ value falls within the range [0, 1]. This metric is commonly applied to linear regression models. In fact, $R^2$ Prediction Accuracy determines how well the fitted model approximates the real data points. A $R^2$ prediction accuracy of 1.0 indicates that the forecasting model is a perfect fit.

### 4.3.4 Refine the model

The problem with the model, however, is that linearity does not necessarily hold across the constituents, since they do not have homogeneous packet flows; in other words, the number of packets in order to complete a task is not the same. The packet flow (PF) is a significant concept in network protocols which directly affects the number of packets required to complete the task. A sensor determines a PF for each task based on the average number of sent and received packets (both control and data packets) in the first execution of each task. In addition, each sensor has a different model due to different constituents' tasks, for example, sensors near to sinks have more global tasks than those far from sinks.

Table 4-6. Packet flow of different constituents

| Packet Flow | Constituent |
|---|---|
| Packets having sensed data | Individual |
| Packets carrying information of current node and its neighbours | Local |
| Scheduling packets to avoid collision | Local |
| Packets carrying topology information | Global |
| Packets carrying routing information | Global |
| Received data packets | Global |

## 4.4 Experimental results

### 4.4.1 Experiment setting

We have simulated a WSN application to track the energy consumption of constituents as well as overall energy consumption of the system. The application collects information about events that occur. Sensors detect an event in their covered area, create a packet and send it to the nearest sink. Sinks are located as a group in specific location. Generally we assume three phases in our WSN application simulator. In the initialisation phase, a sensor executes its own software, creates a connection with immediate nodes as a neighbour and collects information about the neighbour's resources. Then in the collecting phase, the sensor uses the neighbour's information relay data. Moreover, it collects information from the environment, creates data and sends it, in addition to processing and relaying incoming packets. It performs these tasks when it has enough power, otherwise it ignores them. In the maintenance phase, the sensor monitors its neighbours to update their situation, as well as performing extra global tasks such as reorganising topology and reconfiguring routing tables when it is necessary. These phases may be repeated by a sensor a number of times during the network lifetime.

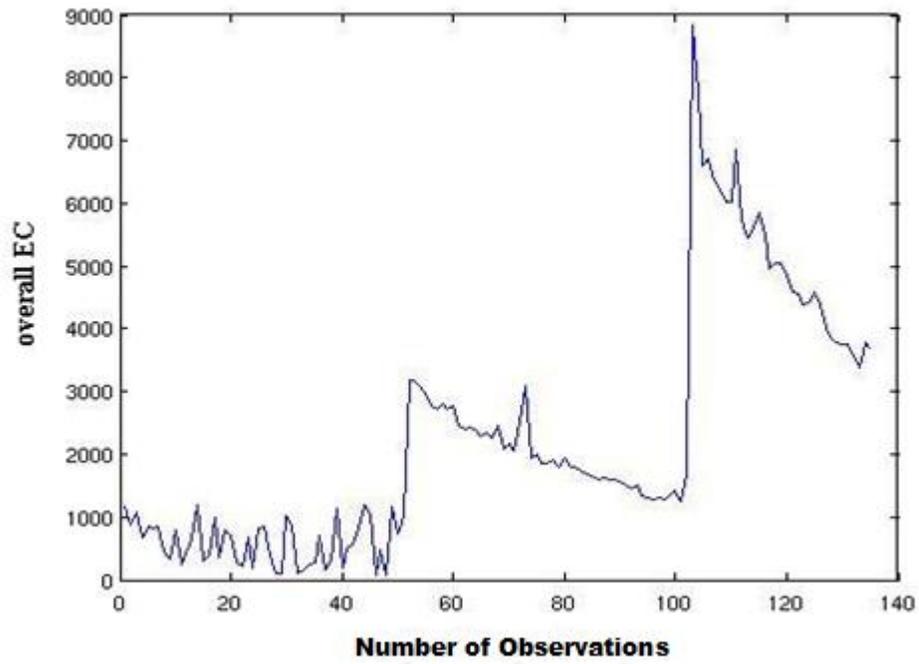

(a)

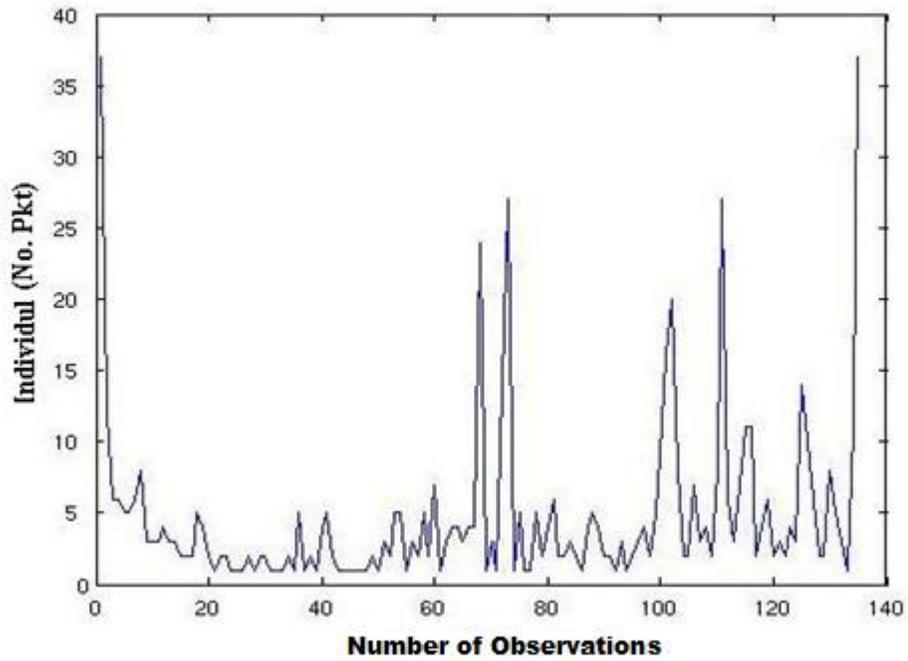

(b)

Figure 4-3.Number of packets of (a) overall EC, (b) individual, (c) local and (d) global constituents against the frequency of observations

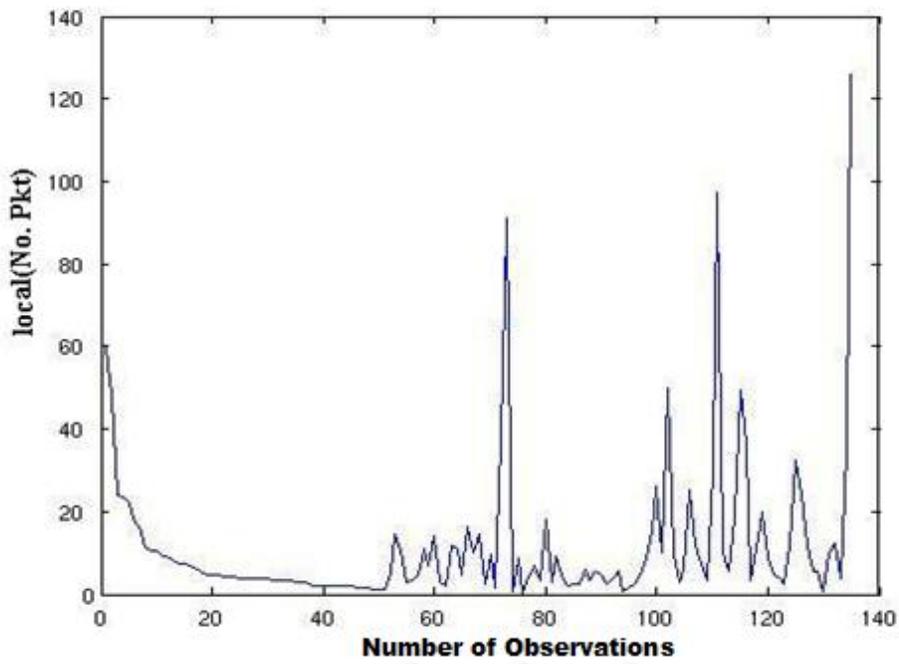

**(c)**

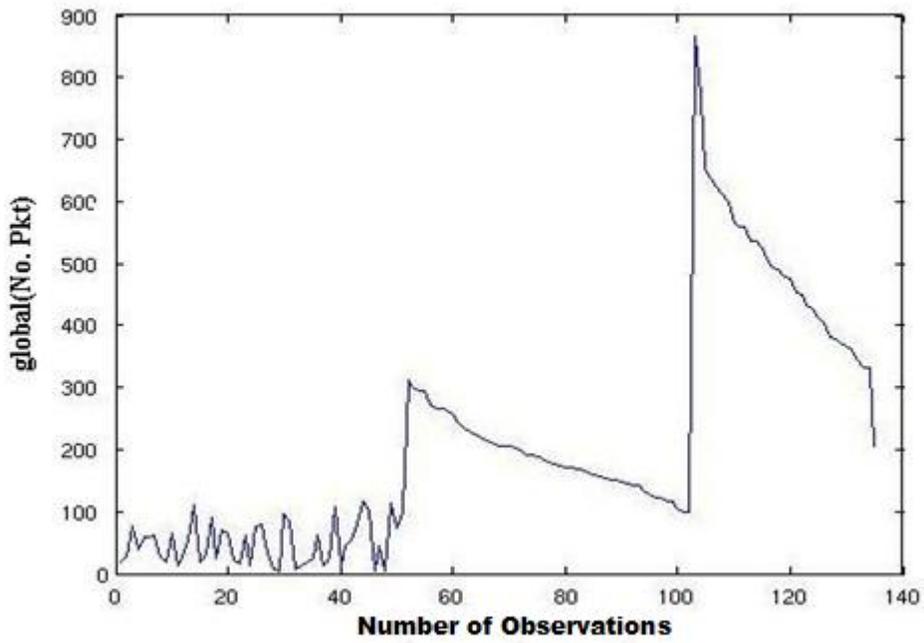

**(d)**

Figure 4-3. (continued)

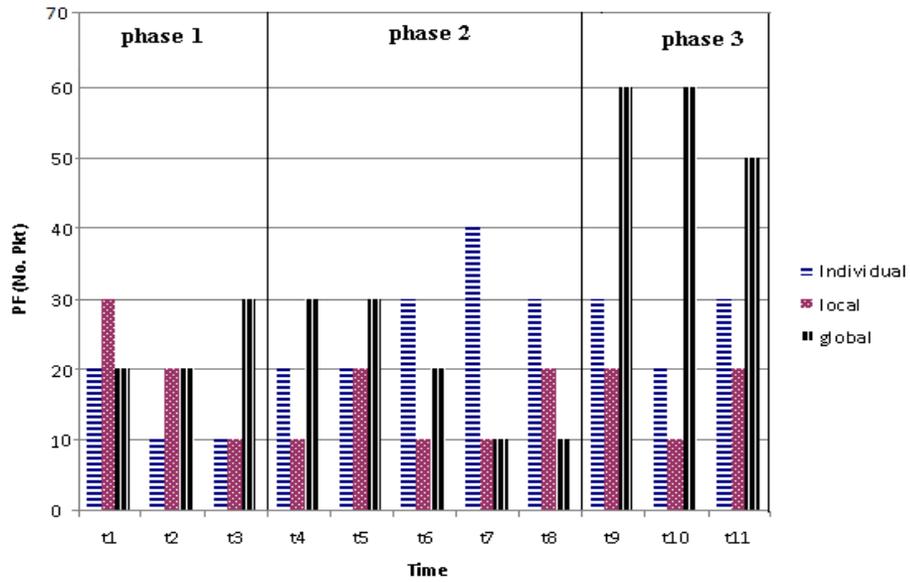

(a)

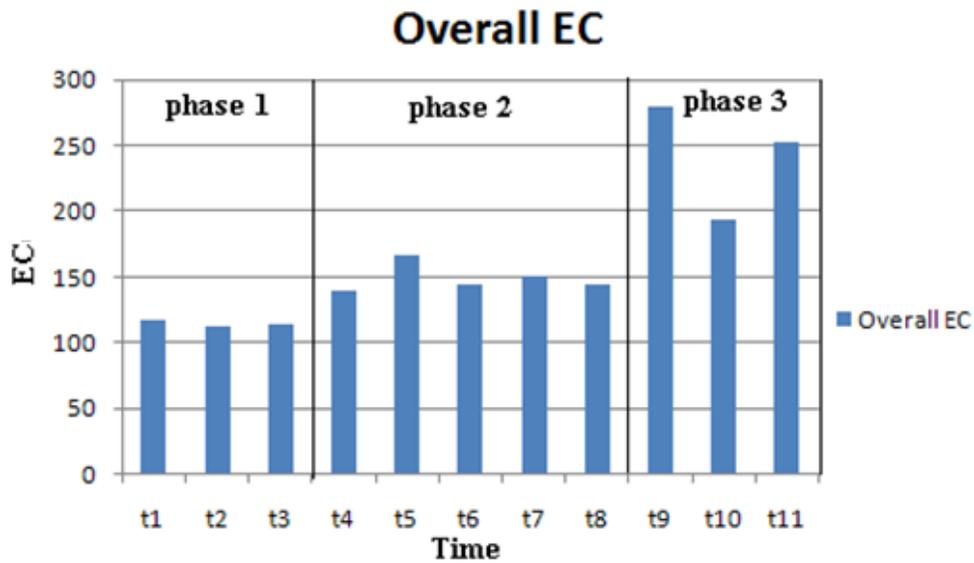

(b)

Figure 4-4. (a) constituent's packet flows in different time slices; (b) overall EC in different time slices.

Table 4-6 shows how packet flows are assigned to constituents in the simulator. In our application, sensors have connection with all immediate nodes in which they always select neighbours based on their residual energy. The sink does not have any roles in the application

and sensors do not harvest energy; hence sink and environment constituents are ignored in our modelling.

### 4.4.2 Results

In this section, we investigate various packet flows and energy consumptions of each constituent in all phases of our WSN application through simulated experiments. The first phase consists of three time slices before a sensor starts sensing, monitoring and relaying data packets. The sensor spends power to start up (individual), transport control packets and initialise connections with neighbours (local). It also sends control packets to set routing tables (global). In the second phase, the sensor starts capturing events and creates data packets and sends them to a sink (individual tasks). Local tasks in this phase involve in monitoring neighbours' resources by sending request packets to its neighbours. Moreover, the sensor is responsible for relaying incoming data packets to their destination by looking at its routing table and choosing a suitable neighbour or path. The third phase starts when the network needs to recover from a disconnected path. In this phase, the sensor performs tasks in the second phase in addition to extra global tasks to maintain the network. These tasks involve capturing information about paths and updating routing tables.

Figure 4-3 shows the PF and the EC of each constituent of a typical sensor in different phases (initialisation, data collection and maintenance) of simulation experiments. In each time slice, the PF of constituents (based on Eqn. 4-10 to Eqn. 4-21) and their energy consumption (based on current power level of a sensor) were recorded. As can be seen from Figure 4-3(b), in phase 3 an increase in the global constituent activities resulted in a drastic increase in the overall energy consumption of the network. However, as indicated in phases 1 and 2, variations of tasks in the individual and local constituents do not have significant effects on overall EC. Referring to Figure 4-3(b), it can be clearly seen that an increase in global tasks results in a peak in the EC, as shown in time slices of phase 3. The global tasks (from the global constituent) are thus very costly in terms of energy consumption in the sensor lifetime and directly affect the overall energy consumption of the WSN application.

Table 4-7. The predication evaluation

|  | values | Boundary and explanation |
|---|---|---|
| **RMSE** | 0.68 | Smaller value means better prediction |
| **MAPE** | 1.31 | Lower value implies better fit |
| $R^2$ **prediction accuracy** | 0.56 | Between 0 and 1 where closer to 1.0 indicates better model |
| **PRED(25)** | 0.43 | Between 0 and 1 where closer to 1.0 represents better prediction |

The routing in the simulation was a simple routing protocol based on the residual energy of neighbours. Ignoring packet loss, overhead of a topology control and security protocols, the global constituent still has a massive influence on the node's energy usage. If more complex protocols are deployed that entail heavy control packet flows to global tasks, the global constituent would become the dominant constituent in term of overall energy usage of the node and hence the energy consumption of the overall WSN application.

By tweaking parameters in Tables 4-1 to 4-5 and running the simulator, different values for individual, local and global constituents are observed; these observations, along with observation of overall energy consumption of the network, are then used in a simplified version of Eqn. 4-10 to find coefficients of individual, local and global constituents. Figure 4-4 shows the overall EC of a sensor and the packet flows of the individual, local and global constituents in different experiments. We recorded the overall energy consumption and packet flow of constituents in order to learn the model's coefficients. We compared the variations in energy consumption with regard to variations of packet flows of constituents (after regularisation) in our model. As may be observed from Figure 4-4, the global constituent is clearly the most dominant constituent, adding a bias to overall EC (compare Figure 4-4(a) and (d)). To test the accuracy of our model, we used it to predict the overall energy consumption of a typical sensor in a number of simulation experiments with random values within the predefined range of a few parameters of the individual, local and global constituents while other parameters were fixed. We then ran experiments on the simulator, captured the actual overall EC and estimated the predicted overall EC (Figure 4-5(a)). Figure 4-5(b) shows the prediction accuracy of our application by comparing the actual EC and its predicted value, and the evaluation criteria are shown in Table 4-7. The values in columns show the accuracy of the fitted models as determined using MAPE, PRED, RMSE and R2 Prediction Accuracy methods, explained in Section 4.3.3. We found that the average square error between the observed EC and the predicted values was about 13%. The errors and low prediction accuracy in Table 4-7 are expected, partly because of the model inaccuracy and partly from the linearity assumption. As may be noticed from Figure 4-5(b), there are some spikes in the prediction errors. These spikes generally occur with high values of energy consumption, which were probably caused by large differences in the energy usage of the global constituent in comparison with other constituents. Figure 4-5(c) shows that larger values of the global constituent imply higher error in the EC prediction. It is expected that the obtained model cannot predict the EC of a sensor perfectly, but it can clearly reveal the relationships between the energy consuming constituents of a sensor.

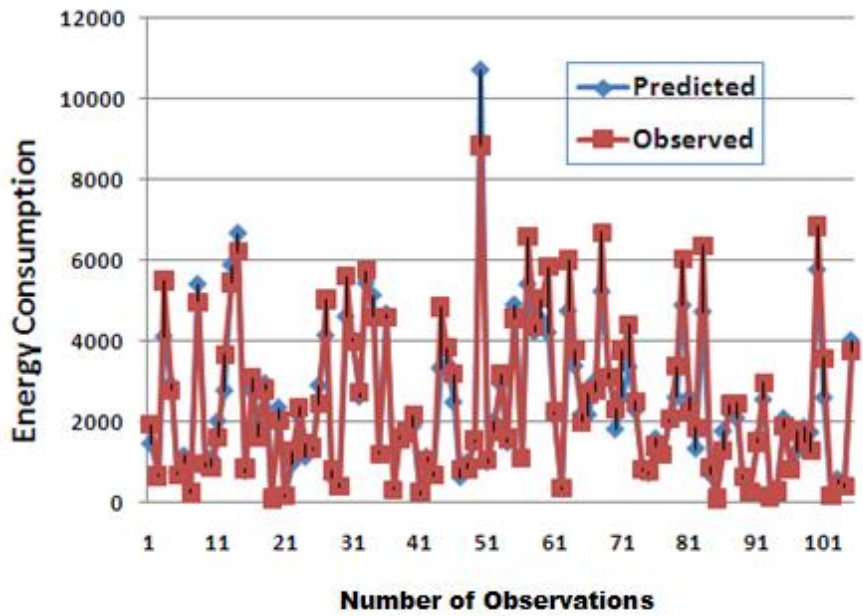

(a)

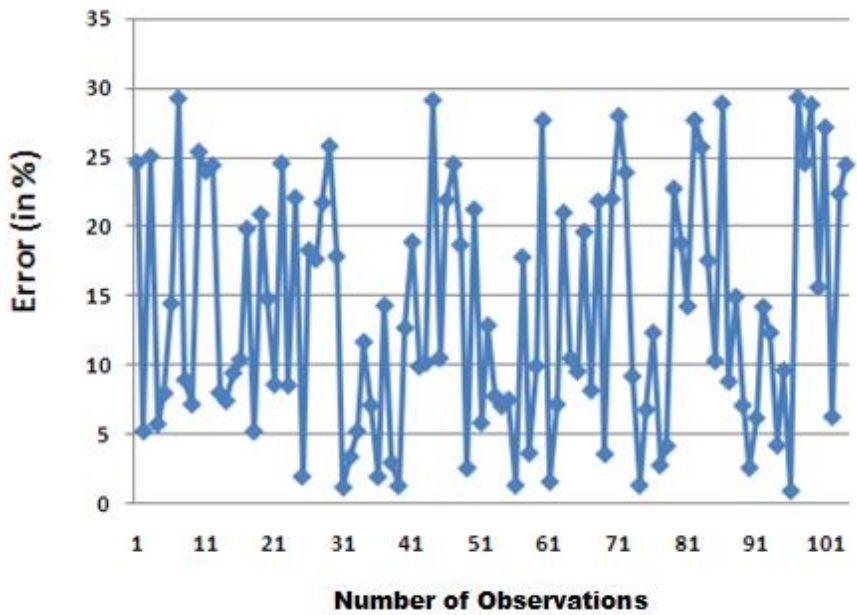

(b)

Figure 4-5. (a) Model-predicted and observed values of EC for several random runs; (b) the error range of the model; (c) comparing the variation of the error value with global packet flows.

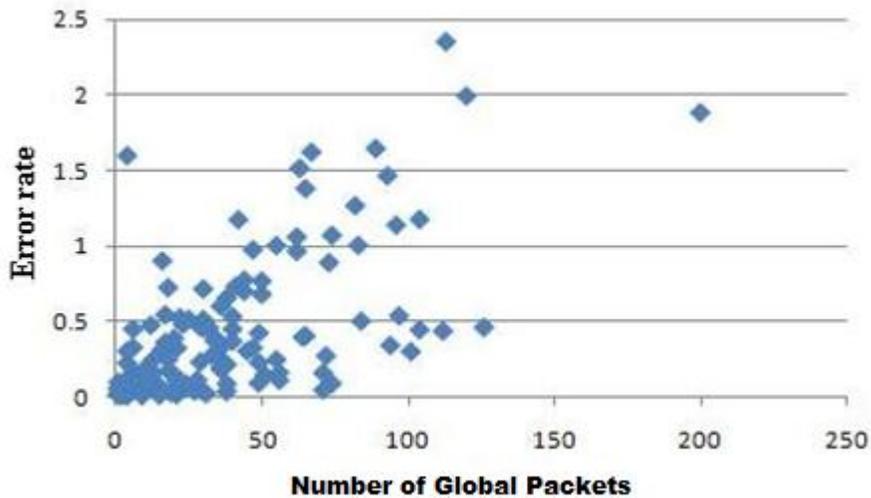

(c)

Figure 4-5 (continued)

## 4.5 Summary and Remarks

The motivation behind this chapter was the need to minimise the overall energy consumption of sensors. We introduced five energy consuming constituents at the sensor level: individual, local, global, environment and sink, where each constituent consists of a set of tasks a sensor is responsible for based on the application characteristics. Our model helps to identify essential EC constituents and their contributions to the overall energy consumption of a sensor. This in turn helps the sensor to spend its energy efficiently. After extracting/profiling the constituents' power usage, linear regression was applied to establish the relationships between constituents' tasks and the overall EC. The model was then utilised by the sensor to prioritise the constituents' tasks in terms of the EC level and their importance, in order to make an appropriate decision such that the sensor can use its power in an effective way and remain alive longer. Using the same model for extracting its power consumption profile, a sensor equipped with an intelligent algorithm can even act appropriately to conserve its energy in power shortage situations. We call these sensors "thrifty sensors", and the idea of thrifty sensors is worth exploring further in the future.

Using the results of this chapter, we will study the parameters of the global constituent, as the most energy consuming constituent, in the next chapter. We will extract prevalent parameters of the global constituent in order to consume energy in an effective way and maximise the network lifetime.

# Chapter 5. Statistical Analysis for Prevalent Parameter Selection for Energy in WSNs

In the previous chapter, we explained our model of energy dissipation at different levels of the network, and concluded that the global constituent has the highest impact on the energy consumption of the network. Although this model gives a complete view of the interaction between different elements of the network and their parameters, modelling of such a system with a high number of parameters is very difficult, and in some cases impossible. Therefore, in this chapter, statistical and machine learning tools are employed to reduce the number of parameters by analysing the dependency between these parameters and the target parameter (i.e., average energy consumption in the network), allowing the most relevant ones to be selected. The applied methods are correlation (Pearson, Spearman and nonlinear second and third degree correlation), Lasso regularisation and p-value. Later, random forest regression is applied to compare the accuracy of prediction for both original and reduced parameters in estimating the average energy consumption of the network.

## 5.1 Mathematical Background

One of the main difficulties in applying machine learning algorithms to model a system occurs when the number of features (here named as parameters) are high compared to the generated data. Thus, when the number of parameters increases, more samples are required; otherwise, the 'curse of dimensionality' (Houle et al., 2010) affects the accuracy of prediction. The curse of dimensionality implies that an increase in the number of parameters (i.e., features or dimensions) results in rapid growth in the volume of the searching space such that available data becomes sparse, which is problematic for methods that require statistical significance. To reach to a statistically reliable result, the required amount of data to support the result often grows exponentially with dimensionality. As an example, assume a system with 5 different values of $N$ parameters. A classification learner needs to distinguish between $5^N$ different configurations of $N$ input parameters. Given that to reach to a good predictor, it must see at least one sample for each configuration, at least $5^N$ distinct samples are required. Moreover, searching in a high volume of (sparse) data makes the convergence of learning algorithm too slow.

Referring to our proposed energy model in the previous chapter, the global constituent, as the dominant constituent in the overall energy consumption of the WSN, includes nine continuous parameters, of which three are set by the user. Although the number of parameters is not high, the continuous nature of the parameters implies that a large number of samples is required. The analytical part of the proposed method in this chapter comprises two mathematical approaches: (1) analyse the dependency between all parameters and the average energy consumption in the network, and consequently reduce the number of parameters by keeping the prevalent parameters with a highly influential impact; and (2) employ random forest regression to model the dependency between energy and the prevalent parameters, and then determine how much prediction accuracy is lost because of this reduction.

### 5.1.1 Parameter reduction

Since a wireless sensor network is a complex system, due to its spatial-temporal nature in both the state of the sensors and their interactions in the network, it is very unlikely that a clean dependency between WSN parameters and average energy consumption would be found via scatter graph. Also, in most cases, scatter graphs are too noisy and full of outliers, making detection of dependency harder. Therefore, a collection of statistical tools is required to automatically capture most of the relationships; these tools should be resilient to noise and outliers and also cover different forms of relationship (e.g., linear, convex, ascending/descending, exponential, logarithmic, etc).

P-value and correlation analysis are the most common methods in statistics for analysis of dependency between two random variables, and in machine learning for feature selection (Hall and Smith, 1999, J. Tang, 2014). As the shape of the relationship on a scatter graph between parameters and average energy consumption in a WSN is unknown, the following tools are applied:
- Pearson correlation: simply checks for a linear relationship in a scatter graph.
- Spearman correlation: determines if the relationship between parameter and energy consumption is an ascending / descending function.

- 2-degree correlation: examines if the relationship can be represented as a convex function (i.e., $x^2$).
- 3-degree correlation: checks whether the relationship is in the form of $x^3$.
- Lasso regression: employed as another standard technique to inspect dependency between the parameter and energy consumption, especially when the relationship between the parameter and energy cannot be represented by above the mentioned functions.

After this analysis, the strength of each parameter is verified by p-value analysis as well as its correlation values. If both analyses indicate a high dependency between average energy consumption and the parameter then it is picked as prevalent parameter for further study; otherwise, it is removed. It should be noted that removing a parameter based on the above relationship tests does not imply no relationship at all; in fact, there might be some other forms of relationship which are hidden in the scatter graph and may be detected with complex functions.

### 5.1.1.1 Pearson Correlation analysis

By definition, linear correlation (a.k.a. Pearson correlation (Hastie T, 2009)) is a measure of dependency between two variable sequences on a scale from $-1$ to $1$. In WSNs, if $\boldsymbol{P_1} = (p_{11}, p_{12}, \ldots, p_{1M})$ and $\boldsymbol{E} = (E_1, E_2, \ldots, E_M)$ are values of the first parameter (e.g., transmission radius) and corresponding average energy consumption, respectively, the normalised linear correlation function between these two series can be expressed as follows:

$$Corr_{norm}(\boldsymbol{P_1}, \boldsymbol{E}) = \frac{\sum_{i=1}^{M}\left((p_{1i} - \mu_{p_1})(E_i - \mu_E)\right)}{\sqrt{\sum_{i=1}^{M}(p_{1i} - \mu_{p_1})^2 \sum_{i=1}^{M}(E_i - \mu_E)^2}} \qquad (5-1)$$

where $\mu_{p_1}$ and $\mu_E$ are the mean of $\boldsymbol{P_1}$ and $\boldsymbol{E}$ series, respectively. A correlation value of $0$ indicates a random or independent relationship between the parameter and average energy consumption, and the correlation values of $1$ and $-1$ denote positive and negative perfect linear associations between them, respectively.

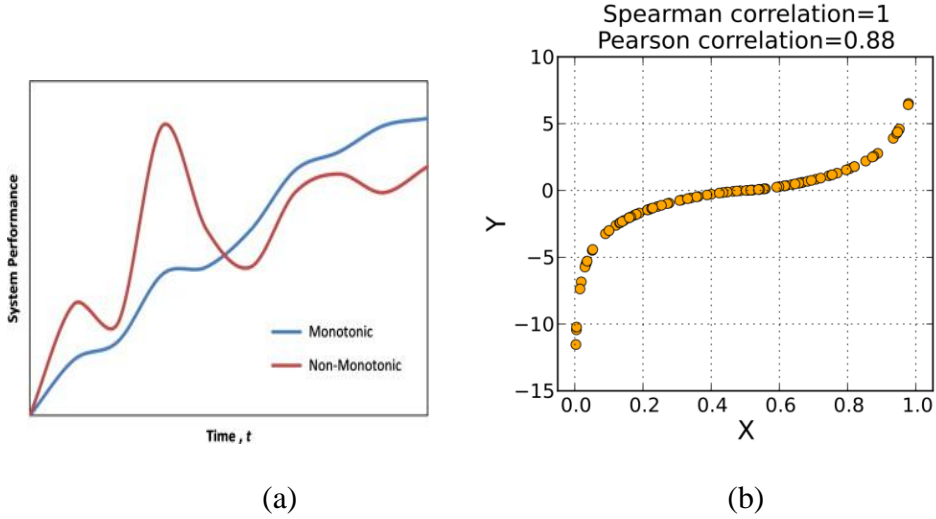

(a)                                            (b)

Figure 5-1. (a) Monotonic function used in Spearman correlation [ http://goo.gl/Ol81E], (b) Pearson vs. Spearman correlations [source:

### 5.1.1.2 Spearman Correlation

As a non-parametric version of Pearson correlation, Spearman correlation measures the degree of association between a parameter and average energy consumption without any assumption about the distribution of data. It evaluates how well the relationship between the two can be described using a monotonic function (Figure 5-1), which is defined as an ascending or descending function. To calculate this correlation, the samples in $P_1$ and $E$ should separately be ranked (i.e., sorted) from smallest to largest. Then, the Spearman correlation is defined as (Zwillinger, 2000):

$$\rho(P_1, E) = \frac{M \sum_{i=1}^{M} p_{1i} E_i - (\sum_{i=1}^{M} p_{1i})(\sum_{i=1}^{M} E_i)}{\sqrt{[M \sum_{i=1}^{M}(p_{1i} - (\sum_{i=1}^{M} p_{1i})^2][M \sum_{i=1}^{M}(E_i)^2 - (\sum_{i=1}^{M} E_i)^2]}}$$

$$= 1 - \frac{6 \sum_{i=1}^{M} distance(p_{1i}, E_i)}{M(M^2 - 1)} \qquad (5-2)$$

where $distance(p_{1i}, E_i)$ is the difference between the ranks of corresponding values $p_{1i}$ and $E_i$. In absence of repeated data, a high Spearman correlation of $+1$ or $-1$ indicates that the parameter is a perfect monotone function of average energy consumption. It is worth noting that the sign of the Spearman correlation represents the direction of association between the parameter and energy consumption. For example, a positive coefficient implies that an increase in $P_1$ leads to an increase in $E$.

### 5.1.1.3 Second and third degree correlation analysis

Since the nature of dependency between a WSN parameter and average energy consumption is unknown, non-linearity between them is highly probable. In (Billings and Voon, 1983) it was indicated that Pearson correlation is not sufficient for nonlinear cases; therefore the same authors of (Mao and Billings, 2000) and (Billings and Zhu, 1994) proposed a higher degree correlation between parameter and target (i.e., $x^n$ instead of $x$ in Pearson correlation). In summary, the normalised higher order is as follows:

$$Corr_{norm}^2 (P_1, E) \qquad (5-3)$$

$$= \frac{\sum_{i=1}^{M} \left( \left((p_{1i})^2 - \frac{1}{M}\sum_{k=1}^{M}(p_{1k})^2\right) \left((E_i)^2 - \frac{1}{M}\sum_{k=1}^{M}(E_k)^2\right) \right)}{\sqrt{\sum_{i=1}^{M} \left( \left((p_{1i})^2 - \frac{1}{M}\sum_{k=1}^{M}(p_{1i})^2\right) \right)^2 \sum_{i=1}^{M} \left( \left((p_{1i})^2 - \frac{1}{M}\sum_{k=1}^{M}(E_k)^2\right) \right)^2}}$$

It should be noticed that $Corr_{norm}^2 (P_1, E) = 1$ represents perfect dependency and $Corr_{norm}^2 (P_1, E) = 0$ represents complete independence between a parameter and average energy consumption in our case. Obviously for $M = 1$, it is identical to Pearson correlation.

### 5.1.1.4 Stability parameter selection with Lasso regularisation

The aim in Lasso regression (as a penalised regression model) is to identify a set of parameters with non-zero (or small) coefficients in the model. The difference between normal regression and Lasso is in the definition of the cost function. In the former, there is no penalty on coefficients, while in the latter, coefficients with high values are penalised; therefore, the estimation of coefficients $\beta^{lasso}$ is defined as (Hastie T, 2009):

$$\beta^{lasso} = \operatorname{argmin}\left( \frac{1}{2} \sum_{i=1}^{M} \left( E_i - \beta_0 - \sum_{j=1}^{N} p_{ij}\beta_j \right)^2 + \lambda \sum_{j=1}^{N} |\beta_j| \right) \qquad (5-4)$$

in which $p_{ij}$ is the selected value for the $i^{th}$ parameter in the $j^{th}$ experiment, and $\lambda$ is a regularisation parameter. It is worth mentioning that Lasso regression between normalised parameters and average energy consumption implies that parameters corresponding to higher value coefficients have more impact on energy.

### 5.1.1.5 P-value

Making a decision about the statistical significance of the above metrics involves the practice of hypothesis testing. In general, the idea is to state a null hypothesis (e.g., that there is no relationship between a parameter and average energy consumption with regards to Pearson correlation) and then to see if the generated data allows us to reject the hypothesis. In statistics, a $Pvalue$ is a numerical measure of the probability that the null hypothesis is true, and therefore indicates the statistical significance of a relationship. In our study, the null hypothesis is that a parameter has no influence on average energy consumption in regards to one of the metrics – e.g., that there is no Pearson correlation between a specific parameter ($\boldsymbol{P_k}$) and the average energy consumption ($\boldsymbol{E}$) – and the $Pvalue$ is a measure of how likely it is that we could have gotten our sample data if the null hypothesis is true. By convention, $Pvalue \leq 0.05$ implies that the null hypothesis can be rejected (i.e., $\boldsymbol{P_k}$ and $\boldsymbol{E}$ are highly correlated); therefore, the relationship between $\boldsymbol{P_k}$ and $\boldsymbol{E}$ is statistically significant.

### 5.1.2 Model Generation with Random Forest Regression

This section describes how the relationship between parameters and average energy consumption in WSN was modelled. Random forests are a type of ensemble learning method in which a multitude of decision trees are constructed at training time; outputs of these trees are then combined on a test data set (i.e., mode of classes for classification and mean prediction for regression). Since each tree employs a small portion of the parameters, random forest regression is resilient to over-fitting.

The first step is to understand how a decision tree works for regression. In a regression tree the aim is to predict real valued numbers at the leaf nodes, as shown in Figure 5-2 for a tree of five features/parameters. Since the target variable (e.g., average energy consumption) is a real valued number, a regression model was fitted from each parameter (e.g., transmission radius) to the target variable (Sharma, 2014); for each parameter, the data was divided at several split points (i.e., tree nodes) followed by calculation of *Sum of Squared Error (SSE)* at each node between the predicted values of the target variable and its actual values. For a node, the variable with minimum SSE was selected. This procedure was recursively repeated until the whole training data was covered.

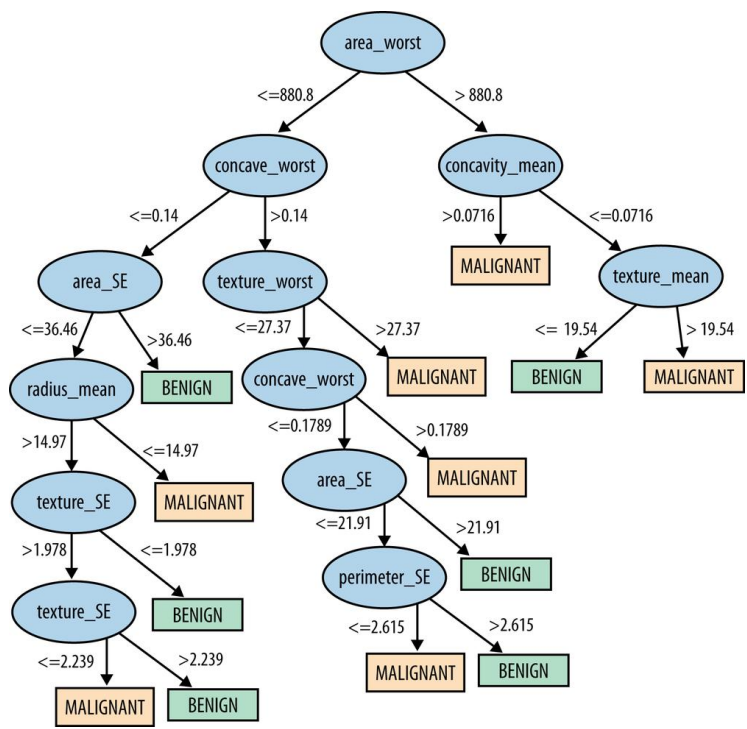

Figure 5-2. Decision tree based on the Wisconsin Breast Cancer dataset
(Foster Provost, 2013).

## 5.2 Experimental Evaluation

In this section, we evaluate the relationship between WSN parameters and average global energy consumption using the described statistical tools, and then model this relationship using random forest regression.

Table 5-1. Configuration parameters in the global constituent and their values

| Paremeter | Range |
|---|---|
| Transmission radius | 30–250 |
| Network size | 10–200 |
| Number of sinks | 1–50 |

## 5.2.1 Experimental Setting

Through simulation of a wireless sensor network application, sensors randomly detected generated events in their covered area, created data packets and afterwards sent these packets to their closest sinks, which were located as a group in a specific location. These sensors performed tasks such as sensing, neighbour monitoring, and relaying data to create packets when they had enough power, otherwise they stopped.

To generate enough data to feed the previous mentioned statistical tools, the simulation was run 1000 times with different values of configuration parameters (i.e., transmission radius, network size and number of sinks) involved in the global constituent, followed by measurement of energy consumption and number of delivered packets to sinks after $\Delta t$ seconds. Meanwhile, a few other parameters influencing energy consumption in global constituent were observed during the experiment; these parameters were transmission cost, transmission delay, average distance between nodes and their neighbours, average number of neighbours, receive cost and average number of hops (e.g., number of neighbours of each node was logged during an experiment and calculated after one run of the simulator: the number of neighbours was averaged to form a new parameter named the average number of neighbours). For each experiment, configuration parameters were randomly assigned a value in their pre-defined ranges (Table 5-1) and at the end of simulation the values of other parameters along with energy consumption were observed; to overcome temporal changes, each experiment was repeated five times and the values of each of the parameter and energy consumption were averaged (which increased the total number of experiments to 5000).

In the parameter selection phase, all of the experiments were used, while in the modelling phase, the data was split into training and test sets using 5-fold cross-validation. Then, two random forest regression models were fitted on the training data: one using all parameters and another using only prevalent parameters; the goal was to show a low effect of removed parameters on the prediction accuracy of the target variable.

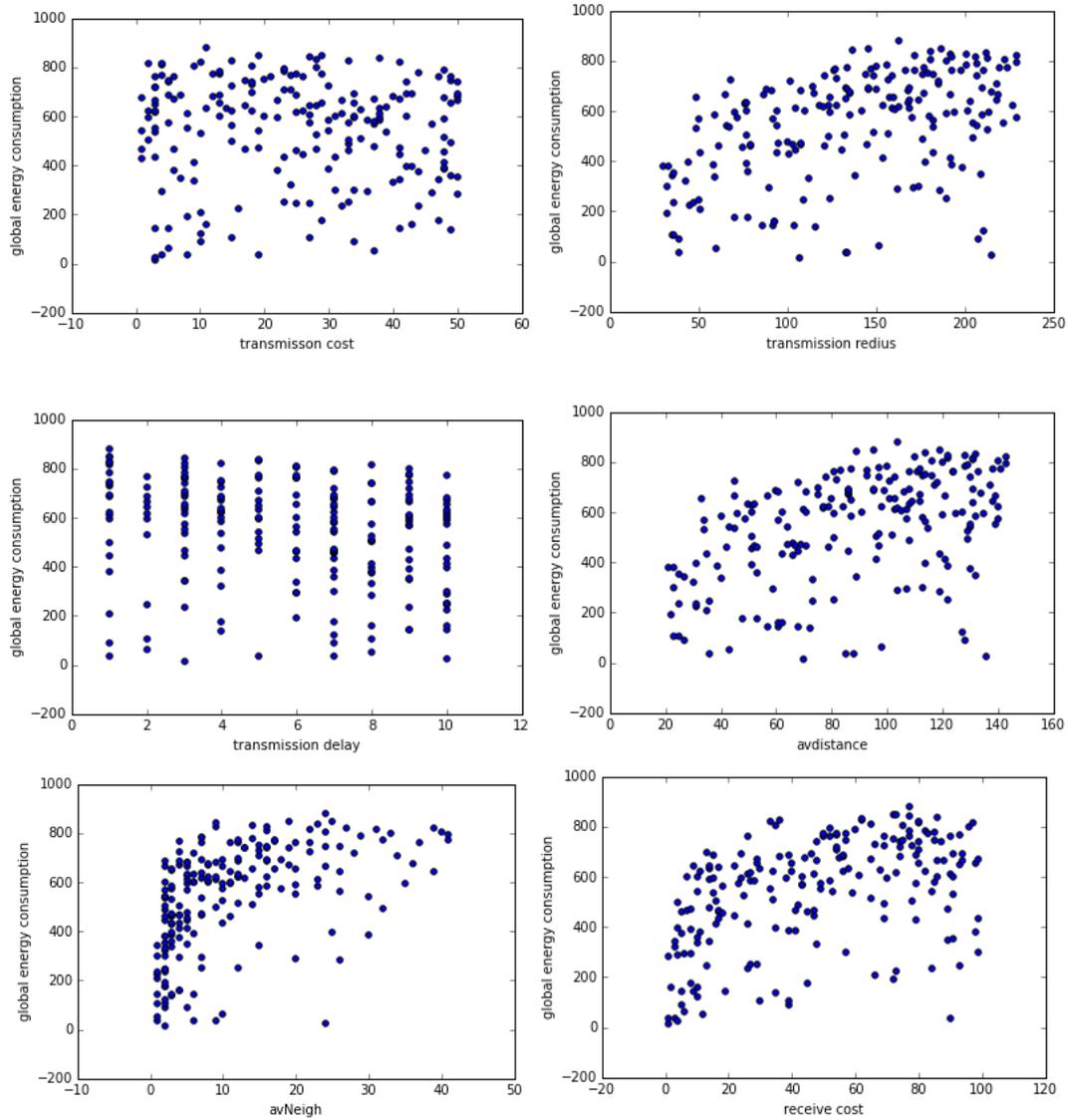

Figure 5-3. (continued on next page) Scatter graphs to display relationships between the parameters and average energy consumption; the x-axis is the parameter value for an experiment and y-axis is the average energy consumption of the global constituent for that experiment. Configuration parameters (i.e., transmission radius, size of network, number of sinks) are set by user at the beginning of the experiment, while other parameters are observed at the end of the experiment from the simulator.

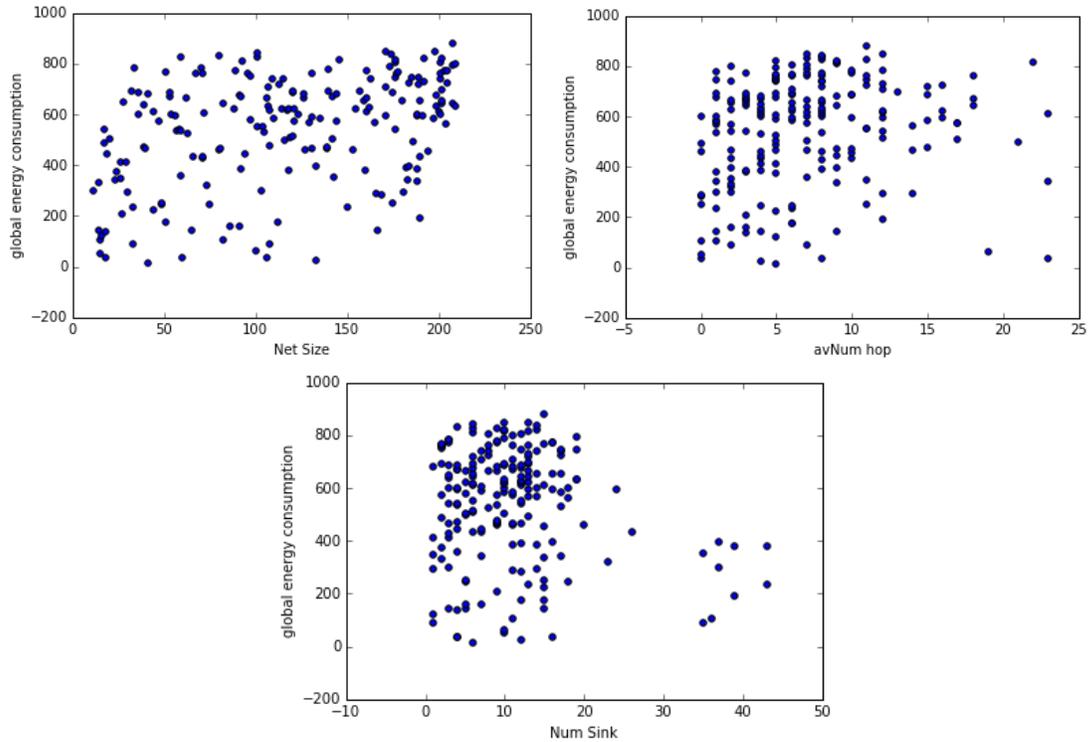

Figure 5-3. (continued)

### 5.2.2 Results

Figure 5-3 shows a series of scatter graphs displaying the relationship between parameters (both configuration and observed) and average energy consumption from our experiments. On the x-axis are the values of a specific parameter in an experiment, while the y-axis shows the average energy consumption of the global constituent for that experiment. As can be observed, it is difficult to find a clear relationship between them; however, in some graphs, by removing noises and outliers, it becomes possible to observe a structural relation. For instance, one may observe a sharp ascending connection between average energy consumption and the average number of neighbours in the network. Generally speaking, it is difficult to visually detect a structural relation between these parameters and average energy consumption, which is the main reason to employ different kinds of correlation analysis in order to detect weak relationships.

The correlation analysis between parameters and average energy consumption in the global constituent is summarised in Table 5-2. The italic-bold numbers in each row are the highest value of correlation between the parameter and average energy consumption. The results show that the connections between parameters and energy consumption are difficult to explain by only one type of correlation; for instance, 3-degree correlation better describes the connection of transmission cost, while 2-degree correlation best suits the relationship between energy consumption and average transmission radius, average distance, number of hops, and number of sinks. Except for network size, which has its highest value in Pearson correlation, the rest of the relationships can be explained by Spearman correlation. From Table 5-2, it is worth noting the following:

- The correlation between energy consumption and both transmission radius, defined as the maximum communication distance, and average distance, is 2th degree, which is confirmed by theory (i.e., $E \propto d^2$).
- Average number of neighbours, defined as the average number of neighbours the sensors have during experiments, is an ascending function, since more neighbours means extra communication by relaying packets through these nodes, and therefore more energy is consumed.
- Receive cost, defined as the energy a sensor's receiver consumes to receive a packet, is directly related to the sensor's radio. Obviously, a higher receive cost implies greater energy consumption.
- Network size implies the number of sensors inside the experiment. When sensors are given a set initial energy, the total energy of the network rises with the size of the network. The absolute value of consumed energy will also increase with the size of the network since more sensors mean more engagement and therefore more energy consumption.
- As we know, a higher number of hops in a network leads to lower energy consumption; the negative correlation between average number of hops and energy consumption confirms this relationship.

Table 5-2. Selection of relevant parameters on average consumed energy per delivered packet. The bold-italic numbers in each row are the highest values of correlation between the parameters and average energy consumption. The grey rows indicate parameters with not enough evidence to be prevalent.

|  | P-value | Pearson Correlation | Spearman Correlation | 2th and 3rd degree Correlation | Lasso regression |
|---|---|---|---|---|---|
| **Transmission cost** | 0.757 | −0.02 | -0.0645 | −0.09, ***−0.113*** | 0.99 |
| **Transmission radius** | 2.054e-11 | 0.45 | 0.443 | ***0.466***, 0.418 | 0.635 |
| **Transmission delay** | 0.013 | 0.175 | ***0.24*** | 0.07, 0.01 | 1 |
| **Average distance** | 1.55e-11 | 0.45 | 0.45 | ***0.464***, 0.428 | 0.895 |
| **Average # of neighbours** | 5.79e-14 | 0.5 | ***0.61*** | 0.435, 0.38 | 0.965 |
| **Receive cost** | 1.4e-11 | 0.455 | ***0.479*** | 0.38, 0.32 | 1 |
| **Network size** | 5.96e-9 | ***0.398*** | 0.39 | 0.373, 0.356 | 1 |
| **Average # of hops** | 0.00157 | −0.273 | -0.302 | ***−0.363***, −0.323 | 1 |
| **Number of sinks** | 0.027 | −0.156 | -0.028 | ***−0.23***, −0.228 | 1 |

The acceptable lower bound for correlation coefficient and p-value, to identify a connection as statistically significant, depends on the application; in general, a correlation value more than 0.4 with $pValue \leq 0.05$ is considered a strong dependency . In our application, we lowered the pass mark for correlation to 0.35 while keeping the same lower bound for $pValue$; therefore, for a specific parameter, if at least one of its correlation coefficients with average energy consumption is higher than 0.35 with $pValue \leq 0.05$, we are confident to name it as a prevalent parameter. As a result, all parameters except transmission cost, transmission delay and number of sinks, shown with grey rows in Table 5-2 and excluded from further modelling, are labelled as prevalent parameters for our WSN application.

A closer look at the scatter graphs of the excluded parameters reveals that the variation in average energy consumption has no structural relationship with variations in values of these parameters. An educated guess to explain this result is that the relationships are too complex to be explained with our set of correlations, or that the level of noise is too high for detection of the employed correlations. In regards to the former, as previously mentioned, a small set of relationship tests are examined in our work (e.g., linear, Spearman and 2th and 3th degree correlations) but many other forms of relationships remain (e.g., logarithmic, exponential, polynomial, higher degree and so on); this applies particularly to the relationship between the number of sinks and average energy consumption. The latter, however, points to a known problem in modelling of complex systems, which has been well-studied in the machine learning domain (Atla et al., 2011, Twala, 2014); this problem cannot be solved by employing complex relations, and implies the need for a pre-processing step to clean data and lessen noise (i.e., outliers). Looking at the corresponding graphs confirms the high noisiness of both transmission cost and transmission delay plots, leading us to make an educated guess that noise is behind the low correlation coefficients between these two parameters and average energy consumption.

To determine how much accuracy is lost by removing non-relevant parameters, random forest regression was employed to model energy consumption once with all parameters and then with only prevalent parameters. The number of trees in both models was 20; both models used *Mean Square Error* to measure the quality of a split. All parameters were considered to look for the best split. The regression also used bootstrap to sample data for the training of each tree (i.e., random sampling with replacement). Figure 5-4 shows the comparison between predicted energy consumption, as the outcome of these models, and actual energy consumption. The data implies in most cases that removing transmission delay and number of sinks in the modelling has a negligible impact on prediction accuracy, and therefore supports neglecting these two parameters. More detailed insight of implementation of these statistical tools and random forest regression (in Python, Scikit-learn and SciPy) can be found in Appendix A.

Table 5-3. Evaluation of random forest regression of models with all parameters and models with only prevalent parameters, with actual data in regard to four crietria defined in Section 4.3.3

|  | **All parameters** | **Only relevant parameters** | **Best fit** |
|---|---|---|---|
| **RMSE** | 0.189 | 0.204 | Smaller value means better prediction |
| **MAPE** | 0.225 | 0.237 | Lower value implies better fit |
| $R^2$ **prediction accuracy** | 0.65 | 0.618 | Between 0 and 1 where closer to 1.0 indicates better model |
| **PRED(25)** | 0.74 | 0.712 | Between 0 and 1 where closer to 1.0 represents better prediction |

Two questions that arise from modelling are as follows: how well do these models describe actual energy consumption, and is it possible to use this model as a true estimate of the WSN system? The prediction accuracy of these models is important. In the previous chapter (Section 4.3.3) we introduced four evaluation criteria to study the degree of fitness in regression models. Table 5-3 reveals the effectiveness of both models (with all parameters and with prevalent parameters) in fitting actual data; as can be seen, the R2 prediction, as the main criteria in regression, and PRED(25) indicate a poor fit, although RMSE and MAPE show a better result. One explanation for the poor fit is that both models suffer from an insufficient number of useful independent parameters. As an example, one should never expect to get a good model by using only one parameter. Adding more parameters (i.e., features) brings into account other sources for variations of energy consumption, resulting in better cover in modelling. However, creating and adding more parameters should be limited to avoid curse of dimensionality and overfitting.

## 5.3 Summary

In this chapter, we proposed a few statistical tools to study the relationships between various parameters and global energy consumption (as the most consuming constituent) in WSNs and consequently select the most important ones. These applied tools were: p-value, correlation (Pearson, Spearman and nonlinear second and third degree correlation) and Lasso regularisation. After parameter reduction, random forest regression was applied to compare the accuracy of prediction for both original and reduced parameters in estimating global energy consumption of the network. Our evaluation on 1000 simulated experiments showed that the number of sinks and transmission delay have a low impact on global energy consumption compared to the other six parameters. It also indicated that our random forest regression model can predict actual data with good accuracy and therefore can be used as an approximation of a network in further study (e.g., optimise energy with regard to values of the parameters).

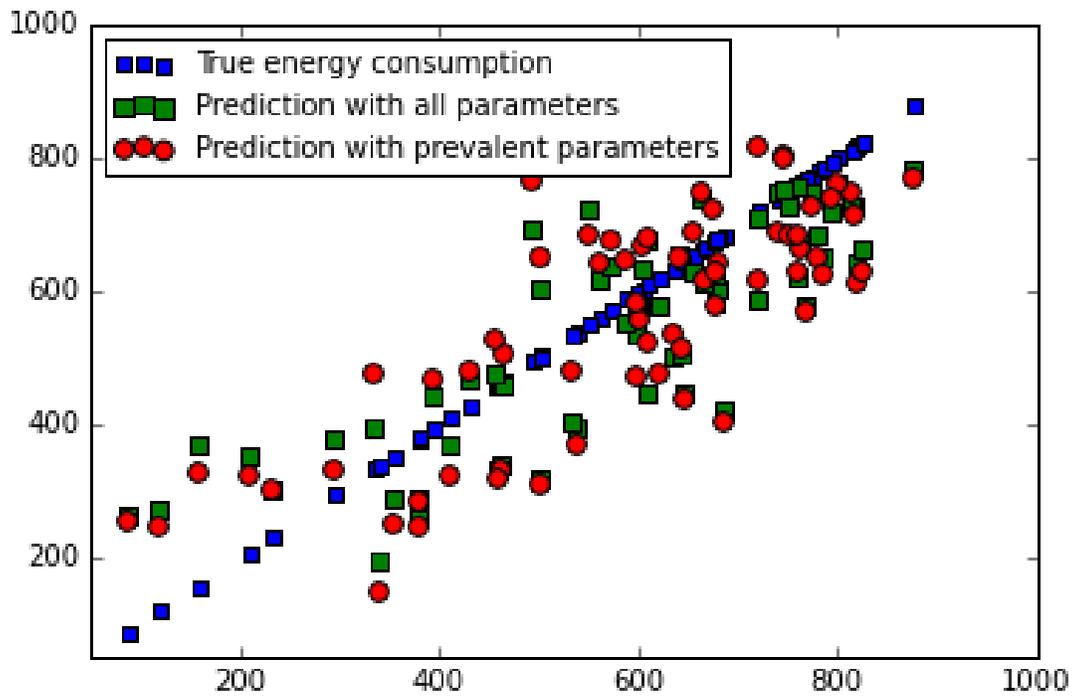

**Figure 5-4.** The prediction accuracy (error) of random forest regression models compared with actual energy consumption (blue): with all parameters (green), only relevant parameters (red).

# Chapter 6. Parametric Dijkstra-based Topology Management and Routing Algorithm

In this chapter, the concept behind the EDA model and energy-based constituents in the previous chapters will be used to create an energy effective topology management algorithm aiming to increase the overall lifetime and performance of various mesh topologies in WSNs. The new algorithm, hereafter called Parametric Dijkstra-based Topology Management Algorithm (PDTM), will be compared with a distance-based method based on Dijkstra. The experiments section shows how PDTM outperforms the other method in terms of increasing lifetime and performance. The aim of the algorithm is to dynamically manage the network topology and associated routing paths from each sensor to its sink, taking into account identified parameters to minimise the energy consumption of the global constituent and hence the overall energy consumption of the whole network.

The first step in developing our new topology management algorithm was to extract prevalent parameters for energy consumption in the individual, local and global constituents. In previous chapters, we investigated various parameters. Since some of the parameters are systemic (e.g., number of neighbours) and some are analytic (e.g., packet flows generated by routing algorithm), combining all of them together to form a mathematical formula is not meaningful, and is difficult as they address different concerns yet they affect one another. For example, the network density alters the average number of neighbours of a sensor, the average distance between nodes and the packet flow between nodes. Moreover, the curse of dimensionality implies that an increase in the number of parameters (i.e., features or dimensions) results in a rapid growth in the volume of searching space such that available data becomes sparse, which is problematic for methods that require statistical significance. To reach to a statistically reliable result, the required amount of data to support the result often grows exponentially with dimensionality. In addressing these difficulties, a three-step algorithm is proposed where in each step only few of these parameters are involved; therefore, a new concept is defined which can be used for optimisation.

Following this concept and due to the large number of parameters, a class of algorithms can be designed; depending on which parameters are used and how precisely they are used, a more sophisticated algorithm and consequently better results in terms of performance and lifetime are achieved.

## 6.1 Network model and priliminary study

We assume all sensor nodes in a WSN are similar in terms of hardware, initial power, transmission radius and so on. The connection among nodes is modelled by a communication graph $G = (V, E)$ where $V$ and $E$ are a set of wireless nodes in the network and their directed links, respectively. Weight between nodes $u$ and $v$ is denoted by $w(u, v)$, which implies a connection cost from node $u$ to node $v$. More precisely, the connection cost is the amount of energy node $u$ spends to send its packet to node $v$, and as expected it is always non-negative. In Chapter 4 and 5 we utilised statistical analyses (i.e., p-value, linear correlation, and non-linear correlation) on 800 experiments to find the most prevalent parameters that contribute to the overall energy consumption of a typical wireless sensor network. Based on this result and our knowledge of these dominant parameters, we consider 'weight' as a function of four parameters:

- Number of hops between a node and sink as a result we achieved in Chapter 5, this factor determines the eligibility of a node to be selected as a relay node. A node with a lower number of hops demonstrates a better option to be in a route path to the sink.
- Distance between two neighbours (geographical distance between nodes by knowledge of coordination of points). Distance has a direct effect on energy consumption.
- Number of neighbours as a parameter which determines the load of local and global tasks
- Node available energy as a dominant parameter on node lifetime

Moreover, we assume two parameters which are important in topological management algorithms :

- Number of nodes in the network; this factor affects the average number of neighbours and hops. Based on our experiments, Figure 6-1(a) and 6-1(b) show that a larger number of nodes results in a higher number of hops to the destination and a higher average number of neigbours per sensor.
- Transmission radius (assumed to be similar for all nodes in the network); this factor changes the network behaviour by changing the average number of neighbours. Based on our experiments, Figure 6-1(c) shows that an increase in transmission radius results in a higher number of neigbours in most of the experiments. We assume locations of nodes are known to the sink. If the network is self-organised, nodes need to inform the sink of their locations in the initialisation phase; however, this does not apply when nodes are deployed in pre-specified locations.

## 6.2 The PDTM algorithm

In this section, we present a description of the PDTM. The algorithm uses our EDA to model the energy consumption of all nodes in the network. In general, sending packets to a lower level node (i.e., a node with a lower number of hops to the sink) is an effective strategy. However, if a higher level node has more energy, it may be considered for the sake of energy balancing and prolonging the life of the whole network. In this case our algorithm increases the connection cost of the node with the lower level. That is, a neighbour node with higher level and more residual energy may be chosen over a neighbour with lower level and less residual energy. Furthermore, a higher value of number of senders implies greater cost in choosing a node as a relay, and hence has the effect of deterring other neighbours from selecting it. In fact, maximum number of senders for a relay node equals the number of neighbours. Also the distance between nodes is an important parameter that directly affects energy consumption. With this strategy, energy consumption is distributed among all possible relay neighbours. As a result, overall network connectivity becomes more stable, inducing improvements in network lifetime and performance.

This algorithm consists of three phases: (1) assigning level and relay degree to nodes in the graph, as two important factors to be used for calculating the cost of links between nodes; (2) computing connection costs between nodes, which will be used to compute the lowest cost value to the sink; and (3) applying the Dijkstra algorithm to find paths with the lowest connection cost (i.e., paths with lowest energy consumption) – this phase assigns a unique path to the sink for each node, which is the lowest cost. In the following we explain each phase in detail.

**Phase 1: Assigning the level and the relay degree to nodes in the graph**

After connecting each node to its neighbourhood and constructing network graph $G = (V, E)$, a node is assigned a level and a relay degree. Starting from the sink node (labelled "0"), we proceed as follows.

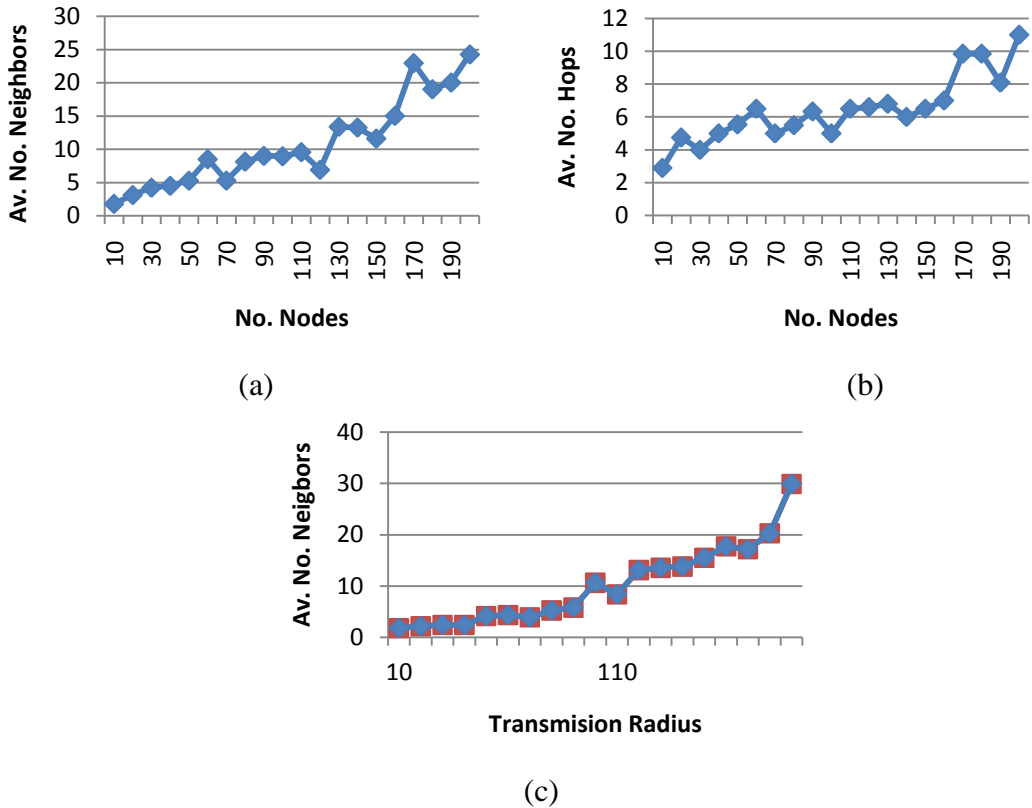

Figure 6-1. The effect of network size on (a) the average number of hops, and (b) the average number of neighbours; (c) the effect of transmission radius on average number of neighbours.

First, all sensors within the transmission radius of the sink are labelled "1" ($L_u = 1$); all sensors within the transmission radius of level 1 sensors are labelled "2" ($L_u = 2$); all sensors within the transmission radius of level 2 sensors are labelled "3". The level is the lowest number of hops between the node and the sink. The relay degree is defined as the number of neighbours that send data to the node for relaying. In other words, a data packet of a node with level $k$ must pass at least another $k$ relaying nodes to reach the sink. A node may become a relay node for its neighbours if its level is lower than theirs. The number of neighbours of a relay node is its relay degree. A node with relay degree of zero is called 'leaf' node. With these definitions, the level and the relay degree of sink node are both zero (as its power is assumed to be infinite). An example is shown in Figure 6-2.

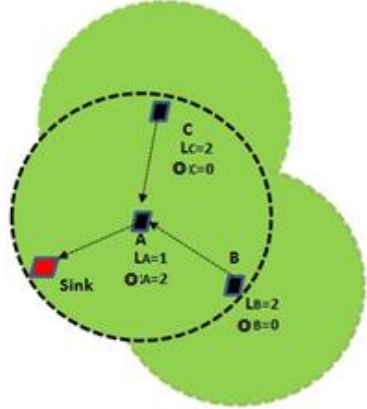

Figure 6-2. Illustration of level of a node and relay nodes

**Phase 2: Defining and evaluating the connection cost between nodes**

The key idea of PDTM is to find the best paths among nodes, subject to distributing energy consumption across all nodes in the network, and consequently minimise energy consumption of the entire network. Therefore, defining a proper function for the connection cost between nodes is critical. Towards this goal, we assume the connection cost between a node and a relay node is proportional to the number of neighbours of the relay node. Based on this assumption, the algorithm always selects nodes with smaller relay degrees to reduce the cost of transporting data to the destination and minimise traffic congestion along the selected path.

Regarding the scatter plots (Figure 5-3) and the dependency table (Table 5-2), it can be expressed that energy consumption is a function of the prevalent parameters. As the values of these parameters are chosen by the user, they are independent. Therefore, the relationship between energy and these parameters can be rewritten as

$$E \propto \mathcal{F}(Transmission\,radius) \times \mathcal{F}(\text{Average distance}) \qquad (6-1)$$
$$\times \mathcal{F}(\text{Average \# of neighbours}) \times \mathcal{F}(NetworkSize)$$
$$\times \mathcal{F}(\text{Average \# of hops})$$

From the other side, global energy is a summation of link cost between all nodes inside the network:

$$E = \sum_i \sum_{j \neq i} W_{ij} \qquad (6-2)$$

```
G (V, E) is the given directed graph
Sink's level =0
For each node
        Initialise an array of neigbor with all accessible nodes
        Initialize node's level to infinity except sink
End For
For each node
  For each node's neigbor
        If node is not dead
                Choose lowest level among neigbors as node's level
        If node's level <= neigbors level
                Compute and Assign weight edge e(v,vj) between node's and
                neigbor using the cost  the function (5)
                Increase degree of the node
        End if
  End For
End For
For each node
        Find the shortest path to Sink using Dijsktra Algorithm
End For
```

Figure 6-3. Pseudo-code for PDTM

where $W_{ij}$ is the link cost between nodes $i$ and $j$. The parameters $Transmission\ radius$ and $Network\ Size$ are specified at the beginning of the network by the user and do not change during the experiment, so they do not change the link cost between two nodes; as a result, the link cost between two nodes may be defined as

$$W_{ij} \propto f(\text{distance}) \times f(\text{\# of neighbours}) \times f(\text{\# of hops}) \qquad (6-3)$$

Also, as indicated in (Liu et al., 2012), the link cost depends on the residual energy of node i; thus, the link cost is rewritten as

$$W_{ij} = K \frac{f(\text{distance}) \times f(\text{\# of neighbours}) \times f(\text{\# of hops})}{residual\ energy\ of\ node\ i} \qquad (6-4)$$

The connection cost (i.e., energy consumption) between node $i$ and the higher level node $j$ in its neighbour (i.e., $L_j \geq L_i$) at time $t$ is defined as

$$W_{i,j}(t) = \frac{O_i(t) * d_{i,j}^{\alpha} * (L_i(t) + 1)}{E_i(t)} \qquad (6-5)$$

```
G (V, E) is the given directed graph
Sink's level =0
For each node
        Initialise an array of neigbor with all accessible nodes
        End For
For each node
  For each node's neigbor
        If node is not dead
                Choose lowest level among neigbors as node's level
        If node's level <= neigbors level
                Compute and Assign weight edge e(v,vj) between node's
                and neigbor using the cost = $d_{i,j}^2$

        End if
   End For
End For
For each node
        Find the shortest path to Sink using Dijsktra Algorithm
End For
```

Figure 6-4. pseudo-code for DDTM

where $d_{i,j}$ is the Euclidean distance between these two nodes, $1 < \alpha \leq 2$ is a constant (Heinzelman et al., 2002), $O_i(t)$ and $L_i(t)$ are the relay degree and the level of node $i$, respectively and $E_i(t)$ is the residual energy of node $i$ at time $t$. More precisely, the connection cost (or weight) is as follows:

- proportional to a power of the Euclidean distance between the node and its neighbour.
- proportional to $O_i(t)$, the number of nodes considering node $i$ as the relay node. A higher value of this number implies a greater cost in choosing this node as a relay, which has the effect of deterring other neighbours from selecting it. The node thus has more chance to lengthen its lifetime.
- inversely proportional to the residual energy of the node. This implies nodes with less energy have a higher weight (i.e., connection cost) and therefore are not likely to become a relay node.
- proportional to the level of the node. This implies that nodes that are closer to the sink (in terms of hops) are more likely to be selected to be relay nodes.

**Phase 3: Dijskstra-based lowest cost paths**

The last part of the proposed algorithm employs Dijkstra to search for the shortest paths from nodes to the sink, subject to connection cost expressed in Eq. 6-5. It may be noted that the connection cost is a function of time and this implies that these costs must be updated at end of each timeslot and Dijkstra must again be applied on the updated connection costs. In our experiment, we reapply the algorithm after the number of packets received by the sink is equal to the number of nodes in the network. The reason for repeating the algorithm is that as the residual energy of a node is reduced in each timeslot, some nodes may cease to exist and as a result alive nodes must find a new path to send their packets to the sink. Recalculating dynamic connection costs and reapplying Dijkstra incur negligible computing overhead for networks with few nodes (such as our simulation), but in large networks, shortest paths algorithms on dynamic costs (Nannicini and Liberti, 2008) may be used to reduce the computation costs. The detail of our algorithm is shown in Figure 6-3.

## 6.3. Experimental Result and Discussion

In this section we present the results of residual energy and lifetime of a WSN obtained from simulating PDTM in comparison with Distance Dijkstra-based Topology management algorithm similar to the method in (Mao et al., 2011, Yin et al., 2008, Zhang et al., 2013), hereafter called DDTM.

### 6.3.1 Distance Dijkstra-based Topology Management algorithm

Distance Dijkstra-based Topology Management algorithm (DDTM) determines the shortest path from each node to the sink. The only difference between DDTM and PDTM is the costs that are assigned to the edges of the network graph. In DDTM, $d_{i,j}^2$ is assigned to the edge between two nodes. The Dijkstra algorithm is used to find shortest paths from each node to the sink. Figure 6-4 shows the pseudo-code of DDTM.

### 6.3.2. Simulation Settings

We evaluated the performance of the proposed algorithm by simulating the topology management algorithm developed to minimise energy consumption paths between nodes. Network topologies were prepared with a two-dimensional uniform distribution generator. We randomly placed 100, 150, 200, 250 and 300 nodes in a rectangular region of 600 by 300 units, where nodes were deployed on a mesh with the granularity of one unit and had the same characteristics, such as battery power. The transmission radius of each sensor node was set to 100, 200, and 300 units. For each set of nodes and transmission radius we generated 30 random graphs.

The simulations were carried out using the simulator we developed as a part of this study. This simulator enables us to study and examine all the concerned parameters of our model. The simulator is written in C# and can generate sensor networks with different density and topology. The simulation has the following main steps:

1. Define the node's characteristics
2. Organise the network by assigning the *x*- and *y*-axes into a two-dimensional rectangular coordinate system for each node.
3. Assign transmission radius
4. Draw the link between each node and its neighbours as the network graph's edges
5. Assign a level to each node
6. Assign a degree to each node
7. Assign weight on each edge of network graph using Eqn.5 for PDTM and $d_{i,j}^2$ for DDTM
8. Apply Dijkstra on the network graph
9. Inform each node of its parent by distributing a control packet which includes the node's position and its parent. Start the time slot timer.
10. Each node consumes energy for generating, receiving, sending a packet and relaying other packets to its parent
11. Stop timeslot timer when the number of received packets in the sink equals the number of nodes in the network.
12. Repeat steps 4 to 11 until the sink does not have any alive neighbours and exit.

The lifetime of an experiment is calculated in time units (milliseconds in our simulator), and it is the duration of running step 4 to 11 until the sink cannot be reached by any node. A node which does not have the energy to send or relay a packet is called a dead node. A node which is disconnected is also counted as a dead node.

A timeslot is defined in the network manager. The timeslot starts when the manager publishes the new paths determined by the algorithm and ends when the prevalent parameters change. In our simulator the nodes frequently generate packets so the manager assumes the end of the timeslot when the number of received packets at the sink is equal to the number of the nodes in the network. The manager then calculates and publishes the new paths to nodes in the network. The packets that are on their way to the sink will continue travelling to the sink regardless of the paths just assigned. Basically, the relay nodes must accept an incoming packet and are not allowed to drop it. The nodes just send the received packets based on the latest assigned path.

All nodes will be informed about their new parent after each timeslot. We use an information packet which includes each node's ID (in our simulator their axis coordinates) and the parent node. This packet will be distributed in the network by the sink. The energy consumption of the information packet can be minimised in different ways in different applications, such as by lengthening timeslots.

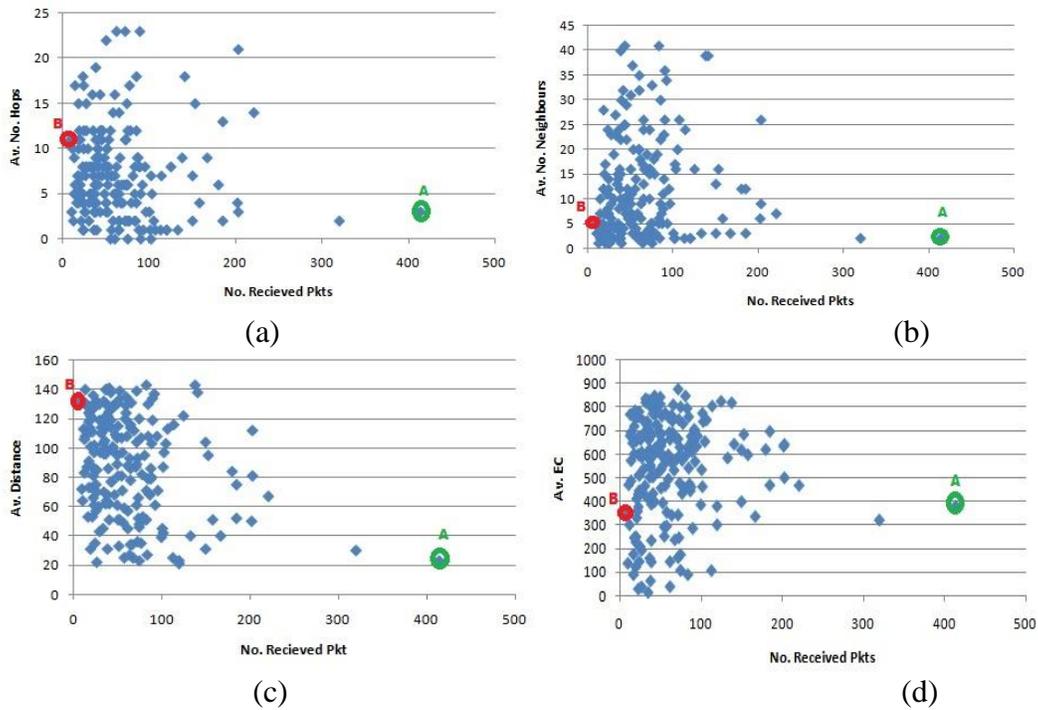

Figure 6-5. The effect of PDTM parameters on performance (a)Average number of hops, (b)Average number of Neighbours, (c) Average Distance (d), Average Energy Consumption

### 6.3.3. Results and Discussion

A comparison between the DDTM and the PDTM is presented in the following sections.

#### 6.3.3.1. PDTM parameters and Network Performance

We define network performance as the number of packets received by the sink at the end of an experiment. A higher number of received packets at the sink implies a better network performance. Figure 6-5 shows the effects of the four dominant parameters in PDTM on the performance of 200 networks with a random number of nodes, position, and transmission radius running same routing method. We studied the best and worst performing networks. Network A (Figure 6-5) is the best performing network in terms of the number of received packets with a low number of average hops (3 hops), neighbours (8 neigbours), and distance (23 units). In contrast, the worst performance is network B, which has an average of 11 hops per experiment and 5 neighbours, but a very high distance (131 units).

It should be noted that the energy consumption of network B is lower than that of network A, as can be seen in Figure 6-5(d), but this does not mean that network B is better than A. This is because network B becomes disconnected earlier, and this results in the lowest number of received packets at the sink. This result shows that a single parameter does not guarantee the minimum overall energy consumption and/or the best performance of a network. We need to consider the play-off between the parameters.

**6.3.3.2. Comparison Based on Network Topology and Routing**

A comparison between DDTM and PDTM is shown in Figures 6-6 and 6-7, respectively, for a small network of 10 nodes. For DDTM, the weight of $d_{i,j}^2$ is assigned to each edge and the traditional Dijsktra shortest path algorithm is then employed.

In Figures 6-6 and 6-7, the green number below each node is the cost from each node to the sink, which is the minimum cost based on Dijkstra. For DDTM, this value is the summation of the edges' weight $d_{i,j}^2$ from a note to its sink, and for PDTM, Eq. 6 is used to calculate the edge weight. The black numbers indicate the nodes' levels. In each time slot, nodes send one packet to the sink (node 0):

- Figures 6-6(a) and 6-7(a) show the networks after the first time slot, both networks produced 9 packets because there are 9 alive nodes in the network and we expect they all generate a packet and send it to the sink from the path generated by Dijkstra. In DDTM, one of the sink's neighbours is the relay node for three other nodes, while in PDTM it serves as the relay for only one neighbour.

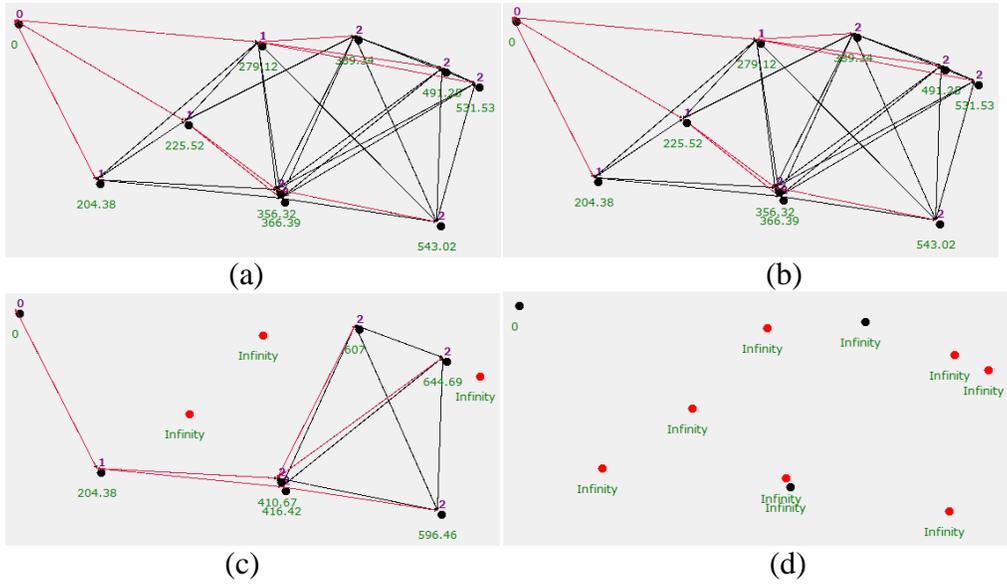

Figure 6-6. Applying DDTM to a network of 10 nodes in four timeslots.

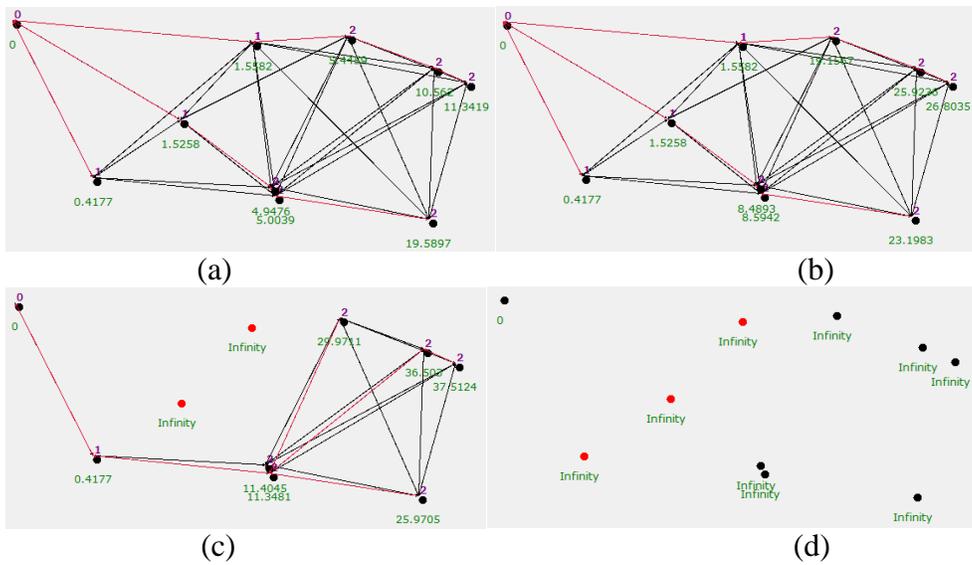

Figure 6-7. Applying PDTM on the same network as in Figure 6-6.

- At the second time slot (Figures 6-6(b) and 6-7(b), all nodes in the PDTM network are still active but in DDTM some nodes died during the experiment, resulting in packets lost in this timeslot. This is the reason why DDTM delivered only 5 successful packets to the sink while PDTM delivered 7 successfully. Dropping packets is mainly due to dead relay nodes in a timeslot.

- Figures 6-6(c) and 6-7(c) show the networks in the third time slot. All nodes except one in the first level of both networks are dead, so nodes in other levels have to send their packets to this node. The number of remaining nodes in PDTM is higher than in DDTM.

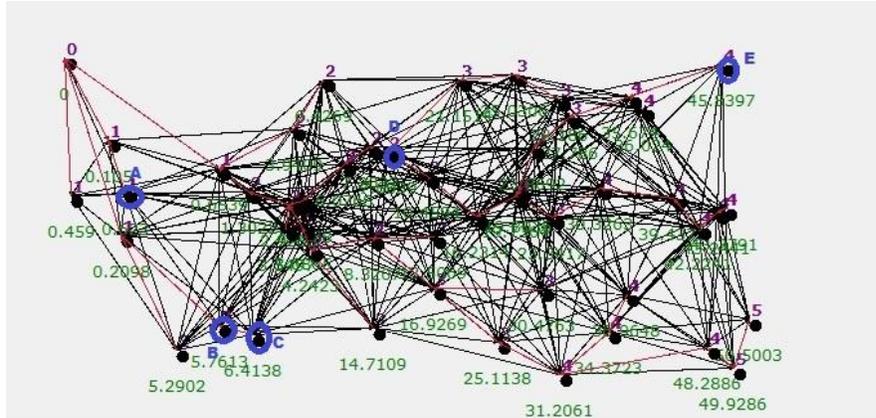

(a)

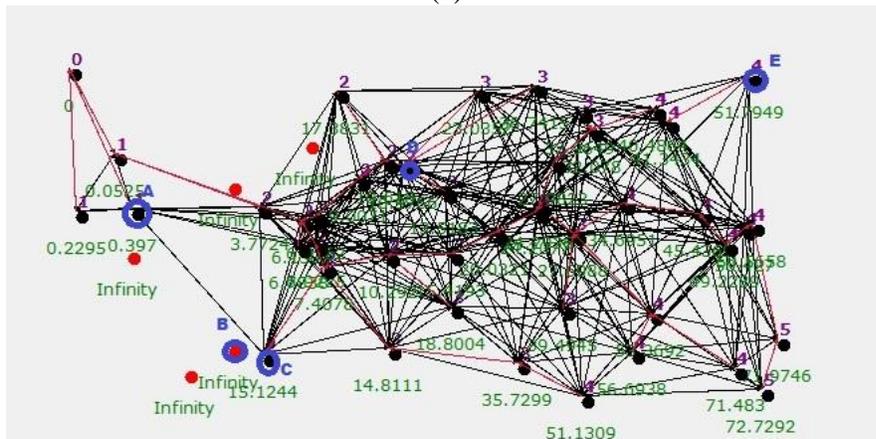

(b)

Figure 6-8. Applying PDTM on the same network in the previous figure:(a) the first time slot, (b) after nine time slots

- As all first level nodes died, both networks cease to exist in Figure 6-6(d) and 6-7(d). The overall number of packets lost and successful packets are 7 and 17, respectively, in DDTM, while these numbers are 3 and 22 for PDTM. In DDTM, all nodes except one are dead, while in PDTM nodes in the second level are still alive (i.e., better distribution of energy consumption between nodes) and can continue operating if at least one of the relay nodes in the first level returns back to action. To conclude, if relay nodes (i.e., the first level nodes) perform longer (e.g., are recharged) the lifetime and performance of the network under PDTM will improve.

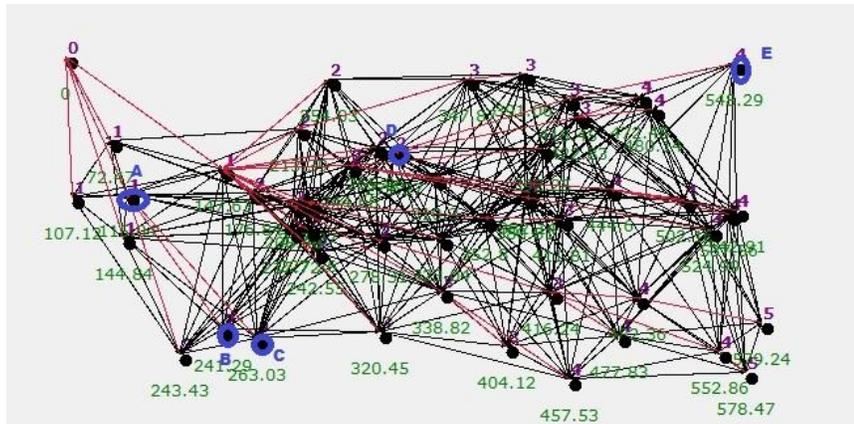

(a)

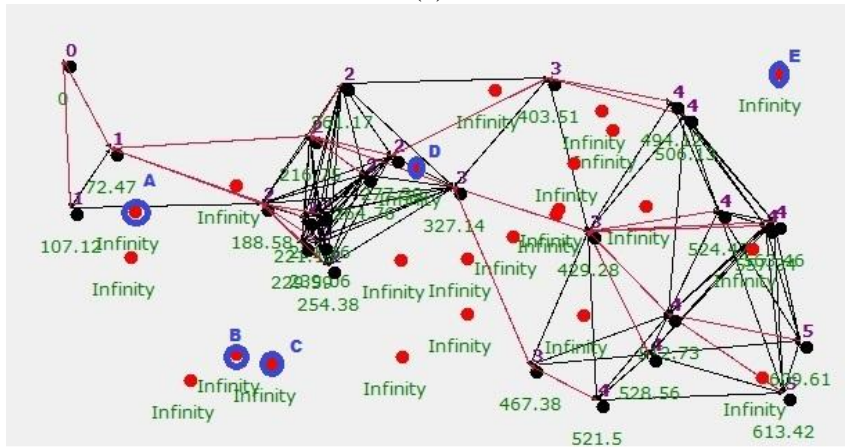

(b)

Figure 6-9. Applying DDTM on a network of 50 nodes: (a) the first time slot, (b) after nine time slots.

One of the main factors impacting energy is the density of a network. Figures 6-8 and 6-9 show the results of our investigation of a denser network with 50 nodes.

Due to the fact that the direct geographical distance between two nodes is always shorter than indirectly via a third node, the shortest path algorithm in DDTM naturally tries to connect two nodes directly if possible; therefore, in this algorithm, communication between a node and sink with a smaller number of middle nodes is preferable. The PDTM is, however, more constrained; it chooses neighbours based on their residual energy, the number of neighbours, and the number of hops from each neighbour to the sink to diversify paths among nodes. In terms of our EDM model, global tasks (i.e., relay packets to the sink) are more distributed in PDTM than DDTM.. Figure 6-8(a) and Figure 6-9(a), at the end of the first timeslot, show that DDTM connects nodes B and C to node A but PDTM connects B to A and C to B, which saves two nodes: node A as a key node to connect to the sink, and node C which can do node B jobs after it is dead in the ninth timeslot. In DDTM node D (level 2) received packets from node E (level 4) by its long distance neighbour (level 3). Node D and most of its neigbours run out of energy in the ninth timeslot. In contrast, in PDTM the task of relaying level 4 nodes like E is distributed in nodes in level 3 and 2; therefore, a long distance breaks into small connections and a level 4 node consumes less energy than sending packets direct to level 2. As Figure 6-9(b) shows in the ninth timeslot, PDTM has more alive level 2 nodes like node D in comparison with DDTM. Moreover, the energy of level 4 nodes also will be saved due to sending packets to a more suitable node.

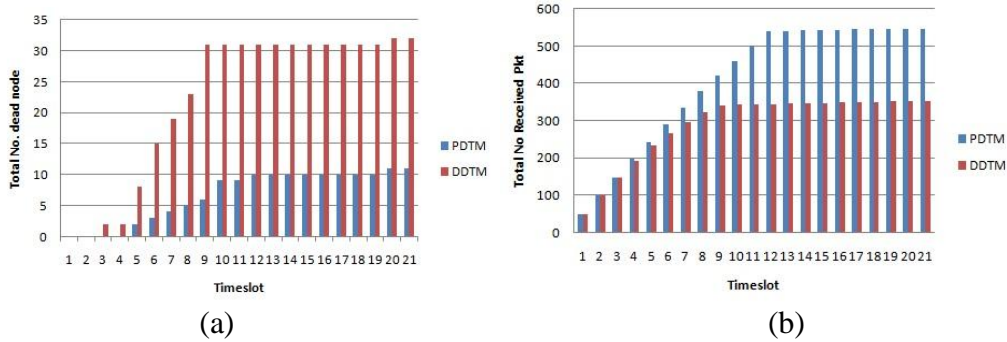

Figure 6-10. Applying DDTM and PDTM on a 50-node network: (a) the total number of dead nodes in each timeslot; (b) the total number of received packets in the sink.

Figure 6-8(b) and 6-9(b) show the status of both networks in the ninth timeslot. The number of dead nodes in DDTM is clearly much higher than PDTM (23 vs. 5, respectively). Figure 6-10(a) shows the total number of dead nodes in both algorithms in each timeslot.

Table 6-1. Comparison between PDTM and DDTM at the end of the experiment for a network with 50 nodes

|  | DDTM | PDTM |
| --- | --- | --- |
| Number of dead nodes | 32 | 11 |
| Number of received packet | 351 | 547 |
| lifetime (in milliseconds) | 551 | 5432 |

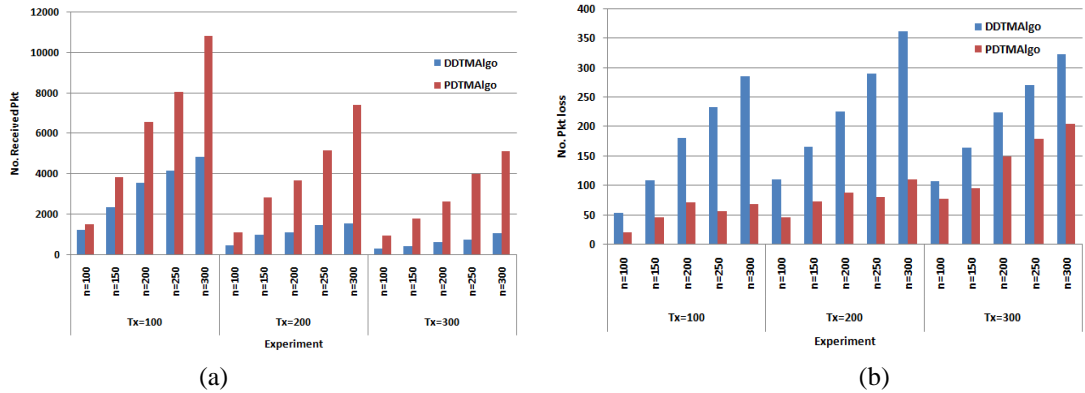

Figure 6-11. Comparison between PDTM and DDTM: (a) number of received packets by sink; (b) packet loss.

A network is assumed dead (i.e., end of the experiment) if its connection to sink is disrupted. Referring to Table 6-1, death of the 50-node network occurs after 5432 milliseconds for PDTM, which is 9.8 times longer than for DDTM. The table also indicates that, compared to DDTM, the number of packets lost in PDTM is 40% lower while it delivered 2.5 times more packets to the sink. As a result, the network with PDTM is more effective than with DDTM. Figure 6-10(b) shows the total number of received packets in each timeslot; the number of successful packets using PDTM is considerably higher than with DDTM.

#### 6.3.3.3. Comparison Based on Network Performance

We compare the performance of PDTM with DDTM with a large number of variations in the transmission radius of sensors and number of nodes in our simulations. The randomly created networks consisted of five graph sizes of 100, 150, 200, 250 and 300 nodes, together with three different transmission radii of 100, 200, and 300 units. Figure 6-11(a) shows the number of received packets by the sink and Figure 6-11(b) shows the packet loss for the simulated networks. It can be seen from these figures that the PDTM constantly delivers a higher number of packets and suffers lower packet loss values than DDTM. Increasing the value of the transmission radius results in a drop in the number of received packets of both algorithms. This is mainly because with a higher value of the transmission radius, each node can communicate with nodes much further from itself, resulting in a wider covering, wider area and more neighbours, and this implies a network with a higher degree of connectivity. As a result, the nodes are able to find more multiple and longer paths (more hops). Using longer paths to the sink or in other words using more hops results in higher overall energy consumption solutions.

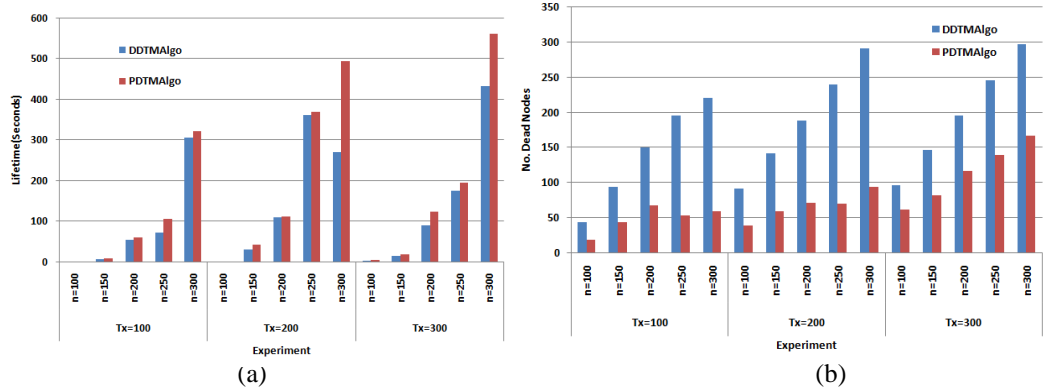

Figure 6-12. Comparison between PDTM and DDTM: (a) lifetime, (b) dead nodes at end of network

#### 6.3.3.4. Comparison Based on number of dead nodes and lifetime

Two other metrics for further study are the impact of the new algorithm on the lifetime of the entire network and the number of dead nodes at end of the network lifetime. There are a few definitions (Champ et al., 2009) for lifetime in WSNs, but it is hard to generalise them. In our algorithm, we define lifetime as the time until the sink is disconnected from other nodes; this happens when the nodes in the first level (neighbours of sink) are all dead. Figure 6-12 shows that our algorithm (PDTM) is superior to DDTM in both lifetime and number of dead nodes in all cases. It can be observed that the PDTM reaches longer lifetime values than DDTM. Moreover, the PDTM shows a sharper rising trend in lifetime in denser networks; therefore, its improvement gap over the DDTM increases with the size of the network.

Table 6-2. Comparison between PDTM and DDTM on all simulations

|  | DDTM | PDTM |
|---|---|---|
| Av. Number of dead nodes | 176 | 76 |
| Av. Number of packet loss | 207 | 92 |
| Av. lifetime(in seconds) | 128 | 161 |
| Av. Number of received packets | 1674 | 4380 |

Table 6-2 summarises the comparison between these two algorithms. This table clearly demonstrates the superior improvement in all four indicators (i.e., number of received packets by sink, packet loss, lifetime and number of dead nodes) of PDTM compared to DDTM in all simulated networks.

### 6.3.3.5. Comparison Based on Energy Consumption

The next comparison is the impact of PDTM on energy consumption of the network as time progresses. Figure 6-13(a) shows the energy consumption and Figure 6-13(b) shows the number of received packets delivered to the sink for a sample network (number of nodes and transmission radius are 150 and 100, respectively). Figure 6-13(a) also shows the percentage of overall energy saving for PDTM compared to DDTM.

It may be observed that the PDTM consistently has lower overall energy consumption than the DDTM from the start of the network until its death. The energy consumption of DDTM increases sharply up to 12 seconds, when it becomes flat (i.e., dead network) while the energy consumption rate is slower, and the flat level is lower for the PDTM. This trend can be explained as follows: in DDTM, key nodes die early because they are closer to the sink and are used as relay nodes for more nodes without considering their limited energy.

## 6.4 Summary and remarks

In this chapter we proposed a new parametric topology management algorithm to manage effectively the energy consumption of sensors in wireless sensor networks. Based on a proposed Energy Driven Model (EDM), the most prevalentrelevant parameters in a typical wireless sensor network were extracted to feed a parametric topology management algorithm. These parameters were the residual energy of nodes, the number of neighbours, the number of neighbours a node acts as their relay, the number of hops, the transmission radius and the distance between nodes. Separate parameters were considered in one or other previous research efforts. Taking all previous researches together, all parameters were covered but not in one particular work. More importantly, Our work is significant in that it exposed the interplays between dominant parameters that affect the overall energy consumption and this opens the door for new energy optimisation methods.

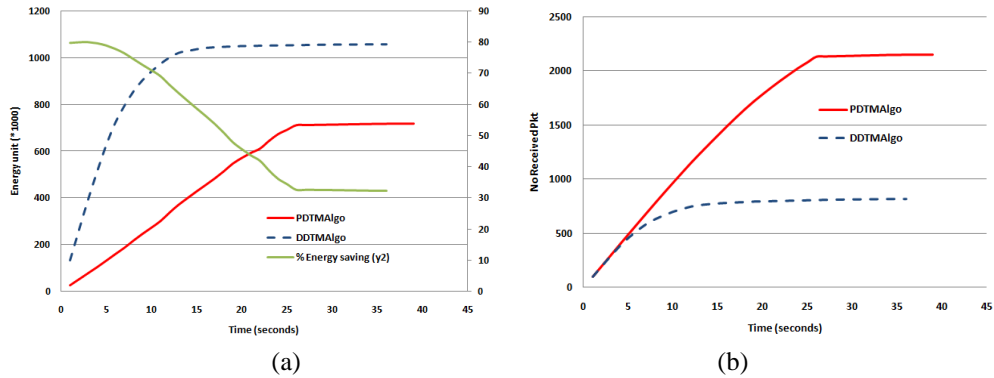

Figure 6-13. Comparison between PDTM and DDTM over time for a 100-node network: (a) energy consumption, (b) number of received packets.

After generating a time-variant connection cost function between nodes in the network based on these parameters, the algorithm employs Dijkstra to search for shortest paths from nodes to their sink with least energy consumption.

Compared with the standard Dijkstra algorithm used in most networking communication algorithms and through extensive simulation, our parametric topology management algorithm showed superior improvement in terms of the number of successfully delivered packets, number of packets lost (in different network topology and network density), and energy consumption of the entire network.

We studied various sensor network parameters; simulators such as NS2 or Omnet++ (Korkalainen et al., 2009) cannot cover all the parameters. Therefore, we had to develop a WSN simulator that allowed us to investigate these parameters and their relationship with the energy consumption.

In this chapter the PDTM algorithm was applied to a typical mesh topology WSN; however, the algorithm can be used in similar sensor applications with similar characteristics such as agricultural fields, or rainforest sensor applications. The selected parameters may be different depending on the characteristics of the application. As a result of our study, taking into account the prevalent parameters and the interplay between them will result in better performance in terms of more work done and longer lifetime through an effective energy consumption strategy.

# Chapter 7. Conclusion and future work

Through a number of contributions, this thesis has advanced the state of the art in wireless sensor networks in terms of energy issues. The main contributions of this thesis were presented in four chapters. In this chapter, we summarise the conclusions and directions for future work on a chapter by chapter basis.

## 7.1 Conclusions

In Chapter 3, a new architecture to study the relationship among parameters and overall energy consumption of wireless sensor network was proposed. This architecture deals with all common aspects of energy consumption in all types of WSNs and identifies constituents that play major roles in energy consumption. Designing wireless sensor networks with this architecture in mind will enable designers to balance the energy dissipation and optimise the energy consumption among all network constituents and sustain the network lifetime for the intended application.

By categorising the overall WSN system into a few constituents, components of each constituent were extracted in terms of their dominant factors (a.k.a parameters), followed by a mathematical formula as a total energy cost function in terms of their constituents. Through simulation of sample networks with different sensor radius, transmission radius and random/selective networks, we showed the proposed model and formulation can be used to optimise lifetime and residual energy of these networks with respect to the contribution of each constituent and its relative significance. It was discussed that optimising the energy of the general model with respect to parameters of all constituents enables one to engineer a balance of energy dissipation among constituents, optimise the energy consumption among them and sustain the network lifetime for the intended application.

In Chapter 4, we proposed a generic model incorporating various energy consumption constituents and components of sensors in a wireless network. This model, while being independent from the underlying network architecture, helps identify essential energy consumption constituents and their contributions to the overall energy consumption of a sensor. Such capability, coupled with the interplay of the sensors with the network, facilitates the realisation of various strategies while fulfilling the individual sensors' constraints in terms of energy. Employing linear regression to model relationships between sensors' functionalities and the overall energy of the network, the model can then be utilised by the sensor to prioritise the constituents' tasks in term of energy level and importance in order to make an appropriate decision. As a result, the sensor can use the power in an effective way and remain alive longer. Through this chapter, it was concluded that the global constituent has the highest impact on the overall energy consumption of a WSN.

In Chapter 5, it was stressed that due to the high number of parameters impacting energy consumption of a network, it is almost impossible to reach a comprehensive model. This reason for this is the curse of dimensionality, a well studied issue in high dimensional systems. To escape from this problem, the generic step is to reduce the dimensionality of systems. Therefore, in this chapter a few statistical tools (e.g., p-value, linear/non-linear correlation and regression with L1-norm) were applied to a list of parameters obtained in previous chapters in regards to the energy consumption of the global constituent in the network, as the dominant constituent in the energy consumption of the network (Chapter 4). After reduction, random forest regressions were applied to both all and relevant lists of parameters to, first, model the relationship between global parameters and energy consumption, and second, to determine how much accuracy we lose due to parameter reduction. Our extensive experiments showed not only the importance of network parameters to energy consumption in the network, but also that removing less important parameters has a minor affect on prediction of energy consumption.

In Chapter 6, prevalent parameters (i.e., reduced parameters in the previous chapter) in a typical wireless sensor network were applied to create a new parametric topology management algorithm aiming at reducing energy consumption of sensors in the network. After using these parameters to generate a time-variant connection cost function between sensors in the graphs, energy-efficient paths between sensors and their associated sinks were observed by employing a Dijkstra algorithm. While the idea of employing Dijkstra in routing of packets in networks is not new, in this chapter we proposed a new complex function of energy-related communication costs taking into account the prevalent parameters. Through extensive simulation, the algorithm with new costs showed superior improvement in terms of the number of successfully delivered packets, number of packets lost (in various different network topologies and network densities), and energy consumption of the entire network compared to the standard Dijkstra algorithm.

**7.2 Future Directions**

- **Chapter 3:** while our aim in this chapter was to extract energy-consuming constituents and their relationships in a generic model, we ignored application-related interplay between these constituents and energy. Therefore, one extension to the architecture would be to explore the patterns and shape of the energy consumption for a generic application (e.g., health monitoring applications) and produce a comprehensive map of energy consumption relative to a specific application. Another goal is to come up with a single overall formulation of the energy consumption of the entire wireless sensor network; preliminary investigation assumed a weighted linear combination of energy consumption of the constituents. Interplay among the components can be taken into account in terms of their weights as some function of the design of the WSN and the application; in the future we plan to produce a more accurate energy cost function which accurately places due emphasis on parameters, components and the playoff factors among components. We believe that a non-linear cost function rather than a simple linear combination would allow the model to adapt better to a specific WSN application. Another important aim, which is being pursued in the next stage of our research, is to model comprehensively the components of each of the five energy constituents of the architecture. The aim is to provide an accurate account of all functional aspects of a constituent and their salient energy-wise parameters. These parameters will allow us to evaluate the performance of WSNs, optimise their operations, and design more energy-efficient applications.

- **Chapter 4:** with the limitations of linear regression (and other similar methods), it was only possible to obtain a linear relationship between the sensors' constituents and energy consumption. Such model is not adequate for obtaining a complex non-linear relationship. Another realistic but more difficult formulation expresses the energy consumption model as a non-linear function of its constituents. As a positional future research, this approach requires more extensive exploration as we do not understand well enough the metric associated with the energy of each constituent, and we need to investigate mathematical models that can handle such a non-linear relationship. Regardless of the approach taken, the aim of the application has to be taken into account as this will determines the 'shape' of the overall energy consumption. For example, the requirements of the application may dictate the topology of the deployed sensor network, its routing mechanisms, or even the characteristics of the employed sensors.

- **Chapter 5:** our finding on the accuracy issue in extracting prevalent parameters calls for further work on employing advanced statistical and machine learning techniques to detect nonlinear correlation/causal relationship between parameters and energy consumption in global constituents in WSNs. Due to the high number of parameters, plotting the relationship between a parameter and energy consumption and subsequently choosing a proper kernel to explain the relationship is not a preferable option. Besides p-value, extra advanced analysis is required to rank parameters (a.k.a model selection) based on their statistical significance (e.g., Bayesian Information Criteria, Akaike Information Criterion). It is worth trying the same analytical tools on other constituents to detect their prevalent parameters. This will give a map of useful parameters impacting the energy consumption of the entire WSN system.

    Random forest regression, used to model the relationship between parameters and energy consumption, acts like a black-box with low information about the process. Employing techniques like deep learning, although they require a large number of training samples, gives a better insight into the system. As explained before, the models do not show a good fitness, probably because of a lack of useful independent parameters. Finding a better set of parameters along with usefulness analysis of them is another plan for future work.

- **Chapter 6:** the shortest path algorithm was developed with the assumption of a static network during each time slot and the evaluation of the topology management algorithm was also performed based on this assumption. This assumption has a drawback, however: if the time slot is wide, then the effect of a dead (or rejoined) node does not appear until the beginning of the next time slot when Dijkstra is rerun to find the shortest path among live nodes. But if the time slot is small, recalculating Dijkstra at the start of each time slot adds an extra load to the system. Although extremely challenging, a future extension can be to incorporate a dynamic state in the network into the shortest path algorithm, by which we mean insert or remove vertices in the graph during the execution.

# Appendix A. Statistical tools for parameter selection

```python
import array
from collections import Counter, defaultdict, OrderedDict
import matplotlib
import matplotlib.pyplot as plt
import numpy as np
import scipy.sparse
from scipy.stats import spearmanr, pearsonr, ttest_ind
from sklearn.metrics import mean_absolute_error, r2_score, mean_squared_error
from sklearn.linear_model import RandomizedLasso, LassoLarsCV
from sklearn.linear_model import LinearRegression, Lasso, Ridge
from sklearn.metrics import mean_squared_error
from sklearn import cross_validation
from sklearn.ensemble import RandomForestRegressor
from sklearn.cross_validation import cross_val_score

raw_data = 'effectiveparameters3.csv'
with open(raw_data, 'r') as f:
    data = f.readlines()
parameters_name = data[0].replace('\n','').replace('\r','').split(',')

param = list()
for d in data[1:]:
    param.append([float(x) for x in d.replace('\n','').replace('\r','').split(',')])

parameters = np.array(param)
x_data = parameters[:,3:-5]
y_data = parameters[:,-4]

parameters_name = parameters_name[3:]

def t_test(X,Y):
    return ttest_ind(X,Y)

def regression(X, Y):
    # is used to calculate p-value for regression analysis
    return f_regression(X,Y)

# calculate and plot linear correlation between these parameters and target
def perason_correlation(X, Y):
    return pearsonr(X,Y)

# calculate and plot linear correlation between these parameters and target
def nonlinear_correlation(X, Y, N=2):
    mX2 = np.mean(X**N)
    mY2 = np.mean(Y**N)
    correlations = np.true_divide(np.sum((X**N-mX2)*(Y**N-mY2)),
                    np.sqrt(np.sum((X**N-mX2)**2)*np.sum((Y**N-mY2)**2)))
    return correlations

def spearman_correlation(X,Y):
    return spearmanr(X,Y)

def linear_regression_L2(X, Y):
    N = len(Y)
    X = np.array(X).reshape(N, 1)
```

```python
    Y = np.array(Y).reshape(N, 1)
    model = LinearRegression(fit_intercept=False).fit(X, Y)
    return (model.intercept_ , model.coef_ )

def randomize_losso(X,Y):
    return RandomizedLasso().fit(X,Y).scores_

def evaluation(original, predicted):
    MAPE = np.sum(np.absolute(original-predicted)/original)/float(len(original))
    normalized_RMSE = np.sqrt(np.mean((original-predicted)**2))/np.mean(original)
    R2 = r2_score(original, predicted)
    PRED = len(np.where(np.absolute(original-predicted)<0.25*original)[0])/float(len(original))
    print('MAPE: {} , RMSE: {}, R2: {}, PRED: {}'.format(MAPE, normalized_RMSE, R2, PRED))
    return MAPE, PRED, normalized_RMSE, R2

def caluclate_relation_power(X, Y):
    N = len(Y)
    X = np.array(X).reshape(N, 1)
    Y = np.array(Y).reshape(N, 1)
    X_train, Y_train = X[:2./3*N], Y[:2./3*N]
    X_test, Y_test = X[2./3*N+1:], Y[2./3*N+1:]

    power = [1, 2, 3, 5]
    for p in power:
        model = LinearRegression().fit(X_train**p,Y_train)
        y_test_prediction = model.predict(X_test)
        print('X^{}: {}'.format(p, evaluation(Y_test, y_test_prediction)))

x_data_normalized = x_data
y_data_normalized = y_data
f_score,p_value = regression(x_data_normalized,y_data_normalized)
noTmask = list()

for index, xd in enumerate(x_data_normalized.T):
    print('{} --> {}: \n    corr: {}, spearman-corr: {}, non-corr(2 degree): {}, non-corr(3 degree): {},  pValue: {}, FScore: {}\n'
        .format(parameters_name[index],
            'average energy/delivered packet',
            perason_correlation(xd,y_data_normalized),
            spearman_correlation(xd,y_data_normalized),
            nonlinear_correlation(xd,y_data_normalized, N=2),
            nonlinear_correlation(xd,y_data_normalized, N=3),
            p_value[index],
            f_score[index]))
    data_sorted = [(x,y) for (x,y) in sorted(zip(xd,y_data_normalized), key=lambda x:x)]
    line_corr = perason_correlation(xd,y_data_normalized)
    if (np.fabs(line_corr[0]) > 0.25 and line_corr[1] < 0.05):
        noTmask.append(index)
        print('Feature {} is effective'.format(parameters_name[index]))
    xd_sorted = [item[0] for item in data_sorted]
    y_data_sorted = [item[1] for item in data_sorted]

    fig,ax = plt.subplots()
    ax.scatter(xd_sorted, y_data_sorted)
    ax.set_xlabel('{}'.format(parameters_name[index]))
    ax.set_ylabel('{}'.format('average energy/delivered packet'))
    fig.show()
    caluclate_relation_power(xd, y_data)
```

```python
model = RandomForestRegressor(random_state=0, n_estimators=20)

cv = cross_validation.KFold(len(x_data), n_folds=5, shuffle=True, indices=True)
for train_index, test_index in cv:
    model.fit(x_data[train_index], y_data[train_index])
    prediction = model.predict(x_data[test_index])

x_reduced = x_data[:,noTmask]
cv = cross_validation.KFold(len(x_reduced), n_folds=5, shuffle=True, indices=True)
for train_index, test_index in cv:
    model.fit(x_reduced[train_index], y_data[train_index])
    prediction_reduced = model.predict(x_reduced[test_index])

N = len(y_data)
model.fit(x_data[:2./3*N,:], y_data[:2./3*N])
prediction = model.predict(x_data[2./3*N+1:,:])

model.fit(x_reduced[:2./3*N,:], y_data[:2./3*N])
prediction_reduced = model.predict(x_reduced[2./3*N+1:,:])

pred_eval = evaluation(y_data[2./3*N+1:], prediction)
pred_reduced_eval = evaluation(y_data[2./3*N+1:], prediction_reduced)
print('prediction: {}\n'.format(evaluation(y_data[2./3*N+1:], prediction)))
print('prediction_reduced: {}\n'.format(evaluation(y_data[2./3*N+1:], prediction_reduced)))

fig,ax = plt.subplots()
plt.axis((0,100,0,100))
ax.scatter(y_data[2./3*N+1:], y_data[2./3*N+1:], s=25, c='b', marker="s", label='prediction')

ax.scatter(y_data[2./3*N+1:], prediction, s=45, c='g', marker="s", label='prediction')
ax.scatter(y_data[2./3*N+1:], prediction_reduced, s=45, c='r', marker="o", label='prediction_reduced')
ax.legend(['True energy consumption', 'Prediction with all parameters', 'Prediction with reduced parameters'], loc='upper left', fontsize='medium')
fig.show()
```

# Appendix B. Simulator detail of Topology Management Algorithm

### A.1 Define node characteristics

```csharp
public class nodeClass
{
        /*Specify number of hops to sink or Level number*/
    public int level;
        /*Array of neighbours who send their packets to this node or in other words
        nodes play as a relay node for these neighbours*/
    public int[] incoming = new int[MAX];
        /*Array of neighbours*/
    public int[] Neighbours = new int[2000];
        //Energy level or buttery power
    public int TotalEnergy = 100000;
    //Node status
    public bool IsDead = false;
    //nodes has a packet to sent as a default
    public int NoPkt = 1;
    //nodes coordination or position
    public int x;
    public int y;

}
//Define Edge between two nodes characteristics
public class Edge
{
        public int from { get; private set; }
    public int to { get; private set; }
        //the cost of sending packet on the edge
    public Double weight { get; private set; }
    public Edge(int from, int to, Double weight)
    {
       this.from = from;
       this.to = to;
       this.weight = weight;
    }
 }
```

### A.2 Define wireless sensor network characteristics

```csharp
public partial class WSN
{
        /* Lists of all nodes and edges */
    private nodeClass[] nodes;

    // specify Transmission Radios
    public int Tx = 300;
        //default value of cost between two typical nodes at the begining
    public int INFINIT = 10000;
```

```csharp
//Number of nodes in the network
public static int MAX =400;
    //Two methods has been tested in this simlation
private static enum method = { PDTMAlgo , DDTMAlgo};
    //Results will be saved in a file
private static String fileResults = "..\\".Method.".txt";
    //specify Number of experiments
private static int experimentCounter = 0;
    //specify Number of packet loss
private static int dropPKt = 0;
    //specify Timer
private System.Diagnostics.Stopwatch ws;

    /*Initialize network */
public WSN()
{
   InitializeComponent();
}

/*Generate A Random positions for nodes in the page*/
public static int GetRandomNumber(int min, int max)
{
   lock (syncLock)
   { // synchronize
      return getrandom.Next(min, max);
   }
}
public void initialize()
{
        //place Sink in a pre specify position as the first node installed in
    the network
   nodes[i].x = 50;
   nodes[i].y = 50;
        /*specify Sink level to be zero*/
   nodes[i].level = 0;
        /*specify number of neighbours who send their packets*/
   nodes[i].outdegree = 0;
   for (i = 1; i < MAX; i++)
   {
            /*place nodes in random position in a 600*400 unit area by
            coordination (X,Y) */
      nodes[i].x = GetRandomNumber(55, 600);//x[i];
      nodes[i].y = GetRandomNumber(55, 400);//y[i];
            //At the beginning we assume node cannot reach to the sink by
            assigning INFNIT to Level
      nodes[i].level = INFINIT;
   }
}
```

## A.3 Create graph of sensors

/*this function organizes the sensor network or in other words a graph of sensor nodes by connecting all nodes that can reach each other based on their transmission radios*

```
/*this function will return False when there is not any node to communicate with sink*/
public int makeGH()
{
   for (i = 0; i < MAX; i++)
   {
      for (int j = 0; j < MAX; j++)
      {
         if (i != j)
         {
            //Calculate distance between nodes i and j
            xij = Math.Pow((nodes[i].x - nodes[j].x), 2);
            yij = Math.Pow((nodes[i].y - nodes[j].y), 2);
                Distance = Math.Sqrt(xij + yij);
            //check if node i can reach node j
            if (Distance<= Tx)
            {
               // j is node i neighbour save j in i's neighbour array
               nodes[i].Neighbours [k] = j;
            }
         }
      }
   }
   //In this loop Level of each node is calculated
   while (track_counter > current)
   {
      // check if node is alive
      if (nodes[i].dead == false)
      {
         for (l = 0; l < nodes[i].neig_counter; l++)
         {
         //check if neighbour is still alive
         if (nodes[nodes[i].Neighbours [l]].dead == false)
         {
            //check if neighbours level is greater

            if(nodes[nodes[i].Neighbours[l]].level>=(nodes[i].level))
            {
               /*choose the lowest level among neigbours as node i level */
               if (nodes[nodes[i].Neighbours [l]].level > (nodes[i].level))
                  nodes[nodes[i].Neighbours [l]].level = nodes[i].level + 1;
               //Method PDTMAlgo
               if (method == PDTMAlgo)
                  nodes[i].outgoing[nodes[i].Neighbours [l]] =
                     Math.Round(Math.Pow(nodes[i].level + 1, 2) *
                     Math.Pow(nodes[i].outgoingDis[nodes[i].Neighbours[l]], 2)
                     * nodes[i].outdegree / (nodes[i].totalEnergy), 4);
               //Method DDTMAlgo
               else
                  nodes[i].outgoing[nodes[i].Neighbours [l]] =
                  Math.Round(nodes[i].outgoingDis[nodes[i].Neighbours [l]], 2);
            } else {
                  nodes[i].incoming[nodes[i].Neighbours [l]] = 1;
```

```csharp
                nodes[i].indegree++;
            }
          }
        }
      }
    }

    return TRUE;
}
```

## A.4  Start Network

```csharp
private void btnRun_Click(object sender, EventArgs e)
{
    while (true)
    {
        //run experiments for density of 100 150 200 250 and 300 nodes in the network
        MAX = MAX + 50;
        if (MAX > 300) break;
        while (true)
        {
            //set transmision radios of nodes for each experiment to 100, 200, 300
            Tx = Tx + 100;
            while (true)
            {
                //run each experiment for N times
                if (exp_counter++ > N)
                    break;
                while (true)
                {
                    //start the timer
                    ws = System.Diagnostics.Stopwatch.StartNew();
                    while (true)
                    {

                        //organize network
                        initialize();
                        makeGH();
                        //if all sink neigbors are dead
                        //stop timer
                        ws.Stop();
                        //Write the results
                        break;
                    }
                    foreach (Edge edge in Edges)
                        /* Runs dijkstra */
                        try
                        {
                            Dijkstra dijk = new Dijkstra(G, 0);
                        }
                        catch (ArgumentException err)
                        {
                            MessageBox.Show(err.Message);
                        }
                        /* every time decrease unit of energy of node with
                           respect of distance  */
```

```csharp
                    nodes[i].totalEnergy -=
                     (int)(Math.Pow((nodes[dpath[i]].x - nodes[i].x), 2) +
                     Math.Pow((nodes[dpath[i]].y - nodes[i].y), 2) *
                            nodes[i].NoPkt);
                    nodes[dpath[i]].NoPkt++;
                    if (nodes[i].totalEnergy <= 0)
                    {
                       dropPKt += (nodes[i].NoPkt - m);
                       nodes[i].dead = true;
                       nodes[i].NoPkt = 0;
                       break;
                       //check if node is dead otherwise generate a
                         packet for the node
                       if (nodes[i].totalEnergy > 0)
                          nodes[i].NoPkt = 1;
                       else nodes[i].IsDead = true;
                          //experiment finished
                          Reset();
                    }
                }
            }
        }
```

## A.5  Dijkstra Algorithm

```csharp
namespace Dijkstra
{
   public class Dijkstra
   {
      /* Takes adjacency matrix in the following format, for a directed graph
         (2-D array)
         * Ex. node 1 to 3 is accessible at a cost of 4
         *     0 1 2 3 4
         * 0 { 0, 2, 5, 0, 0},
         * 1 { 0, 0, 0, 4, 0},
         * 2 { 0, 6, 0, 0, 8},
         * 3 { 0, 0, 0, 0, 9},
         * 4 { 0, 0, 0, 0, 0}
        */

      /* Resulting arrays with distances to nodes and how to get there */
      public double[] dist { get; private set; }
      public int[] path { get; private set; }

      /* Holds queue for the nodes to be evaluated */
      private List<int> queue = new List<int>();

      /* Sets up initial settings */
      private void Initialize(int s, int len)
      {
         dist = new double[len];
         path = new int[len];

         /* Set distance to all nodes to infinity - alternatively use
            Int.MaxValue for use of Int type instead */
```

```csharp
      for (int i = 0; i < len; i++)
      {
         dist[i] = Double.PositiveInfinity;
         queue.Add(i);
      }
      /* Set distance to 0 for starting point and the previous node to null
         (-1) */
      dist[s] = 0;
      path[s] = -1;
   }
   /* Retrives next node to evaluate from the queue */
   private int GetNextVertex()
   {
      double min = Double.PositiveInfinity;
      int Vertex = -1;
      /* Search through queue to find the next node having the smallest
         distance */
      foreach (int j in queue)
      {
         if (dist[j] <= min)
         {
            min = dist[j];
            Vertex = j;
         }
      }
      queue.Remove(Vertex);
      return Vertex;
   }
   /* Takes a graph as input an adjacency matrix (see top for details) and a
      starting node */
   public Dijkstra(double[,] G, int s)
   {
      /* Check graph format and that the graph actually contains something */
      if (G.GetLength(0) < 1 || G.GetLength(0) != G.GetLength(1))
      {
         throw new ArgumentException("Graph error, wrong format or no nodes
                     to compute");
      }
      int len = G.GetLength(0);
      Initialize(s, len);
      while (queue.Count > 0)
      {
        int u = GetNextVertex();
        /* Find the nodes that u connects to and perform relax */
        for (int v = 0; v < len; v++)
        {
          /* Checks for edges with negative weight */
          if (G[u, v] < 0)
          {
             throw new ArgumentException("Graph contains negative
                         edge(s)");
          }

          /* Check for an edge between u and v */
          if (G[u, v] > 0)
          {
            /* Edge exists, relax the edge */
            if (dist[v] > dist[u] + G[u, v])
            {
```

```
                    dist[v] = dist[u] + G[u, v];
                    path[v] = u;
                }
            }
        }
    }
  }
}
```

# References

CodeBlue healthcare project. [Online]. Available: http://fiji.eecs.harvard.edu/CodeBlue/.

http://www.mobilab.unina.it/TinySAN.html [Online].

https://cs.wmich.edu/wsn/ [Online].

Mica2datasheet. [Online]. Available: https://www.eol.ucar.edu/rtf/facilities/isa/internal/CrossBow/DataSheets/mica2.pdf.

Spearman's correlation. http://www.statstutor.ac.uk/resources/uploaded/spearmans.pdf.

April 28, 2013. ZigBee, [Online]. Available: http://www.zigbee.org/.

December 2011. National Center for Health Statistics [Online]. Available: http://www.cdc.gov/nchs/Default.htm.

AKYILDIZ, I. F., SU, W., SANKARASUBRAMANIAM, Y. & CAYIRCI, E. 2002a. Wireless sensor networks: a survey. Comput. Netw., 393-422.

AKYILDIZ, I. F., WEILIAN, S., SANKARASUBRAMANIAM, Y. & CAYIRCI, E. 2002c. A survey on sensor networks. Comm. Mag., 102-114.

AL-KARAKI, J. N. & KAMAL, A. E. 2004. 'Routing techniques in wireless sensor networks: a survey'. IEEE Transaction on wireless communications, 11, 6-28.

ALZOUBI, K. M., WAN, P. & FRIEDER, O. 2002. Distributed Heuristics for Connected Dominating Sets in Wireless Ad Hoc Networks. Journal of Communications and Networks, 4, 22-29.

AMAN KANSAL, FENG ZHAO, JIE LIU, NUPUR KOTHARI & BHATTACHARYA, A. 2010. Virtual Machine Power Metering and Provisioning. ACM Symposium on Cloud Computing (SOCC).

ANASTASI, G., CONTI, M., FRANCESCO, M. D. & PASSARELLA, A. 2009. Energy conservation in wireless sensor networks: A survey. Ad Hoc Netw., 7, 537-568.

ATLA, A., TADA, R., SHENG, V. & SINGIREDDY, N. 2011. Sensitivity of different machine learning algorithms to noise. J. Comput. Sci. Coll., 26, 96-103.

BACCOUR, N., KOUB, A., YOUSSEF, H., JAM, M. B., ROS, D. D., RIO, ALVES, R. & BECKER, L. B. 2010. F-LQE: a fuzzy link quality estimator for wireless sensor networks. Proceedings of the 7th European conference on Wireless Sensor Networks. Coimbra, Portugal: Springer-Verlag.

BERZOSA, J., MABE, J., GRABHAM, N. J. & TUDOR, M. J. 2012. Social issues of power harvesting as key enables of WSN in pervasive computing. 2012 IEEE: International Conference on Pervasive Computing and Communications Workshops (PERCOM Workshops).

BHATTACHARYA, A. & KUMAR, A. 2014. A shortest path tree based algorithm for relay placement in a wireless sensor network and its performance analysis. Comput. Netw., 71, 48-62.

BILLINGS, S. A. & VOON, W. S. F. 1983. Structure detection and model validity tests in the identification of nonlinear systems. Proceeding of the institution of electronic engineers. UK

BILLINGS, S. A. & ZHU, Q. M. 1994. Nonlinear model validation using correlation tests. Int. J. Contr., 60, 1107–1120.

BISWAS, S. & MORRIS, R. 2005. ExOR: opportunistic multi-hop routing for wireless networks. SIGCOMM Comput. Commun. Rev., 35, 133-144.

C.BURATTI, A. C., D.DARDARI, R.VERDONE 2009. An Overview on Wireless Sensor Networks Technology and Evolution. sensors, 9.

CERPA, A. & ESTRIN, D. 2002. ASCENT: Adaptive Self-Configuring sEnsor Network Topologies. Computer Communication Review, 32, 62.

CHACHULSKI, S., JENNINGS, M., KATTI, S. & KATABI, D. 2007. Trading structure for randomness in wireless opportunistic routing. SIGCOMM Comput. Commun. Rev., 37, 169-180.

CHAMP, J., SAAD, C. & BAERT, A.-E. 2009. Lifetime in Wireless Sensor Networks. Complex, Intelligent and Software Intensive Systems, 2009. CISIS '09. International Conference on.

CHU, H.-T., HUANG, C.-C., LIAN, Z.-H. & TSAI, J. J. P. 2006. A ubiquitous warning system for asthma-inducement. IEEE Computer Society.

D. WEI, Y. J., S. VURAL, K. MOESSNER, AND R. TAFAZOLLI. November 2011. An energy-efficient clustering solution for wireless sensor networks. IEEE Transactions on Wireless Communications, 10, 1-11.

DONG, Q., BANERJEE, S., ADLER, M. & MISRA, A. 2005. Minimum energy reliable paths using unreliable wireless links. Proceedings of the 6th ACM international symposium on Mobile ad hoc networking and computing. Urbana-Champaign, IL, USA: ACM.

E. JOVANOV, A. L., D. RASKOVIC, P. COX, R. ADHAMI, F. ANDRASIK, 2003. Stress monitoring using a distributed wireless intelligent sensor system,. IEEE Engineering in Medicine and Biology Magazine.

FENGYUAN REN, J. Z., TAO HE, CHUANG LIN, SAJAL K. DAS, December, 2011 EBRP: Energy-Balanced Routing Protocol for Data Gathering in Wireless Sensor Networks. IEEE Transactions on Parallel and Distributed Systems, , 22, 2108-2125

FOSTER PROVOST, T. F. 2013. Data Science for Business: What you need to know about data mining and data-analytic thinking, O'Reilly Media.

G. VIRONE, A. W., L. SELAVO, Q. CAO, L. FANG, T. DOAN, Z. HE, AND J. A. STANKOVIC 2006. An advanced wireless sensor network for health monitoring. Distributed Diagnosis and Home Healthcare.

GARCIA-LUNA-ACEVES, J. J., MOSKO, M. & PERKINS, C. E. 2006. A new approach to on-demand loop-free routing in networks using sequence numbers. Comput. Netw., 50, 1599-1615.

GHAFFARI, A. 2014. An Energy Efficient Routing Protocol for Wireless Sensor Networks using A-star Algorithm. Journal of Applied Research and Technology, 12, 815-822.

GOLDSMITH, A. J. & WICKER, S. B. 2002. Design Challenges for Energy-Constrained Ad Hoc Wireless Networks. IEEE Wireless Communications Magazine,, 8–27.

H.EDGAR & CALLAWAY, J. 2004. Wireless Sensor Networks.

HALKES, G. P., DAM, T. & LANGENDOEN, K. G. 2005. Comparing energy-saving MAC protocols for wireless sensor networks. Mobile Networks and Applications Journal of Kluwer Academic, 10, 783-791.

HALL, M. A. & SMITH, L. A. 1999. Feature Selection for Machine Learning: Comparing a Correlation-Based Filter Approach to the Wrapper. Proceedings of the Twelfth International Florida Artificial Intelligence Research Society Conference. AAAI Press.

HANDE, A. & CEM, E. 2010. Wireless sensor networks for healthcare: A survey. Comput. Netw. %@ 1389-1286, 54, 2688-2710.

HASTIE T, T. R., FRIEDMAN J 2009. The Elements of Statistical Learning: Data Mining, Inference, and Prediction, Springer.

HEINZELMAN, W. 2000. Application Specific Protocol Architectures for Wireless Networks. Ph.D Thesis, Massachusetts Institute of Technology.

HEINZELMAN, W., KULIK, J. & BALAKRISHNAN, H. Adaptive Protocols for Information Dissemination in Wireless Sensor Networks.  5th ACM/IEEE Mobicom Conference August 1999a Seattle, WA. 174-85.

HEINZELMAN, W. R., CHANDRAKASAN, A. & BALAKRISHNAN, H. 2000. Energy-efficient communication protocol for wireless microsensor networks. 33rd Annual Hawaii International Conference on System Sciences.

HEINZELMAN, W. R., CHANDRAKASAN, A. P. & BALAKRISHNAN, H. 2002. An application specific protocol architecture for wireless microsensor networks. IEEE transactin on Wireless Communications.

HEINZELMAN, W. R., KULIK, J. & BALAKRISHNAN, H. 1999b. Adaptive protocols for information dissemination in wireless sensor networks. Proceedings of the 5th annual ACM/IEEE international conference on Mobile computing and networking. Seattle, Washington, USA: ACM.

HOANG, D. B. Wireless Technologies and Architectures for Health Monitoring Systems. Digital Society, 2007. ICDS '07. First International Conference on the, 2-6 Jan. 2007 2007. 6-6.

HOULE, M. E., KRIEGEL, H.-P., KR, P., GER, SCHUBERT, E. & ZIMEK, A. 2010. Can shared-neighbor distances defeat the curse of dimensionality? Proceedings of the 22nd international conference on Scientific and statistical database management. Heidelberg, Germany: Springer-Verlag.

HSEIN-PING, K. & DO-UN, J. 2009. Wearable patch-type ECG using ubiquitous wireless sensor network for healthcare monitoring application. Proceedings of the 2nd International Conference on Interaction Sciences: Information Technology, Culture and Human. ACM.

HUSSAIN., M. P. S. A. M. Z. 2010. A top-down hierarchical multi-hop secure routing protocol for wireless sensor networks. International Journal of Ad Hoc, Sensor and Ubiquitous Computing,, 1.

IBRAHIEM M. M. EL EMARY, S. R. 2013. Wireless Sensor Networks: From Theory to Applications, CRC Press.

INTANAGONWIWAT, C., GOVINDAN, R. & ESTRIN, D. Directed dicusion: a scalable and robust communication paradigm for sensor networks. Proceedings of ACM MobiCom '00, 2000 Boston, MA. 56-67.

ISLAM, S., KEUNG, J., LEE, K. & LIU, A. 2012. Empirical prediction models for adaptive resource provisioning in the cloud. Future Generation Comp. Syst., 28, 155-162.

J. TANG, A. S. A. H. L. 2014. Feature selection for classification: A review. Data classification: Algorithms and Applications. CRC Press.

J.N.M.VALDEZ. 2011. Wireless Technologies for Indoor Asset Positioning. Master, Tempere University of Technology.

JAYANTHY., M. R. E. J. A. T. 2010. An analysis of various parameters in wireless sensor networks using adaptive FEC technique. . International Journal of Ad Hoc, Sensor and Ubiquitous Computing,, 1, 33-34.

JEONGGIL, K., JONG HYUN, L., YIN, C., RV, ZVAN MUSVALOIU, E., ANDREAS, T., GERALD, M. M., TIA, G., WALT, D., LEO, S. & RICHARD, P. D. 2010. MEDiSN: Medical emergency detection in sensor networks. ACM Trans. Embed. Comput. Syst, 10, 1-29.

JOAQU, J., GARZ, E. & BOUSO, C. 2007. An energy-efficient adaptive modulation suitable for wireless sensor networks with SER and throughput constraints. EURASIP J. Wirel. Commun. Netw., 2007, 23-23.

K. W. GOH, J. L., Y. KIM, E. K. TAN, AND C. B. SOH 2005. A PDA-based ECG beat detector for home cardiac care. In IEEE Engineering in Medicine and Biology Society. Shanghai, China,.

KAMYABPOUR, N. & HOANG, D. B. 2010. A hierarchy energy driven architecture for wireless sensor networks. 24th IEEE International Conference on Advanced Information Networking and Applications (AINA-2010). Perth, Australia: IEEE Computer Society.

KONRAD, L., DAVID, J. M., THADDEUS, R. F. F.-J., ALAN, N., ANTONY, C., VICTOR, S., GEOFFREY, M., MATT, W. & STEVE, M. 2004. Sensor Networks for Emergency Response: Challenges and Opportunities. IEEE Pervasive Computing, 3, 16-23.

KORKALAINEN, M., SALLINEN, M., K, N., RKK, INEN & TUKEVA, P. 2009. Survey of Wireless Sensor Networks Simulation Tools for Demanding Applications. Proceedings of the 2009 Fifth International Conference on Networking and Services. IEEE Computer Society.

KUMAR, S. 2011. NIH GEI project [Online]. The University of Memphis. Available: http://www.cs.memphis.edu/~santosh/AutoSense.html.

KURTIS KREDO, I. & MOHAPATRA, P. 2007. Medium access control in wireless sensor networks. Comput. Netw., 51, 961-994.

LANGENDOEN., T. D. A. K. An adaptive energy-efficient MAC protocol for wireless sensors networks. . In Proceedings of SenSys'03, 2003.

LE, H., HOANG, D. & POLOAH, R. 2008. S-Web: An Efficient and Self-organizing Wireless Sensor Network Model. Lecture Notes in Network-Based Information Systems, 5186, 179-188.

LI, X.-Y., SONG, W.-Z. & WANG, W. 2005. A unified energy-efficient topology for unicast and broadcast. Proceedings of the 11th annual international conference on Mobile computing and networking. Cologne, Germany: ACM.

LIANG, J., WANG, J. & CHEN, J. 2009. A Delay-Constrained and Maximum Lifetime Data Gathering Algorithm for Wireless Sensor Networks. MSN. IEEE Computer Society.

LIU, A., REN, J., LI, X., CHEN, Z. & SHEN, X. 2012. Design principles and improvement of cost function based energy aware routing algorithms for wireless sensor networks. Computer Networks, 56, 1951-1967.

LORINCZ, K., WELSH, M. 2006. MoteTrack: A robust, decentralized approach to RF-based location tracking. . Pers. Ubiquit. Comput., 11.

MAO, K. Z. & BILLINGS, S. A. 2000. Multi-directional model validity tests for non-linear system identification. Int. J. Contr., 73, 132–143.

MAO, X., TANG, S., XU, X. & LI, X.-Y. 2011. Energy Efficient Opportunistic Routing in Wireless Sensor Networks. Parallel and Distributed Systems, IEEE Transactions, 22, 1934-1942.

MEHMET C. VURAN & AKYILDIZ, I. F. XLP: A Cross-Layer Protocol for Efficient Communication in Wireless Sensor Networks. IEEE TRANSACTIONS ON MOBILE COMPUTING, 9.

MIN, R., BHARDWAJ, M., CHO, S.-H., SHIH, E., SINHA, A., WANG, A. & CHANDRAKASAN, A. 2001. Low-power wireless sensor networks. Fourteenth International Conference on VLSI Design.

MINHAS, M. R., GOPALAKRISHNAN, S. & LEUNG, V. C. M. 2009. An Online Multipath Topology management Algorithm for Maximizing Lifetime in Wireless Sensor Networks. Proceedings of the 2009 Sixth International Conference on Information Technology: New Generations. IEEE Computer Society.

MISRA, A. & BANERJEE, S. 2002. MRPC: maximizing network lifetime for reliable routing in wireless environments. WCNC. IEEE.

MUSZNICKI, B., TOMCZAK, M. & ZWIERZYKOWSKI, P. Dijkstra-based Localized Multicast Topology management in Wireless Sensor Networks. 8th IEEE, IET International Symposium on Communication Systems, Networks and Digital Signal Processing, 2012. IEEE.

N.B.RIZVANDI, J.TAHERI & ZOMAYA, A. Y. 2011. Some observations on optimal frequency selection in DVFS-based energy consumption minimization. Journal of Parallel and Distributed Computing, 71, 1154–1164.

NANNICINI, G. & LIBERTI, L. 2008. Shortest paths on dynamic graphs. International Transactions in Operational Research, 15, 551-563.

NORMAN, J., JOSEPH, J. P. & ROJA, P. P. 2010. A Faster Routing Scheme for Stationary Wireless Sensor Networks - A Hybrid Approach. CoRR, abs/1004.0421.

NOURCHENE, B., LAMIA, C. & LOTFI, K. 2011. A Comprehensive Overview of Wireless Body Area Networks WBAN. Int. J. E-Health Med. Commun., 1-30.

OCTAV, C., CHENYANG, L., THOMAS, C. B. & GRUIA-CATALIN, R. 2010. Reliable clinical monitoring using wireless sensor networks: experiences in a step-down hospital unit. Proceedings of the 8th ACM Conference on Embedded Networked Sensor Systems ACM.

P. MOHANTY, S. P., N. SARMA, AND S. S. SATAPATHY. 2010. Security issues in wireless sensor network data gathering protocols: a survey. . Journal of Theoretical and Applied Information Technology, 14–27.

P. SAMUNDISWARY, D. S., AND P. DANANJAYAN 2010. Secured greedy perimeter stateless routing for wireless sensor networks. International Journal of Ad Hoc, Sensor and Ubiquitous Computing, , 1, 9–20.

P.KUMAR, S. G. L., N.J.LEE 2012. E-SAP: Efficient-Strong Authentication Protocol for Healthcare Applications Using Wireless Medical Sensor Networks. sensors, 2, 1625-1647.

POTTIE, G. J. & KAISER, W. J. 2000. Wireless integrated network sensors. Commun. ACM, 43, 51-58.

QIONG, H., BO, D. & SUBIR, B. 2013. Pulse Switching. Wireless Sensor Networks. CRC Press.

RABAEY, J. M., AMMER, M. J., SILVA, J. L. D., PATEL, D. & ROUNDY, S. 2000. PicoRadio Supports Ad Hoc Ultra-Low Power Wireless Networking. Computer, 33, 42-48.

RABBAT, M. & NOWAK, R. D. 2004. Distributed optimization in sensor networks. Proceedings of the Third International Symposium on Information Processing in Sensor Networks (IPSN 2004). Berkeley, CA, USA: ACM.

RAGHUNATHAN, V. & CHOU, P. H. 2006. Design and power management of energy harvesting embedded systems. Proceedings of the 2006 international symposium on Low power electronics and design. Tegernsee, Bavaria, Germany: ACM.

RAGHUNATHAN, V., SCHURGERS, C., PARK, S. & SRIVASTAVA, M. B. 2002. Energy Aware Wireless Microsensor Networks. IEEE Signal Processing Magazine.

RAHMAN, M. A., EL SADDIK, A. & GUEAIEB, W. 2008. Wireless Sensor Network Transport Layer: State of the Art. In: MUKHOPADHYAY, S. & HUANG, R. (eds.) Sensors: Advancement In Modeling, Design Issues, Fabrication And Practical Applications. Springer %8 July.

RAISINGHANI, V. T. & IYER, S. 2004. Cross-Layer Design Optimizations in Wireless Protocol Stacks,. Computer Communications, 27, 720-724.

RAMAKRISHNAN, I. M. M. E. E. A. S. 2013. Applications. Wireless Sensor Networks. CRC Press.

RAMESH, M. V., SREEDEVI, A. G., KANDASAMY, K. & RANGAN, P. V. 2012. Energy comparison of balanced and progressive sensor networks. WOCC. IEEE.

RIZVANDI, N. B., TAHERI, J., MORAVEJI, R. & ZOMAYA, A. Y. Network Load Analysis and Provisioning of MapReduce Applications.  Parallel and Distributed Computing, Applications and Technologies (PDCAT), 2012 13th International Conference on, 14-16 Dec. 2012 2012. 161-166.

S. K. NARANG, G. S., AND A. ORTEGA. March 2010. Unidirectional graph-based wavelet transforms for efficient data gathering in sensor networks. IEEE International Conference on Acoustics Speech and Signal Processing (ICASSP 2010).

S. KUMAR, A. September 8, 2011. http://www.cs.memphis.edu/~santosh/AutoSense.html: NIHGEI project at The University of Memphis, 2007.

S. PARK, W. L., AND D.-H. CHO. 2012. Fair clustering for energy efficiency in a cooperative wireless sensor network. In 75th IEEE Conference on Vehicular Technology.

SALHIEH, A., WEINMANN, J., KOCHHAL, M. & SCHWIEBERT, L. 2001. Power Efficient Topologies for Wireless Sensor Networks. International Conference on Parallel Processing.

SANKARASUBRAMANIAM, Y., ZG,AKAN, R. B. & AKYILDIZ, I. F. 2003. ESRT: event-to-sink reliable transport in wireless sensor networks. Proceedings of the 4th ACM international symposium on Mobile ad hoc networking & computing. Annapolis, Maryland, USA: ACM.

SCHAAR, M. V. D. & SHANKAR, N. S. 2005. Cross layer wireless multimedia transmission challenges, principles and new paradigms. IEEE Wireless Communications Magazine,, 12, 50-58.

SCHEUERMANN, B., LOCHERT, C. & MAUVE, M. 2008. Implicit hop-by-hop congestion control in wireless multihop networks. Ad Hoc Netw., 6, 260-286.

SHAH, R. C. & RABAEY, J. M. 2002. Energy aware routing for low energy ad hoc sensor networks. IEEE Conference on Wireless Communications and Networking.

SHAKKOTTAI, S., RAPPAPORT, T. S. & KARLSSON, P. C. 2003. Cross-Layer Design for Wireless Networks. IEEE Communications Magazine, 41, 74-80.

SHAOQING, W. & JINGNAN, N. 2010. Energy efficiency optimization of cooperative communication in wireless sensor networks. EURASIP J. Wirel. Commun. Netw., 2010, 1-8.

SHARMA, A. 2014. How does random forest work for regression? [Online]. Available: https://www.quora.com/How-does-random-forest-work-for-regression-1.

SRAVAN, A., KUNDU, S. & PAL, A. 2007. Low Power Sensor Node for a Wireless Sensor Network. 20th International IEEE Conference on VLSI Design.

SVEDA, M. & TRCHALIK, R. ZigBee-to-Internet Interconnection Architectures.  Systems, 2007. ICONS '07. Second International Conference on, 22-28 April 2007 2007. 30-30.

TIAN, Y., JARUPAN, B., EKICI, E. & ÖZGÜNER, F. 2006. Real-time task mapping and scheduling for collaborative in-network processing in DVS-enabled wireless sensor networks. . IPDPS. IEEE.

TWALA, B. 2014. Reasoning with Noisy Software Effort Data. Appl. Artif. Intell., 28, 533-554.

WAN, P.-J., CALINESCU, G., LI, X.-Y. & FRIEDER, O. 2002. Minimum-energy broadcast routing in static ad hoc wireless networks. ACM Wireless Networks.

WANG, Q., HEMPSTEAD, M. & YANG, W. A Realistic Power Consumption Model for Wireless Sensor Network Devices. Third Annual IEEE Communications Society Conference on Sensor, Mesh and Ad Hoc Communications and Networks (IEEE SECON 2006), 2006a VA, USA. IEEE, 286-295.

WANG, Q., HEMPSTEAD, M. & YANG, W. A Realistic Power Consumption Model for Wireless Sensor Network Devices. Third Annual IEEE Communications Society Conference on Sensor, Mesh and Ad Hoc Communications and Networks (IEEE SECON 2006), 2006b VA, USA. IEEE, 286-295.

WEDDELL, A. S., HARRIS, N. R. & WHITE, N. M. Alternative Energy Sources for Sensor Nodes: Rationalized Design for Long-Term Deployment. IEEE Instrumentation and Measurement Technology Conference Proceedings, 12-15 May 2008. 1370-1375.

WERNER-ALLEN, G., LORINCZ, K., WELSH, M., MARCILLO, O., JOHNSON, J., RUIZ, M. & LEES, J. 2006. Deploying a Wireless Sensor Network on an Active Volcano. IEEE Internet Computing, 10, 18-25.

WOOD, A., VIRONE, G., DOAN, T., CAO, Q., SELAVO, L., WU, Y., FANG, L., HE, Z., LIN, S., STANKOVIC, J., 2006. ALARM-NET: wireless sensor networks for assisted-living and residential monitoring; . In: CS-2006-01, T. R. (ed.). Department of Computer Science, University of Virginia: Charlottesville, VA, USA.

XU, S. & SAADAWI, T. 2001. Does the IEEE 802.11 MAC protocol work well in multihop wireless ad hoc networks? Comm. Mag., 39, 130-137.

XU, Y., BIEN, S., MORI, Y., HEIDEMANN, J. & ESTRIN, D. 2003. Topology Control Protocols to Conserve Energy in Wireless Ad Hoc Networks. California: Center for Embedded Networked Computing.

YIN, G., YANG, G., WU, Y. & ZHANG, B. 2008. Energy-Efficient Routing Algorithms for Wireless Sensor Networks. Internet Computing in Science and Engineering, 2008. ICICSE '08. Harbin: IEEE.

YOUNG HAN NAM , INTERDISCIPLINARY PROGRAM IN MED., BIOL. ENG. MAJOR, SEOUL NAT. UNIV, SOUTH KOREA , CHEE, Z. H. Y. J. & PARK, K. S. 29 Oct-1 Nov 1998. Development of remote diagnosis system integrating digital telemetry for medicine. Engineering in Medicine and Biology Society, 1998. Proceedings of the 20th Annual International Conference of the IEEE  3

YOUNIS., K. A. A. M. 2005. A survey on routing protocols for wireless sensor networks. The Journal of Ad Hoc Networks, 3, 325-349.

Z. TAN, Y. L., AND Z. ZHANG. March 2011. Performance requirement on energy efficiency in WSNs. In 3rd International IEEE Conference on Computer Research and Development (ICCRD).

ZHANG, L., YANG, W., RAO, Q., NAI, W. & DONG, D. 2013. An Energy Saving Topology management Algorithm Based on Dijkstra in Wireless Sensor Networks. Journal of Information & Computational Science. IEEE Computer Society.

ZWILLINGER, D. A. K., S 2000. CRC Standard Probability and Statistics Tables and Formulae, New York, Chapman & Hall.